\documentclass{article}

\usepackage{hyperref}       
\usepackage{tikz}

\usepackage{url}            
\usepackage{booktabs}       
\usepackage{amsfonts}       
\usepackage{nicefrac}       
\usepackage{microtype}      
\hypersetup{colorlinks,linkcolor={blue},citecolor={blue},urlcolor={blue}}
\usepackage{natbib}
\usepackage{amsfonts}
\usepackage{graphicx,amssymb,mathrsfs,amsmath,color,fancyhdr}
\usepackage{amsthm}
\usepackage{cases}
\usepackage{mathbbol}
\usepackage{geometry}
\usepackage{cite}

\usepackage{latexsym,amsopn,amstext,amsthm,amsxtra,bm,calc,ifpdf}
\usepackage{enumerate}
\usepackage{subfigure}
\usepackage{listings}
\usepackage{enumitem}
\newcommand{\black}{\color{black}}

\newcommand{\indep}{\perp \!\!\! \perp}
\usepackage{threeparttable} 
\def\O{\mathcal{O}}
\def\pr{\textnormal{pr}}
\usepackage[toc,page]{appendix}
\usepackage{authblk}
\hypersetup{
	colorlinks=true,
	linkcolor=black
}
\newcommand\Sset{{\mathcal{S}}} 
 
\usepackage{amsthm,cite,url,graphicx,booktabs,lipsum,color,bm,caption,soul}
\usepackage{colortbl}
\definecolor{c1base}{HTML}{E0115F}   
\definecolor{c2base}{HTML}{50C878}   
\definecolor{c3base}{HTML}{0F52BA}   
\definecolor{c4base}{HTML}{9966CC}   
\definecolor{c5base}{HTML}{FFD700}   
\definecolor{c6base}{HTML}{B9F2FF}   
\definecolor{c7base}{HTML}{FF7F50}   
\definecolor{c8base}{HTML}{40E0D0}   
\definecolor{c9base}{HTML}{DA70D6}   
\definecolor{c10base}{HTML}{FFE4B5}  
\definecolor{c11base}{HTML}{C0C0C0}  
\definecolor{c12base}{HTML}{00A86B}  
\definecolor{c13base}{HTML}{4682B4}  
\definecolor{c14base}{HTML}{DDA0DD}  
\definecolor{c15base}{HTML}{F0E68C}  
\definecolor{c16base}{HTML}{B87333}  

\colorlet{c1}{c1base!60}    
\colorlet{c2}{c2base!60}    
\colorlet{c3}{c3base!60}    
\colorlet{c4}{c4base!60}    
\colorlet{c5}{c5base!60}    
\colorlet{c6}{c6base!60}    
\colorlet{c7}{c7base!60}    
\colorlet{c8}{c8base!60}    
\colorlet{c9}{c9base!60}    
\colorlet{c10}{c10base!60}  
\colorlet{c11}{c11base!60}  
\colorlet{c12}{c12base!60}  
\colorlet{c13}{c13base!60}  
\colorlet{c14}{c14base!60}  
\colorlet{c15}{c15base!60}  
\colorlet{c16}{c16base!60}  
\newtheorem{theorem}{Theorem}
\newtheorem{corollary}{Corollary}

\newtheorem{example}{Example}
\newtheorem{assumption}{Assumption}

\usepackage{caption}
\hypersetup{
  colorlinks=true,
  linkcolor=blue,
  citecolor=blue,
  urlcolor=blue,
  filecolor=blue
}
\def\T{{ \mathrm{\scriptscriptstyle T} }}

\title{{\bf  Assessing Interactive Causes of an Occurred Outcome   Due to Two Binary Exposures}}

\author[1]{Shanshan Luo}
\author[2]{Wei Li}
\author[1]{Xueli Wang}
\author[3]{Shaojie Wei}
\author[1]{Zhi Geng}
\affil[1]{School of Mathematics and Statistics, Beijing Technology and Business University}\vspace{4pt}
\affil[2]{Center for Applied Statistics and School of Statistics, Renmin University of China}\vspace{4pt}
\affil[3]{School of Systems Science and Statistics, Beijing Wuzi University}

\date{} 
\begin{document}

\date{} 
\maketitle
\begin{abstract}In contrast to evaluating treatment effects, causal attribution analysis focuses on identifying the key factors responsible for an observed outcome. For two binary exposure variables and a binary outcome variable, researchers need to assess not only the likelihood that an observed outcome was caused by a particular exposure, but also the likelihood that it resulted from the interaction between the two exposures. For example, in the case of a male worker who smoked, was exposed to asbestos, and developed lung cancer, researchers aim to explore whether the cancer resulted from smoking, asbestos exposure, or their interaction. Even in randomized controlled trials, widely regarded as the gold standard for causal inference, identifying and evaluating retrospective causal interactions between two exposures remains challenging. In this paper, we define posterior probabilities to characterize the interactive causes of an observed outcome. We establish the identifiability of posterior probabilities by using a secondary outcome variable that may appear after the primary outcome.  We apply the proposed method to the classic case of smoking and asbestos exposure. Our results indicate that for lung cancer patients who smoked and were exposed to asbestos, the disease is primarily attributable to the synergistic effect between smoking and asbestos exposure.
\end{abstract}
%
\vspace{1cm} 

\addtocontents{toc}{\protect\setcounter{tocdepth}{-10}}
   \section{Introduction}
\label{sec:intro}
{\black Most statistical research on causal inference focuses on evaluating the effects of causes, which is forward-looking and used for predicting the causal effects of interventions. However, few studies focus on inferring the causes of an already occurred outcome, taking a backward-looking perspective to examine retrospectively why certain outcomes happened \citep{pearl2000causality,dawid2000causal,lu2023evaluating,li2024Biometrika,geng2024prospective,luo2024assessing,zhao2023conditional,zhang2025identifying}. This class of problems is commonly referred to as causal attribution. Importantly, causal effects and causal attribution address fundamentally different questions and can yield contrasting conclusions. For instance, \citet{zhang2025identifying} found that job training has a positive causal effect on income at the population level, yet among individuals who achieved high income after receiving training, the high income is {not} primarily attributed to training—indicating that these high earners would likely have achieved their success even without training.}

To derive the causes of effects in cases involving a binary cause or exposure variable and a binary effect or outcome variable, \citet{pearl2000causality} proposed the probability of necessity (PN), the probability of sufficiency (PS), and the probability of necessity and sufficiency (PNS), while \citet{dawid2000causal} defined the probability of causation (PC). More recently, \citet{lu2023evaluating} and \citet{li2024Biometrika} extended these ideas to multiple binary exposures by proposing posterior effects of causes, and \citet{zhang2025identifying} generalized the probability of necessity to ordinal outcomes. However, these studies do not discuss how the causes interact to bring about the observed outcomes. In the case of two binary exposures, various definitions of interactions are available \citep{miettinen1982causal, darroch1994synergism, vanderweele2007identification, rothman2008modern}. For instance, in statistical models that do not utilize potential outcomes, interaction terms describe the associative relationships between exposures, commonly referred to as statistical interactions. In contrast, frameworks involving potential outcomes allow for the definition of biological or causal interactions, as well as synergistic interactions, which characterize the underlying causal mechanisms among multiple variables.

To explore causal interactions from a retrospective perspective, given the existing evidence of the occurred outcome in the case of two binary exposures and one binary outcome, we need to imagine three other counterfactual outcomes \citep{darroch1994synergism}.  For example, consider a male worker who smoked $(Z=1)$, was exposed to asbestos $(M=1)$, and subsequently developed lung cancer $(Y=1)$. Researchers are interested in assessing the likelihood that the synergistic effect of smoking and asbestos exposure was responsible for his lung cancer. To assess this type of causal interaction, we need to consider three counterfactual scenarios for this worker: if he smoked $(Z=1)$ but was not exposed to asbestos $(M=0)$, if he did not smoke $(Z=0)$ but was exposed to asbestos $(M=1)$, and if he neither smoked $(Z=0)$ nor was exposed to asbestos $(M=0)$.  If this worker would not have developed lung cancer in any of these three counterfactual scenarios, this  would suggest that the observed lung cancer is caused by the interaction of smoking and asbestos exposure.

In this paper, we assess the interactive causes of a binary observed outcome resulting from two binary exposures. We introduce posterior probabilities to evaluate the likelihood that interactive causes led to the observed outcome.  We propose a novel method for identifying the posterior probabilities of various interactions between two binary exposures using a secondary outcome variable \citep{zhang2009likelihood,mealli2012refreshing}.  We apply the proposed framework to evaluate the synergistic effects and perform attribution analysis in the classic case of smoking and asbestos exposure \citep{hilt1986previous,darroch1994synergism,vanderweele2014tutorial}. The likelihood based estimation method and extensive simulation experiments are provided in Supplementary Material.

This paper makes two key contributions beyond existing work. First, it formally characterises interactive causal attribution based on evidence involving post-treatment variables, offering a principled framework for identifying causal interactions between two binary exposures. Second, it addresses a previously unexplored identification concerning the recovery of the joint distribution of the four potential outcomes corresponding to two binary exposures. Most existing approaches can only provide bounds for such distributions \citep{miettinen1982causal,darroch1994synergism,vanderweele2014tutorial}, even under the monotonicity assumption. To the best of our knowledge, this causal attribution framework and identification strategy have not been established in the existing literature.

The rest of the paper is organized as follows. Section \ref{sec:not} introduces the necessary notation and formal definitions. Section \ref{sec:secondary-outcome} develops the identification strategy for posterior probabilities, both under the monotonicity assumption and in more general settings. Section \ref{sec:app-lung} examines synergistic effects and causal attribution in the classic lung cancer case. Section \ref{sec:discussion} concludes with a brief discussion of implications and potential extensions.  Additional technical results and simulation studies are provided in Supplementary Material.
\section{Notation and Definitions}
\label{sec:not}

Throughout this paper, we consider the case of two binary exposure variables and one binary outcome variable. {\black Let $\boldsymbol{X}$ denote the baseline covariates.}  Let $Z$ and $M$ denote two binary causes or exposures,  where $Z=1$ represents smoking, $Z=0$ no smoking, $M=1$ asbestos exposure, and $M=0$ no asbestos  exposure. Let $Y$ be a binary outcome, with $Y=1$ indicating lung cancer and $Y=0$ no lung cancer.  Let $Y_{\black z,m}$ denote  the potential outcome of $Y$ under the exposure levels $Z=z$ and $M=m$. For example, if $Z=1$ (the individual smokes) and $M=1$ (the individual is exposed to asbestos), then $Y_{1,1}$ represents the potential outcome of lung cancer under these conditions, while $Y_{1,0}$ represents the potential outcome if the individual smokes but is not exposed to asbestos. We adopt the consistency assumption, meaning that $Y=Y_{\black z,m}$ when $Z=z$ and $M=m$.   Let $\mathcal{O}=(Z=z,M=m,Y=y)$ denote the observed evidence for $z,m,y\in\{0,1\}$. In this paper, we are interested in analyzing the likelihood that two binary exposures jointly contributed to the development of lung cancer, given the observed evidence $\mathcal{O}$.

Different causal interactions in a population can be quantified by the joint probabilities of potential outcomes under four exposure statuses. To formally assess these interactions, let $ G = (Y_{0,0}, Y_{0,1}, Y_{1,0}, Y_{1,1}) $ represent the vector of joint potential outcomes corresponding to the four combinations of exposures. For simplicity, we denote $ G = (Y_{0,0}, Y_{0,1}, Y_{1,0}, Y_{1,1}) $ as $ G = Y_{0,0} Y_{0,1} Y_{1,0} Y_{1,1} $ throughout this paper. 
  Each class $ G = rstu $, where $ r, s, t, u \in \{0, 1\} $, represents a specific type of causal interaction between    $ Z $ and $ M $ with respect to the binary outcome $ Y $.
As described by \citet{miettinen1982causal}, individuals in the entire population can be categorized into $ 2^4 = 16 $ classes based on $ G = 0000, \ldots, G = 1111 $.  
 For any $r, s, t, u \in \{0,1\}$, the probability of the latent class $G = rstu$ in the  population can be expressed as:
\begin{equation}
    \label{eq:pi-inter}
\pi_{rstu}=\pr( G=rstu)= \pr(Y_{0,0}=r, Y_{0,1}=s, Y_{1,0}=t, Y_{1,1}=u) .
\end{equation}
In this context, the joint probability quantifies how different combinations of smoking and asbestos exposure lead to lung cancer in the  population \citep{miettinen1982causal}.  Table \ref{tab:16-types} presents all 16 possible classes and their corresponding descriptions.  
For example, if $ G = 0001 $, it represents a scenario where lung cancer ($ Y = 1 $) occurs only when both smoking ($ Z = 1 $) and asbestos exposure ($ M = 1 $) are present, indicating that the interaction between smoking and asbestos is both necessary and sufficient to cause lung cancer. In contrast, $ G = 0011 $ indicates that lung cancer ($ Y = 1 $) occurs solely in the presence of smoking ($ Z = 1 $), regardless of asbestos exposure, suggesting that smoking is a necessary and sufficient cause of lung cancer in this scenario. Similarly, $ G = 0101 $ indicates that lung cancer ($ Y = 1 $) occurs only when asbestos exposure ($ M = 1 $) is present. Furthermore, $ G = 0111 $ implies that lung cancer ($ Y = 1 $) occurs in the presence of either smoking ($ Z = 1 $) or asbestos exposure ($ M = 1 $), or both, indicating that either alone is sufficient to cause lung cancer, and that both together will also cause it. 

\begin{table}[h]
\centering
\vspace{2mm}
\caption{16 possible classes for the four potential outcome combinations $G=(Y_{0,0},Y_{0,1},Y_{1,0},Y_{1,1})$, where the classes marked by $^\ast$ do not exist under the monotonicity assumption  \ref{ass:monotonicity}, and the term in parentheses {\black represents} simplified descriptions of the class.
}
\label{tab:16-types}
\resizebox{0.9949790\textwidth}{!}{  
\begin{threeparttable}
\begin{tabular}{cccccccc} 
\addlinespace[2mm]
\toprule
{Type} & $G$ & &{Class Description} \\ \addlinespace[1mm]
\hline\addlinespace[1mm]
1$^{~}$ & (0,0,0,0) && \begin{tabular}[c]{@{}c@{}}Lung cancer never occurs (immune).\end{tabular} \\ \addlinespace[1mm]
\hline\addlinespace[1mm]
2$^{~}$ &  (0,0,0,1)  && \begin{tabular}[c]{@{}c@{}}Lung cancer occurs if and only if smoking with asbestos exposure\\ is present (synergistic).\end{tabular} \\ \addlinespace[1mm]
\hline\addlinespace[1mm]
3$^\ast$ &  (0,0,1,0)  &&  \begin{tabular}[c]{@{}c@{}}Lung cancer occurs if and only if smoking without asbestos exposure is present.\end{tabular} \\ \addlinespace[1mm]
\hline\addlinespace[1mm]
4$^{~}$ & (0,0,1,1)  && \begin{tabular}[c]{@{}c@{}}Lung cancer occurs if and only if smoking is present (smoking).\end{tabular} \\ \addlinespace[1mm]
\hline\addlinespace[1mm]
5$^\ast$ &  (0,1,0,0)  && \begin{tabular}[c]{@{}c@{}}Lung cancer occurs if and only if non-smoking with asbestos exposure is present.\end{tabular} \\ \addlinespace[1mm]
\hline\addlinespace[1mm]
6$^{~}$ & (0,1,0,1) && \begin{tabular}[c]{@{}c@{}}Lung cancer occurs if and only if asbestos exposure is present (asbestos).\end{tabular} \\ \addlinespace[1mm]
\hline\addlinespace[1mm]
7$^\ast$ &  (0,1,1,0) && \begin{tabular}[c]{@{}c@{}}Lung cancer occurs if and only if either non-smoking with asbestos exposure,\\  or smoking without asbestos exposure, is present.\end{tabular} \\ \addlinespace[1mm]
\hline\addlinespace[1mm]\addlinespace[1mm]
8$^{~}$ & (0,1,1,1) & & \begin{tabular}[c]{@{}c@{}} 
Lung cancer occurs if and only if either smoking or asbestos exposure \\ is present (parallel).
\end{tabular}\\ \addlinespace[1mm]
\hline\addlinespace[1mm]
9$^\ast$ &  (1,0,0,0)  && \begin{tabular}[c]{@{}c@{}}Lung cancer disappears if and only if either smoking or asbestos exposure is present.\end{tabular} \\ \addlinespace[1mm]
\hline\addlinespace[1mm]
10$^\ast$  &  (1,0,0,1)&& \begin{tabular}[c]{@{}c@{}}Lung cancer disappears if and only if either non-smoking with asbestos exposure,\\  or smoking without asbestos exposure, is present.\end{tabular} \\ \addlinespace[1mm]
\hline\addlinespace[1mm]
11$^\ast$  &  (1,0,1,0)  && \begin{tabular}[c]{@{}c@{}}Lung cancer disappears if and only if asbestos exposure is present.\end{tabular} \\ \addlinespace[1mm]
\hline\addlinespace[1mm]
12$^\ast$  &  (1,0,1,1)  &&  \begin{tabular}[c]{@{}c@{}}Lung cancer disappears if and only if non-smoking with asbestos exposure is present.\end{tabular} \\\addlinespace[1mm] 
\hline\addlinespace[1mm]
13$^\ast$ & (1,1,0,0) && \begin{tabular}[c]{@{}c@{}}Lung cancer disappears if and only if smoking is present.\end{tabular} \\ \addlinespace[1mm]
\hline\addlinespace[1mm]
14$^\ast$ &  (1,1,0,1) && \begin{tabular}[c]{@{}c@{}}Lung cancer disappears if and only if smoking without asbestos exposure is present.\end{tabular} \\ \addlinespace[1mm]
\hline\addlinespace[1mm]
15$^\ast$ &  (1,1,1,0)&& \begin{tabular}[c]{@{}c@{}}Lung cancer disappears if and only if smoking with asbestos exposure is present.\end{tabular} \\ \addlinespace[1mm]
\hline\addlinespace[1mm]
16$^{~}$ &  (1,1,1,1) && \begin{tabular}[c]{@{}c@{}}Lung cancer always occurs (doomed).\end{tabular} \\ \addlinespace[1mm]
\bottomrule
\end{tabular}
\begin{tablenotes}
  \item   $Z=0:$ non-smoking,~  $Z=1:$ smoking,~ $M=0:$ no asbestos exposure,~
  $M=1:$ asbestos exposure.
\end{tablenotes}
\end{threeparttable}}
\end{table}

In practice, smoking ($Z=1$) is more likely to lead to lung cancer than not smoking, and exposure to asbestos ($M=1$) also increases the likelihood of developing lung cancer compared to no exposure.  This type of monotonicity is commonly characterized by the following assumption  \citep{vanderweele2014tutorial,rothman2008modern}.
\begin{assumption}[Monotonicity]  \label{ass:monotonicity}
$Y_{0 0}\leq Y_{0,1}\leq Y_{1,1}$ and $Y_{0,0}\leq Y_{1,0}\leq Y_{1,1}$.
\end{assumption} 

{\black Assumption \ref{ass:monotonicity} requires that each exposure has a non-negative effect on outcomes for all individuals, while leaving the relationship between $Y_{0,1}$ and $Y_{1,0}$ unrestricted. This is a cross-world assumption involving comparisons across different potential outcomes, which cannot be directly validated from observed data. However, this assumption is widely adopted in the causal inference literature for encouragement designs and noncompliance problems \citep{angrist1996identification,ding2017principal,jiang2021identification}, and it imposes testable restrictions on the observed data distribution in certain cases.} 
Under the monotonicity assumption, the number of latent classes is reduced from 16 to 6 by eliminating the classes marked with the symbol $^\ast$ in Table \ref{tab:16-types}.  In the next section, we will discuss the identification of $\pi_{rstu}$ for any $r,s,t,u\in\{0,1\}$ under the cases where Assumption \ref{ass:monotonicity} holds and does not hold, respectively.

Another commonly used no-confounding assumption for causal inference in the literature is deﬁned below \citep{pearl2014interpretation}. 

\begin{assumption}[No confounding] \label{assump:no-confounding}
For any $z,m\in\{0,1\}$, $(Z,M)\indep Y_{z,m} \mid \boldsymbol{X}$.
\end{assumption}
 
{\black Assumption~\ref{assump:no-confounding} naturally holds in randomized experiments where both $Z$ a an important practical scenarind $M$ are randomized exposures, ensuring no unobserved confounders between the exposures $(Z, M)$ and the outcome $Y$. In observational studies, however, this assumption requires that all common causes of $(Z, M)$ and $Y$ are measured and adjusted for through the covariates $\boldsymbol{X}=\boldsymbol{x}$.} 
For any $z,m \in \{0,1\}$, let $\delta_{\black z,m}(\boldsymbol{x}) = \pr(Y_{\black z,m}=1\mid \boldsymbol{X}=\boldsymbol{x})$ represent the probability that the potential outcome $Y_{\black z,m}$ equals one given $\boldsymbol{X}=\boldsymbol{x}$. Under Assumption \ref{assump:no-confounding}, $\delta_{\black z,m}(\boldsymbol{x})$ can be identified from observed data: $\delta_{\black z,m}(\boldsymbol{x}) = \pr(Y=1 \mid Z=z, M=m, \boldsymbol{X}=\boldsymbol{x})$. For simplicity, we condition all statements on $\boldsymbol{X}$ throughout, and write $\delta_{\black z,m}$ instead of $\delta_{\black z,m}(\boldsymbol{X})$ in the main text. {\black In the Supplementary Material, where we provide more complete discussions on estimation and simulation, we explicitly include $\boldsymbol{X}$ in the notation.}

 Table \ref{tab: probabilities} presents the composition of conditional probabilities for $\delta_{z,m}$ and $1 - \delta_{z,m}$, both under the monotonicity assumption and without it. {\black Additionally, in Section \ref{sec:graph-ill} of the Supplementary Material, we provide visual representations of how the observed evidence $(Z=z, M=m, Y=y)$ links to the latent types $G$.} It is evident that the data structure under the monotonicity assumption is simpler. For example, we have:
\[ 
\begin{gathered}
      \pr(Y =0\mid Z=0,M=0)  = \textstyle\sum_{s,t,u=0,1}\pi_{0stu}, ~~~(\text{without monotonicity}), \\
  \pr(Y =0\mid Z=0,M=0)  = \pi_{0000} + \pi_{0001} + \pi_{0011} + \pi_{0101} + \pi_{0111}, ~~~(\text{with monotonicity}),
\end{gathered} 
\]
where the first equation includes all terms from the third column of the seventh row in Table \ref{tab: probabilities} for $\pr(Y=0\mid Z=0,M=0)$ without the monotonicity assumption, while the second equation includes the terms from the second column of the same row with monotonicity.

\begin{table}[h]
\centering
\vspace{1mm}
\caption{The compositions of conditional probabilities $\pr(Y=y \mid Z=z, M=m)$ for $z,m,y \in \{0,1\}$ with and without the monotonicity assumption  \ref{ass:monotonicity}.}
\label{tab: probabilities}
\resizebox{0.99484975\linewidth}{!}{%
\begin{tabular}{ccclclc}
\addlinespace[2mm]
\toprule
Conditional probability   &  & With monotonicity                              &  & Without  monotonicity             \\ \hline\addlinespace[1mm]$\pr(Y=0\mid Z=1,M=1) $     &  & $ \pi_{0000} $             &  &  $ \pi_{0000}, \pi_{0100}, \pi_{1000}, \pi_{1100}, \pi_{0010}, \pi_{0110}, \pi_{1010}, \pi_{1110} $            \\\addlinespace[0.5mm]
$\pr(Y=1\mid Z=0,M=0) $  &  & $ \pi_{1111} $   &  &  $ \pi_{1000}, \pi_{1001},  \pi_{1010}, \pi_{1011}, \pi_{1100},  \pi_{1101},  \pi_{1110}, \pi_{1111} $                        \\\addlinespace[0.5mm]
$\pr(Y=0\mid Z=0,M=1) $   &  & $ \pi_{0000}, \pi_{0001}, \pi_{0011} $   &  &  $ \pi_{0000}, \pi_{0001},  \pi_{0010}, \pi_{0011}, \pi_{1000},  \pi_{1001},  \pi_{1010}, \pi_{1011} $        \\\addlinespace[0.5mm]
$\pr(Y=0\mid Z=1,M=0) $   &  & $ \pi_{0000}, \pi_{0001}, \pi_{0101} $  &  &  $ \pi_{0000}, \pi_{0001},  \pi_{0100}, \pi_{0101}, \pi_{1000}, \pi_{1001},  \pi_{1100}, \pi_{1101} $              \\\addlinespace[0.5mm]

$\pr(Y=1\mid Z=0,M=1) $  &  & $ \pi_{0101}, \pi_{0111}, \pi_{1111} $          &  &  $ \pi_{0100}, \pi_{0101},  \pi_{0110}, \pi_{0111}, \pi_{1100},  \pi_{1101},  \pi_{1110}, \pi_{1111} $     \\\addlinespace[0.5mm]
$\pr(Y=1\mid Z=1,M=0) $  &  & $ \pi_{0011}, \pi_{0111}, \pi_{1111} $       &  &  $ \pi_{0010}, \pi_{0011},  \pi_{0110}, \pi_{0111}, \pi_{1010}, \pi_{1011},  \pi_{1110}, \pi_{1111} $        \\\addlinespace[0.5mm]
$\pr(Y=0\mid Z=0,M=0) $   &  & $ \pi_{0000}, \pi_{0001}, \pi_{0011}, \pi_{0101},  \pi_{0111} $   &  &  $ \pi_{0000}, \pi_{0001},  \pi_{0010}, \pi_{0011}, \pi_{0100},  \pi_{0101},  \pi_{0110}, \pi_{0111} $\\\addlinespace[0.5mm]
$\pr(Y=1\mid Z=1,M=1) $     &  & $\pi_{0001}, \pi_{0011}, \pi_{0101}, \pi_{0111}, \pi_{1111}$ &  & $ \pi_{0001}, \pi_{0101}, \pi_{1001}, \pi_{1101}, \pi_{0011}, \pi_{0111}, \pi_{1011}, \pi_{1111} $   \\  \bottomrule
\end{tabular}

}
\end{table}

In addition to measuring causal interactions at the population level through the probabilities $\pi_{rstu}$, we can also perform  causal attribution analysis   by conditioning on observed evidence. The posterior probability of the latent class $G=rstu$ conditional on the observed evidence $\mathcal{O}=(Z=z,M=m,Y=y)$ for $z,m,y
\in\{0,1\}$ is defined as
\begin{eqnarray*}
\pi_{rstu} (\mathcal{O}) =  \pr(Y_{0,0}=r, Y_{0,1}=s, Y_{1,0}=t, Y_{1,1}=u \mid \mathcal{O}).
\end{eqnarray*} 
If the observed evidence $\mathcal{O}$ is empty, define $\pi_{rstu}(\mathcal{O}) = \pi_{rstu}$ without loss of generality,  which characterizes the causal interactions for the entire population as shown in \eqref{eq:pi-inter}.  More generally, in attribution analysis, if the observed evidence is $\mathcal{O} = (Z=1, M=1, Y=1)$, the posterior probability $\pi_{rstu}(\mathcal{O})$ can be used to analyze the interactive causes of the observed lung cancer ($Y=1$). For example, $\pi_{0001}(\mathcal{O})$ represents the probability that the observed lung cancer in the population $\mathcal{O}$ is due to the synergistic effect of smoking and asbestos exposure, $\pi_{0011}(\mathcal{O})$ represents the probability that the lung cancer in the population $\mathcal{O}$ is due to smoking alone, and $\pi_{0101}(\mathcal{O})$ represents the probability that the lung cancer in the population $\mathcal{O}$ is due to asbestos exposure alone. The following example further illustrates the potential applications of these causal interactions based on observed evidence.

\begin{example} \label{exm:ex3}  
For lung cancer patients who have both smoked and been exposed to asbestos, we  discuss the proportion of responsibility between the tobacco company and the asbestos company. This scenario is consistent with the observed evidence $\mathcal{O} = (Z =1, M = 1,Y = 1)$. We consider the monotonicity assumption \ref{ass:monotonicity}, and Table \ref{tab:payment} shows the values of $\pi_{rstu}(\mathcal{O})$. The setup for the values in the first column is detailed in Table \ref{tab:est-res} of Section \ref{sec:app-lung}. The second column displays the causes of the disease, while the third and fourth columns show the responsibility proportions for the asbestos and tobacco companies, respectively. For example, $G={0001}$ represents cases where both smoking and asbestos exposure are jointly necessary and sufficient causes of lung cancer, as shown in the second row of Table \ref{tab:payment}, with equal responsibility (50\%) for both companies.

  Thus, the responsibilities (or attributions) for lung cancer caused solely by asbestos exposure and solely by smoking are, respectively:
\begin{gather*}
(\textit{asbestos dust})~~71.01\% \times 50\% +  14.16\% \times 100\%  + 4.35\% \times 50\%  =51.84\%,\\(\textit{smoking})
~~71.01\% \times 50\% +  7.84\% \times 100\% +4.35\% \times 50\%  =45.52\%.
\end{gather*}
Thus, the remaining $100\% - 51.89\% - 45.67\% = 2.64\%$ of responsibility is attributed to other risk factors, such as age and genetic factors. When further considering compensation for asbestos workers with lung cancer, $51.84\%$ of the loss due to cancer should be covered by asbestos exposure compensation, based on the responsibility attributed to asbestos exposure.
 
\begin{table}[h]
    \centering
    \caption{The posterior probabilities $\pi_{rstu}(\O)$ given the evidence $\O = (Z=1,M=1,Y=1).$}
\label{tab:payment}
\resizebox{0.94905\textwidth}{!}{
\begin{threeparttable}
\centering
\begin{tabular}{ccccc}
\addlinespace[2mm]
\toprule
  $\pi_{rstu}(\O)$        &  & Causes of lung cancer            &  \begin{tabular}[c]{@{}c@{}}Attributed to\\\addlinespace[-0.25mm]smoking\end{tabular}        & \begin{tabular}[c]{@{}c@{}}Attributed to\\\addlinespace[-0.25mm]asbestos \end{tabular}  \\ \addlinespace[1mm]\hline\addlinespace[1mm]
$\pi_{0000}(\O)=0\%$      &  & \begin{tabular}[c]{@{}c@{}}Lung cancer never occurs (immune)\end{tabular}      & $0\% $                 & $0\% $                \\\addlinespace[1mm]
\hline\addlinespace[1mm]
$\pi_{0001}(\O)=71.01\%$ &  & \begin{tabular}[c]{@{}c@{}}Lung cancer occurs if and only if smoking with\\  asbestos exposure is present (synergistic)\end{tabular} & $50\%$                 & $50\%$                \\\addlinespace[1mm]
\hline\addlinespace[1mm]
$\pi_{0011}(\O)=14.16 \%$ &  &  \begin{tabular}[c]{@{}c@{}}Lung cancer occurs if and only if  smoking\\  is present (smoking) \end{tabular}         & $100\%$                 & $ 0\%$               \\\addlinespace[1mm]
\hline\addlinespace[1mm]
$\pi_{0101}(\O)=7.84\%$ &  &  \begin{tabular}[c]{@{}c@{}}Lung cancer occurs if and only if asbestos exposure\\ is present (asbestos)\end{tabular}          & $0\%$                & $100\%$                 \\\addlinespace[1mm]
\hline\addlinespace[1mm]
$\pi_{0111}(\O)=4.35 \%$  &  & \begin{tabular}[c]{@{}c@{}}Lung cancer occurs if and only if either smoking  \\or asbestos exposure is present (parallel) \end{tabular}      & $50\%$                 & $50\% $               \\\addlinespace[1mm]
\hline\addlinespace[1mm]
$\pi_{1111}(\O)=2.64 \%$  &  & \begin{tabular}[c]{@{}c@{}}Lung cancer always occurs (doomed)\end{tabular}    & $0\% $                 & $0\% $                \\ \addlinespace[1mm]\toprule
\end{tabular}
 \end{threeparttable}}
\end{table}
\end{example}
 
 {\black   Previous literature has extensively discussed the total posterior causal effect conditional on the observed evidence \citep{lu2023evaluating,li2024Biometrika}.  However, these methods cannot decompose the total effect to quantify the specific responsibility that each cause should bear when multiple causes jointly contribute to an outcome. Our proposed framework addresses this gap by providing a quantitative decomposition of responsibility across two causes.} The proposed posterior probabilities have important applications in epidemiology, legal sentencing, fairness assessments, and medical compensation \citep{dawid2014fitting,vanderweele2014tutorial,rothman2008modern,egami2019causal,imai2023principal,ben2024policy}.  Although the synergistic effects have been extensively studied,  for instance, \citet{miettinen1982causal} explored the nonparametric bounds for $\pi_{rstu}$, and \citet{darroch1994synergism} investigated the estimation procedure using the maximum entropy method, neither study established the nonparametric identifiability of $\pi_{rstu}$ or $\pi_{rstu}(\mathcal{O})$.    Even under the monotonicity assumption  \ref{ass:monotonicity} and  non-confounding assumption \ref{assump:no-confounding}, the posterior probabilities $\pi_{rstu}$ and $\pi_{rstu}(\mathcal{O})$ are not identifiable without additional assumptions. 

\section{Identification of Posterior Probabilities} \label{sec:secondary-outcome}
\subsection{Identification with the monotonicity assumption}\label{ssec:mon-results}
In this subsection, we propose a parametric model-based approach for identifying the posterior probabilities, utilizing an additional secondary outcome \citep{Fabrizia2013JASA} under the monotonicity assumption \ref{ass:monotonicity}. In the next subsection, we explore identification without the monotonicity assumption by imposing additional restrictions on the distinguishability of latent classes. For any  $r,s,t,u\in\{0,1\}$, we have
  \begin{align*}   
     \pi_{rstu} (\O)=          \mathbb{I}( Y=1)  \pi_{1stu}\big/\delta_{0,0}+  \mathbb{I}( Y=0)  \pi_{0stu}\big/(1-\delta_{0,0}),~~\text{if}~\O=(Z=0,M=0,Y),\\
     \pi_{rstu} (\O)=   \mathbb{I}( Y=1)  \pi_{r1tu}\big/\delta_{0,1}+  \mathbb{I}( Y=0)  \pi_{r0tu}\big/(1-\delta_{0,1}),~~\text{if}~\O=(Z=0,M=1,Y),\\
     \pi_{rstu} (\O)=   \mathbb{I}( Y=1)  \pi_{rs1 u}\big/\delta_{1,0}+  \mathbb{I}( Y=0)  \pi_{rs0u}\big/(1-\delta_{1,0}),~~\text{if}~\O=(Z=1,M=0,Y ),\\
     \pi_{rstu} (\O)=   \mathbb{I}( Y=1)  \pi_{rst 1}\big/\delta_{1,1}+  \mathbb{I}( Y=0)  \pi_{rst0}\big/(1-\delta_{1,1}),~~\text{if}~\O=(Z=1,M=1,Y ), 
  \end{align*} where the indicator function $\mathbb{I}(Y=y)$ equals 1 when $Y=y$, and 0 otherwise for $y \in \{0,1\}$. Therefore, the identifiability of $\pi_{rstu}$ sufficiently guarantees the identifiability of $\pi_{rstu}(\mathcal{O})$ given the evidence set  $\mathcal{O} = (Z=z, M=m, Y=y)$. It is evident from Table \ref{tab: probabilities} that under Assumptions \ref{ass:monotonicity} and \ref{assump:no-confounding}, both $\pi_{0000}$ and $\pi_{1111}$ are identifiable from $ \pi_{0000}=1-\delta_{1,1}$ and $\pi_{1111}=\delta_{0,0}$. However, the other probabilities remain unidentifiable.  Thus, we introduce a secondary outcome $W$, satisfying the following parametric model assumption.

\begin{assumption}[Normality]  \label{assumption:normal}
(i)
Given $(Z=z, M=m, G=rstu)$, the secondary outcome $W$ has density $\phi(w; \mu_{z,m,rstu}, \sigma^2_{z,m,rstu})$, where $\phi(w;\mu, \sigma^2)$ denotes the normal density with mean $\mu$ and variance $\sigma^2$. (ii) We assume that for any $z$, $m$, and $r_1s_1t_1u_1 \neq r_2s_2t_2u_2$, the parameter pairs $(\mu_{z,m,r_1s_1t_1u_1}, \sigma_{z,m,r_1s_1t_1u_1}^2)$ and $(\mu_{z,m,r_2s_2t_2u_2}, \sigma_{z,m,r_2s_2t_2u_2}^2)$ are distinct.
  
\end{assumption}

The normality condition in Assumption \ref{assumption:normal}(i) is widely used in causal inference for identifying causal estimands, including missing data \citep{miao2016identifiability}, truncation by death \citep{zhang2009likelihood}, non-compliance \citep{jiang2021identification, bia2022assessing}, and unmeasured confounding \citep{miao2023identifying, shuai2024identifying}. {\black In practice,} the secondary outcome variable $W$  required in Assumption \ref{assumption:normal}  is typically collected after the two exposure variables $(Z, M)$ and may even be recorded after the primary outcome variable $Y$, which is often seen in clinical trials \citep{ding2011identifiability,wang2017identification} or labor economics \citep{zhang2009likelihood,mealli2012refreshing}.   For example, in the context of asbestos and smoking, a secondary outcome variable $W$, such as quality of life score or body weight, can be used to reflect the patient's overall health condition.  

{\black As illustrated in Example 3.1.4 of \citet{Titterington1985}, Assumption \ref{assumption:normal}(ii) ensures the sufficient separation of component distributions, which guarantees the separability of conditional distributions $f(W \mid Z=z, M=m, Y=y, G=rstu)$ across latent classes and the identifiability of mixture weights $\pr(G=rstu\mid Z=z,M=m,Y=y)$. When these component distributions are sufficiently distinct, the class probabilities $\pi_{rstu}$ become identifiable through their connection to these mixture weights. Notably, this distinctiveness need not occur simultaneously in both location (means) and scale (variances). Differences in either location or scale alone can suffice for identifiability, provided the parameter pairs remain distinct across mixture components. While we adopt normality for identification, our conclusions extend to general identifiable mixture distributions \citep{Titterington1985}, and numerical studies  on $\pi_{rstu}(\mathcal{O})$ in Section \ref{sec:sim-detail} of the Supplementary Material suggest robustness to normality violations.}

When Assumption \ref{assumption:normal}(ii) is violated, for instance, when two classes share identical parameters, the nominal six classes effectively collapse into fewer components. To assess whether this assumption holds empirically, one can test $H_0$: the number of classes $k_0 < 6$ (fewer classes) versus $H_1$: the number of classes $k_1 = 6$ (six distinct classes). This can be done using finite mixture model tests, such as the likelihood ratio tests of \citet{LoMendellRubin2001} and \citet{Kasahara02102015}, which provide a formal framework for comparing models with different numbers of components. Rejection of the null hypothesis indicates that the data support the full six-class model, suggesting Assumption \ref{assumption:normal}(ii) holds. Conversely, failure to reject the null suggests some classes share identical parameters and should be merged, reducing the effective number of latent classes.

\begin{theorem} \label{thm:identification}
Under Assumptions \ref{ass:monotonicity}, \ref{assump:no-confounding}, and \ref{assumption:normal}, the posterior probabilities $\pi_{rstu}$ and $\pi_{rstu}(\O)$ are identifiable for $r, s, t, u \in \{0,1\}$, where $\O = (Z=z, M=m, Y=y)$ with $z, m, y \in \{0,1\}$.
\end{theorem}

The proof of Theorem \ref{thm:identification} is based on the identifiability of  Gaussian mixture model (see Example 3.1.4 in \citet{Titterington1985}). Although we assume a normal distribution for $f(W\mid Z=z,M=m, G=rstu)$ in Assumption \ref{assumption:normal}, it is important to note that no parametric restrictions are imposed on the posterior probabilities $\pi_{rstu}$ and $\pi_{rstu}(\O)$   for $r, s, t, u \in \{0,1\}$, indicating that these probabilities are, to some extent, nonparametrically identified. The normal distribution in Assumption \ref{assumption:normal} of Theorem \ref{thm:identification} can be replaced by other distributions, such as the Gamma distribution (see Example 3.1.5 in \citet{Titterington1985}) or the Binomial distribution (see Example 3.1.6 in \citet{Titterington1985}).

For some scientific questions, researchers may benefit from incorporating the secondary outcome $W$ in attribution analyses. For example, practitioners may want to retrospectively analyze whether lung cancer patients who have smoked and been exposed to asbestos show heterogeneity in attribution due to differences in the secondary outcome variable, such as body weight. This requires us to consider the identifiability of the causal parameter $\pr(G = {rst1} \mid Z = 1, M = 1,Y = 1, W)$. To simplify the exposition, we define an extended evidence set $\tilde \O$, which includes the secondary outcome $W$ along with the original evidence $\O=(Z=z, M=m, Y=y)$ for $z, m, y \in \{0,1\}$, as $\tilde \O = (O, W)$. Thus, we have:  
\begin{align*}   \begin{aligned}
     \pi_{rstu} (\tilde\O)= \left\{\begin{matrix}
      \dfrac{\sum_{y=0}^1  \mathbb{I}( Y=y)f(W\mid \O,G=ystu)\pi_{ystu}( \O)}{f(W\mid \O)},~ \text{if}~\O=(Z=0,M=0,Y ),\\\addlinespace[1mm]
   \dfrac{\sum_{y=0}^1 \mathbb{I}( Y=y)f(W\mid \O,G=rytu)\pi_{rytu}( \O)} {f(W\mid \O)},~  \text{if}~\O=(Z=0,M=1,Y ),\\\addlinespace[1mm]
  \dfrac{\sum_{y=0}^1  \mathbb{I}( Y=y)f(W\mid \O,G=rsyu)\pi_{rsyu}( \O)}{f(W\mid \O)},~ \text{if}~\O=(Z=1,M=0,Y ),\\\addlinespace[1mm]
   \sum_{y=0}^1   \dfrac{\mathbb{I}( Y=y)f(W\mid \O,G=rsty)\pi_{rsty}( \O) }{f(W\mid \O)},~ \text{if}~\O=(Z=1,M=1,Y ).
     \end{matrix}\right.
 \end{aligned} 
\end{align*}
The above equation shows that, based on Theorem \ref{thm:identification}, to identify  posterior probabilities  $\pi_{rstu}(\tilde\O)$, we only need to identify the conditional probabilities $f(W \mid \O, G = rstu)$. We summarize the identifiability of $\pi_{rstu} (\tilde\O)$ under the monotonicity assumption as follows:

 \begin{corollary}
 \label{coro:equal-probs}
Under the assumptions of Theorem \ref{thm:identification}, for any $r, s, t, u \in \{0,1\}$, the conditional probabilities $f(W \mid \O, G = rstu)$ and posterior probabilities $\pi_{rstu}(\tilde\O)$ are identifiable if $\pi_{r_1s_1t_1u_1} \neq \pi_{r_2s_2t_2u_2}$ for any $r_1s_1t_1u_1 \neq r_2s_2t_2u_2$, where  $\O = (Z=z, M=m, Y=y)$  and $\tilde\O = (Z=z, M=m, Y=y, W=w)$ with $z, m, y \in \{0,1\}$.
 \end{corollary}
 

Under the assumptions of Theorem \ref{thm:identification}, we can identify $\pi_{rstu}$, allowing us to test whether the condition $\pi_{r_1s_1t_1u_1} \neq \pi_{r_2s_2t_2u_2}$ holds. If distinctiveness is established, that is, if no two classes share the same probability, then Corollary \ref{coro:equal-probs} allows us to incorporate the secondary outcome $W$  to investigate heterogeneity in causal attribution across subpopulations.

{\black Based on the identifiability results under the monotonicity assumption \ref{ass:monotonicity}, we provide an Expectation-Maximization (EM) algorithm in Section \ref{ssec: estimation} of the Supplementary Material to obtain maximum likelihood estimates (MLEs), treating the latent class $G$ as missing data \citep{zhang2009likelihood,ding2017principal}. Additional simulation studies for $\pi(\O)$ and $\pi(\tilde\O)$ related to Assumption \ref{assumption:normal}, as well as robustness checks under violations of this assumption, are provided in Sections \ref{ssec:results-pi-O} and  \ref{ssec:results-pi-tildeO} of the Supplementary Material.  The results suggest that estimation of $\pi(\O)$ is generally robust to violations of Assumption \ref{assumption:normal}, while estimation of $\pi(\tilde\O)$ shows more sensitivity to departures from normality.}

\subsection{Identification without  monotonicity assumption} \label{ssec:iden-without-mon}

The monotonicity assumption \ref{ass:monotonicity} requires that neither exposure $Z$ nor $M$ has a negative effect on the outcome variable $Y$ for any individual. However, in many applications, certain exposures may benefit some individuals while harming others. Therefore, the monotonicity assumption \ref{ass:monotonicity} may not hold in these cases. To address this issue, we propose the following alternative assumption for identification.

\begin{assumption}[Distinguishability of latent classes]    \label{assump:equal-prop} 
For any $z,m\in\{0,1\}$, at least {\it one} of the following three conditions holds: (i) $\pi_{r_1s_1t_1u_1} \neq \pi_{r_2s_2t_2u_2}$ for any distinct $r_1s_1t_1u_1 \neq r_2s_2t_2u_2$; (ii) the mean parameters $\mu_{\black z,m,rstu} = \mu_{rstu}$, as defined in Assumption \ref{assumption:normal}, depend solely on $G = rstu$ and satisfy $\mu_{r_1s_1t_1u_1} \neq \mu_{r_2s_2t_2u_2}$ for any $r_1s_1t_1u_1 \neq r_2s_2t_2u_2$; and (iii) the variance parameters $\sigma^2_{\black z,m,rstu} = \sigma^2_{rstu}$, as defined in Assumption \ref{assumption:normal}, depend solely on $G = rstu$ and satisfy $\sigma^2_{r_1s_1t_1u_1} \neq \sigma^2_{r_2s_2t_2u_2}$ for any $r_1s_1t_1u_1 \neq r_2s_2t_2u_2$.
\end{assumption}

{\black The three conditions in Assumption \ref{assump:equal-prop} ensure that posterior probabilities remain identifiable without monotonicity assumptions. These conditions impose constraints on the mixing proportions, mean parameters, and variance parameters, respectively.   {\black A key practical advantage is that identification can be achieved through any single one of these three conditions, allowing researchers to select the most appropriate condition for their specific application.} 
Assumption \ref{assump:equal-prop}(i) requires $\pi_{r_1s_1t_1u_1} \neq \pi_{r_2s_2t_2u_2}$ for $r_1s_1t_1u_1 \neq r_2s_2t_2u_2$, ensuring distinct probabilities for different latent classes in Table \ref{tab:16-types}. While this assumption is testable in Corollary \ref{coro:equal-probs}, here it serves as a prerequisite. 
Assumption \ref{assump:equal-prop}(ii) specifies that mean parameters $\mu_{z,m,rstu}$ depend solely on $G=rstu$, not on $Z$ or $M$, and differ across latent classes. Without monotonicity, there are 16 distinct mean parameters, with no restrictions on variance parameters $\sigma^2_{z,m,rstu}$. 
Assumption \ref{assump:equal-prop}(iii), parallel to (ii), requires that variance parameters $\sigma^2_{z,m,rstu}$ depend solely on $G=rstu$, yielding 16 distinct variances with no restrictions on means.}

{\black Assumption \ref{assump:equal-prop} provides three alternative conditions, offering flexible strategies for identifying latent class proportions and selecting an appropriate secondary outcome $W$. Importantly, only one of these three conditions needs to hold for identification.   
Assumption \ref{assump:equal-prop}(i) requires that the proportions $\pi_{rstu}$ differ across different latent classes. In most practical settings, different causal types naturally occur with different frequencies in the population, making this assumption quite plausible. Conversely, assuming that all or some latent classes have equal proportions would be restrictive and often implausible. Therefore, unless there is strong prior knowledge supporting the equal proportions across latent classes, Assumption \ref{assump:equal-prop}(i) provides a realistic foundation for selection. Under this condition, $W$ can be a downstream variable that is influenced by the exposures $(Z, M)$, as long as it still carries information about the latent class $G$.    
In contrast, Assumptions \ref{assump:equal-prop}(ii) and (iii) are applicable when we lack external information about whether $\pi_{rstu}$ values differ across latent classes, but can identify a proxy variable $W$ that satisfies an exclusion restriction. These conditions require that the conditional distribution of $W$ given $G$ differs across latent classes through either the mean or variance, and critically, that $W$ is independent of the exposures $(Z, M)$ given $G$. In this case, one should choose $W$ as a pure proxy variable that is only directly affected by $G$.}
\begin{theorem}
\label{thm: identification-nonmo}
Given Assumptions \ref{assump:no-confounding},    \ref{assumption:normal} and \ref{assump:equal-prop},  for any $r,s,t,u\in\{0,1\}$, the conditional  probabilities $\pi_{rstu}$,    $\pi_{rstu}(\O) $ and $\pi_{rstu}(\tilde\O) $      are  identifiable, where $\tilde\O = (Z=z, M=m, Y=y, W )$ and $\O = (Z=z, M=m, Y=y)$ with $z, m, y \in \{0,1\}$.
\end{theorem}

Compared to Theorem \ref{thm:identification} and Corollary \ref{coro:equal-probs},   Theorem \ref{thm: identification-nonmo} establishes  identifiability results by leveraging more restrictions in Assumption \ref{assump:equal-prop} on the mixing proportions, mean parameters, or  variances. In addition to the identifiability of  $\pi_{rstu}(\O) $, we also establish the identifiability of  $\pi_{rstu}(\tilde\O) $  in Theorem \ref{thm: identification-nonmo}, which allows us to address more general causal attribution problems. {\black Parallel to Corollary \ref{coro:equal-probs}, we can also establish the identification results for the conditional probabilities $f(W \mid \O, G = rstu)$, we omit it for simplicity.} Based on the identifiability results in Theorem \ref{thm: identification-nonmo},  we also provide the EM algorithm without the monotonicity assumption in Section \ref{ssec: estimation-nomo} of the Supplementary Material.

\section{Analysis of Interactive  Effects and Causal Attribution in Lung Cancer}
\label{sec:app-lung}
\subsection{Descriptive and review analysis} 
{\black In this section, we apply our framework to the classic smoking-asbestos case. This example is particularly significant as it represents an important practical scenario in occupational health and asbestos litigation, where determining cause-specific attribution, rather than just causal effects, is essential for allocating legal liability and workers' compensation (as shown in Example \ref{exm:ex3}). The underlying data of this case are derived from actual studies and supported by an extensive body of research literature \citep{hilt1986previous, darroch1994synergism, vanderweele2014tutorial}, providing a reliable foundation for investigating interactive attribution probabilities.   To demonstrate the broader scalability of the proposed method, we present in Section \ref{sec:example-cov-extend} of the Supplementary Material another example within the same smoking and asbestos context that incorporates covariates such as genotype, drinking history, and age, as well as a secondary outcome variable measuring quality of life. A  real-data analysis concerning the attribution of depression to employment and education is provided in Section \ref{ssec:app-job} of the Supplementary Material.}


During the selective tuberculosis screening conducted in Telemark County, {\black Norway,} from 1982 to 1983, residents from nine municipalities were also screened for occupational asbestos exposure and related diseases.  All eligible males received a questionnaire regarding their exposure to asbestos dust in the workplace. Detailed information on smoking habits, including past and present behavior, was also collected. 
In this study, we classify smokers as $Z=1$ and non-smokers as $Z=0$. Participants exposed to asbestos are classified as $M=1$, while those not exposed are classified as $M=0$. Here, $Y=1$ represents individuals with lung cancer, and $Y=0$ represents those without lung cancer.  

{\black For the summary data from \citet{vanderweele2014tutorial} (Table 1) and \citet{hilt1986previous} (Figure 2), summarized in Table \ref{tab:summary-stat} of the Supplementary Material, where the lung cancer incidence rates are $\delta_{0,0} = 6/5057 \approx 0.12\%$ for non-smokers without asbestos exposure, $\delta_{0,1} = 5/749 \approx 0.67\%$ for non-smokers with asbestos exposure, $\delta_{1,0} = 118/12383 \approx 0.95\%$ for smokers without asbestos exposure, and $\delta_{1,1} = 141/3130 \approx 4.51\%$ for smokers with asbestos exposure. We first consider the nonparametric bounds under Assumption \ref{ass:monotonicity} \citep{miettinen1982causal,vanderweele2007identification}.} 
The resulting bounds on the posterior probabilities $\pi_{rstu}(\O)$, when $\O$ is the empty set and when $\O = (Z=1,M=1,Y=1)$, are presented in Tables \ref{tab:methods-bound}(a) and \ref{tab:methods-bound}(b), respectively. It is observed that the class $G=0000$ has the largest proportion, with $\pi_{0000} = 95.50\%$, indicating that approximately 95.5\% of the population would not develop lung cancer regardless of smoking or asbestos exposure. In contrast, the class $G=1111$ has a proportion of $\pi_{1111} = 0.12\%$, suggesting that approximately 0.12\% of the population will develop lung cancer irrespective of smoking or asbestos exposure. Additionally, the bound $3.00\% \leq \pi_{0001} \leq 3.55\%$ indicates the existence of individuals experiencing a synergistic effect, while the bound $\black 0.28\% \leq \pi_{0011} \leq 0.83\%$ suggests the presence of individuals for whom lung cancer is caused solely by smoking. Given the evidence set $\O = (Z=1, M=1, Y=1)$, we note that the proportion of the latent class $G=0001$ is the highest, indicating that lung cancer in $\O $ is most likely caused by the synergistic effect of smoking and asbestos exposure.

\begin{table}[h]
    \centering
\vspace{1mm}
\caption{The nonparametric  bounds and maximum entropy estimates for posterior probabilities $\pi_{rstu}(\O)$, where $\pi_{0000}$ and $\pi_{1111}$ are identified under the monotonicity assumption \ref{ass:monotonicity} as $\pi_{0000} = 1 - \delta_{1,1}$ and $\pi_{1111} = \delta_{0,0}$, respectively.}
    \label{tab:methods-bound}\label{tab:methods-entropy} 
    \resizebox{1\linewidth}{!}{%
\setlength{\arrayrulewidth}{1.25pt} \begin{tabular}{ccc|ccc}
\multicolumn{3}{c|}{(a) Bound for $\O=\emptyset$}                                                    & \multicolumn{3}{c}{(c) Maximum entropy method for $\O=\emptyset$}         \\
\multicolumn{3}{l|}{}                                                                                & \multicolumn{3}{l}{}                                                      \\\addlinespace[-2.5mm]
Synergistic                           & Smoking                             & Immune                 & Synergistic             & Smoking                & Immune                 \\ 
$\pi_{0001} \in[3.00\%,3.55\%] $      & $\pi_{0011} \in[0.28\%,0.83\%] $    & $\pi_{0000} =95.50\%$    & $\pi_{0001}=3.11\%$       & $\pi_{0011}=0.73\%$      & $\pi_{0000}=95.50\%$     \\ 
\multicolumn{1}{l}{}                  & \multicolumn{1}{l}{}                & \multicolumn{1}{l|}{}  & \multicolumn{1}{l}{}    & \multicolumn{1}{l}{}   & \multicolumn{1}{l}{}   \\\addlinespace[-2.5mm]
Asbestos                              & Parallel                            & Doomed                 & Asbestos                & Parallel               & Doomed                 \\ 
$\pi_{0101} \in[0 \%,0.55\%] $        & $\pi_{0111} \in[0 \%,0.55\%]$       & $\pi_{1111} =0.12\%$     & $\pi_{0101}=0.44\% $      & $\pi_{0111}=0.10\% $     & $\pi_{1111}=0.12\%$      \\  
\multicolumn{1}{l}{}                  & \multicolumn{1}{l}{}                & \multicolumn{1}{l|}{}  & \multicolumn{1}{l}{}    & \multicolumn{1}{l}{}   & \multicolumn{1}{l}{}   \\ \addlinespace[-2.5mm]
\hline
\multicolumn{1}{l}{}                  & \multicolumn{1}{l}{}                & \multicolumn{1}{l|}{}  & \multicolumn{1}{l}{}    & \multicolumn{1}{l}{}   & \multicolumn{1}{l}{}   \\ 
\multicolumn{3}{c|}{(b) Bound for $\O=(Z=1,M=1,Y=1) $}                                               & \multicolumn{3}{c}{(d) Maximum entropy method for $\O=(Z=1,M=1,Y=1) $}    \\  
\multicolumn{3}{l|}{}                                                                                & \multicolumn{3}{l}{}                                                        \\\addlinespace[-2.5mm]
Synergistic                           & Smoking                             & Immune                 & Synergistic             & Smoking                & Immune                 \\ 
$\pi_{0001}(\O)\in[66.66\%,78.85\%] $ & $\pi_{0011}(\O)\in[6.33\%,18.51\%]$ & $\pi_{0000} (\O)=0\%$    & $\pi_{0001}(\O)= 68.97\%$ & $\pi_{0011}(\O)=16.20\%$ & $\pi_{0000}(\O)=0\%$     \\\addlinespace[-2.5mm]
\multicolumn{1}{l}{}                  & \multicolumn{1}{l}{}                & \multicolumn{1}{l|}{}  & \multicolumn{1}{l}{}    & \multicolumn{1}{l}{}   & \multicolumn{1}{l}{}   \\ 
Asbestos                              & Parallel                            & Doomed                 & Asbestos                & Parallel               & Doomed                 \\ 
$\pi_{0101}(\O)\in[0 \%,12.19\%] $    & $\pi_{0111} (\O)\in[0 \%,12.19\%] $ & $\pi_{1111} (\O)=2.64\%$ & $\pi_{0101}(\O)=9.88\% $  & $\pi_{0111}(\O)=2.31\%$  & $\pi_{1111}(\O)=2.64\%$ 
\end{tabular}
}
\end{table}

Next, we consider the maximum entropy estimation method proposed by \citet{darroch1994synergism} under the monotonicity assumption \ref{ass:monotonicity}. The goal of this method is to find  the values that maximize the entropy, $- \sum \pi_{rstu} \log \pi_{rstu}$, subject to the constraints outlined in the second column of  Table \ref{tab: probabilities}. The estimation results for the cases where $\O$ is an empty set and $\O = (Z=1,M=1,Y=1)$ are shown in Table \ref{tab:methods-entropy}(c) and (d), respectively, and clearly fall within the nonparametric bounds in Table \ref{tab:methods-bound}(a) and (b).  For those who were exposed to both smoking and asbestos and developed lung cancer, i.e., the evidence $\O = (Z=1, M=1, Y=1)$ in Table \ref{tab:methods-entropy}(d), 68.97\% of the cases were caused by the synergistic effect of smoking and asbestos, while 16.20\% were attributed to smoking alone. Among the remaining individuals, approximately 2.31\% of lung cancer cases resulted from either smoking or asbestos exposure, and 9.88\% were caused by asbestos alone. Additionally, 2.64\% of the lung cancer cases were attributed to other risk factors, such as age and genetic factors.


\subsection{Analysis with  a secondary outcome}
\label{sec:application-secondary}
While the previous methods adequately explain the summary data on smoking and asbestos exposure, they cannot fully identify the posterior probabilities  $\pi_{rstu}$ and $\pi_{rstu}(\O)$, where $\O = (Z=z, M=m, Y=y)$ with $z,m,y\in\{0,1\}$. In this subsection, we propose a more detailed analysis of lung cancer risk using the method introduced in Section \ref{sec:secondary-outcome}. Based on the summary data presented in Table \ref{tab:summary-stat} of the
Supplementary Material, we sample 21319 individuals. Additionally, we synthetically generated the secondary outcome, body weight (measured in kilograms), for further analysis.  The details of the data generation process are provided in Section \ref{ssec:bodyweight-second} of the Supplementary Material.  


Before the formal analysis, we recommend first considering whether the monotonicity assumption \ref{ass:monotonicity} is supported by the observed data.  For the cases with and without monotonicity, we apply the EM algorithms from Sections \ref{ssec: estimation} and  \ref{ssec: estimation-nomo} of the Supplementary Material to obtain the  MLEs. Under the monotonicity assumption, the latent variable $G$ has 6 classes, whereas without the monotonicity assumption, $G$ has 16 classes. Using the MLEs and the number of parameters in both scenarios, we compute the Akaike Information Criterion (AIC). The results show that the AIC is lower with the monotonicity assumption than without it. Therefore, the data provides stronger support for the monotonicity assumption. This assumption is reasonable in this context because smokers ($Z = 1$) are more likely to develop lung cancer than non-smokers ($Z = 0$), and individuals exposed to asbestos ($M = 1$) are more likely to develop lung cancer than those not exposed ($M = 0$). Therefore, we adopt the monotonicity assumption for further analysis.

\begin{table}[h]
\centering
\caption{The posterior probabilities $\pi_{rstu}(\mathcal{O})$ for different evidence obtained using $W$.}
\vspace{0.15cm}
\label{tab:est-res}
\resizebox{0.99949790\textwidth}{!}{  
  \begin{threeparttable}
\setlength{\arrayrulewidth}{1.25pt} 
\begin{tabular}{ccc|ccc}
\multicolumn{3}{c|}{(a) Proposed method for $\O=\emptyset$}                    & \multicolumn{3}{c}{(d) Proposed method for $\O=(Z=1,M=0,Y=1)$}                  \\\addlinespace[-2.5mm]
\multicolumn{3}{l|}{}                                                          & \multicolumn{3}{l}{}                                                            \\
Synergistic              & Smoking                  & Immune                   & Synergistic              & Smoking                  & Immune                    \\
$\pi_{0001}=$3.20$\%$      & $\pi_{0011}=$0.64$\%$      & $\pi_{0000}=$95.50$\%$     & $\pi_{0001}(\O)=$0$\%$     & $\pi_{0011}(\O)=$66.95$\%$ & $\pi_{0000}(\O)=$0$\%$      \\
\multicolumn{1}{l}{}     & \multicolumn{1}{l}{}     & \multicolumn{1}{l|}{}    & \multicolumn{1}{l}{}     & \multicolumn{1}{l}{}     & \multicolumn{1}{l}{}      \\\addlinespace[-2.5mm]
Asbestos                 & Parallel                 & Doomed                   & Asbestos                 & Parallel                 & Doomed                    \\
$\pi_{0101}=$0.35$\%$      & $\pi_{0111}=$0.20$\%$      & $\pi_{1111}=$0.12$\%$      & $\pi_{0101}(\O)=$0$\%$     & $\pi_{0111}(\O)=$20.57$\%$ & $\pi_{1111}(\O)=$12.49$\%$  \\
                         &                          &                          &                          &                          &                           \\\addlinespace[-1mm] 
\hline\addlinespace[-1mm]
                         &                          &                          &                          &                          &                           \\
\multicolumn{3}{c|}{(b) Proposed method for $\O=(Z=0,M=0,Y=1)$}                & \multicolumn{3}{c}{(e) Proposed method for $\O=(Z=1,M=1,Y=1)$}                  \\
\multicolumn{3}{l|}{}                                                          & \multicolumn{3}{l}{}                                                            \\\addlinespace[-2.5mm]
Synergistic              & Smoking                  & Immune                   & Synergistic              & Smoking                  & Immune                    \\
$\pi_{0001}(\O)=$0$\%$     & $\pi_{0011}(\O)=$0$\%$     & $\pi_{0000}(\O)=$0$\%$     & $\pi_{0001}(\O)=$71.01$\%$ & $\pi_{0011}(\O)=$14.16$\%$ & $\pi_{0000}(\O)=$0.00$\%$   \\
\multicolumn{1}{l}{}     & \multicolumn{1}{l}{}     & \multicolumn{1}{l|}{}    & \multicolumn{1}{l}{}     & \multicolumn{1}{l}{}     & \multicolumn{1}{l}{}      \\\addlinespace[-2.5mm]
Asbestos                 & Parallel                 & Doomed                   & Asbestos                 & Parallel                 & Doomed                    \\
$\pi_{0101}(\O)=$0$\%$     & $\pi_{0111}(\O)=$0$\%$     & $\pi_{1111}(\O)=$100$\%$   & $\pi_{0101}(\O)=$7.84$\%$  & $\pi_{0111}(\O)=$4.35$\%$  & $\pi_{1111}(\O)=$2.64$\%$   \\
                         &                          &                          &                          &                          &                           \\ \addlinespace[-1mm]
\hline\addlinespace[-1mm]
                         &                          &                          &                          &                          & \multicolumn{1}{l}{}      \\
\multicolumn{3}{c|}{(c) Proposed method for $\O=(Z=0,M=1,Y=1)$}                &                          &                          &                           \\\addlinespace[-2.5mm]
\multicolumn{3}{l|}{}                                                          & \multicolumn{1}{l}{}     & \multicolumn{1}{l}{}     & \multicolumn{1}{l}{}      \\
Synergistic              & Smoking                  & Immune                   &                          &                          &                           \\
$\pi_{0001}(\O)=$0$\%$     & $\pi_{0011}(\O)=$0$\%$     & $\pi_{0000}(\O)=$0$\%$     &                          &                          &                           \\ \addlinespace[-2.5mm]
\multicolumn{1}{l}{}     & \multicolumn{1}{l}{}     & \multicolumn{1}{l|}{}    & \multicolumn{1}{l}{}     & \multicolumn{1}{l}{}     & \multicolumn{1}{l}{}      \\
Asbestos                 & Parallel                 & Doomed                   &                          &                          &                           \\
$\pi_{0101}(\O)=$52.88$\%$ & $\pi_{0111}(\O)=$29.36$\%$ & $\pi_{1111}(\O)=$17.83$\%$ &                          &                          &                          
\end{tabular}
  \end{threeparttable}}
\end{table}

Table \ref{tab:est-res} presents the posterior probabilities $\pi_{rstu}(\mathcal{O})$ for different subpopulations, obtained using the secondary outcome variable.  First, we present the estimates of  $\pi_{rstu}$ for $r,s,t,u\in\{0,1\}$ in Table $\ref{tab:est-res}(a)$. Comparing Table \ref{tab:est-res}(a) with the maximum entropy results in Table \ref{tab:methods-entropy}, we find that the posterior probabilities are very close and maintain the same order, though not identical. This difference arises because we use an additional secondary outcome to identify $\pi_{rstu}$, which is not guaranteed by the maximum entropy method.  We also use the results from Table $\ref{tab:est-res}$(a) in the first column of Table \ref{tab:payment} to discuss the proportions of responsibility between the tobacco company and the asbestos company. For the evidence   $\mathcal{O}=(Z=0, M=0, Y=1)$, lung cancer must be caused by other factors. For the evidence   $\mathcal{O}=(Z=1, M=0, Y=1)$, lung cancer is primarily caused by smoking. For the evidence  $\mathcal{O}=(Z=0, M=1, Y=1)$, lung cancer is mainly due to asbestos exposure. For the evidence  $\mathcal{O}=(Z=1, M=1, Y=1)$, lung cancer is primarily caused by the synergistic effects of smoking and asbestos exposure.  {\black Compared to other methods in Table \ref{tab:methods-entropy}, our proposed method offers a clear advantage in identifiability. The bound-based method only provides an identification region, while the maximum entropy method yields a point estimate within that region but does not necessarily converge to the true parameter. In contrast, our method fully identifies the attribution probabilities $\pi_{rstu}(\mathcal{O})$ by incorporating a secondary outcome $W$ under appropriate assumptions, enabling reliable causal attribution.}

\begin{figure}[h]
    \centering 
    \includegraphics[width=1\linewidth]{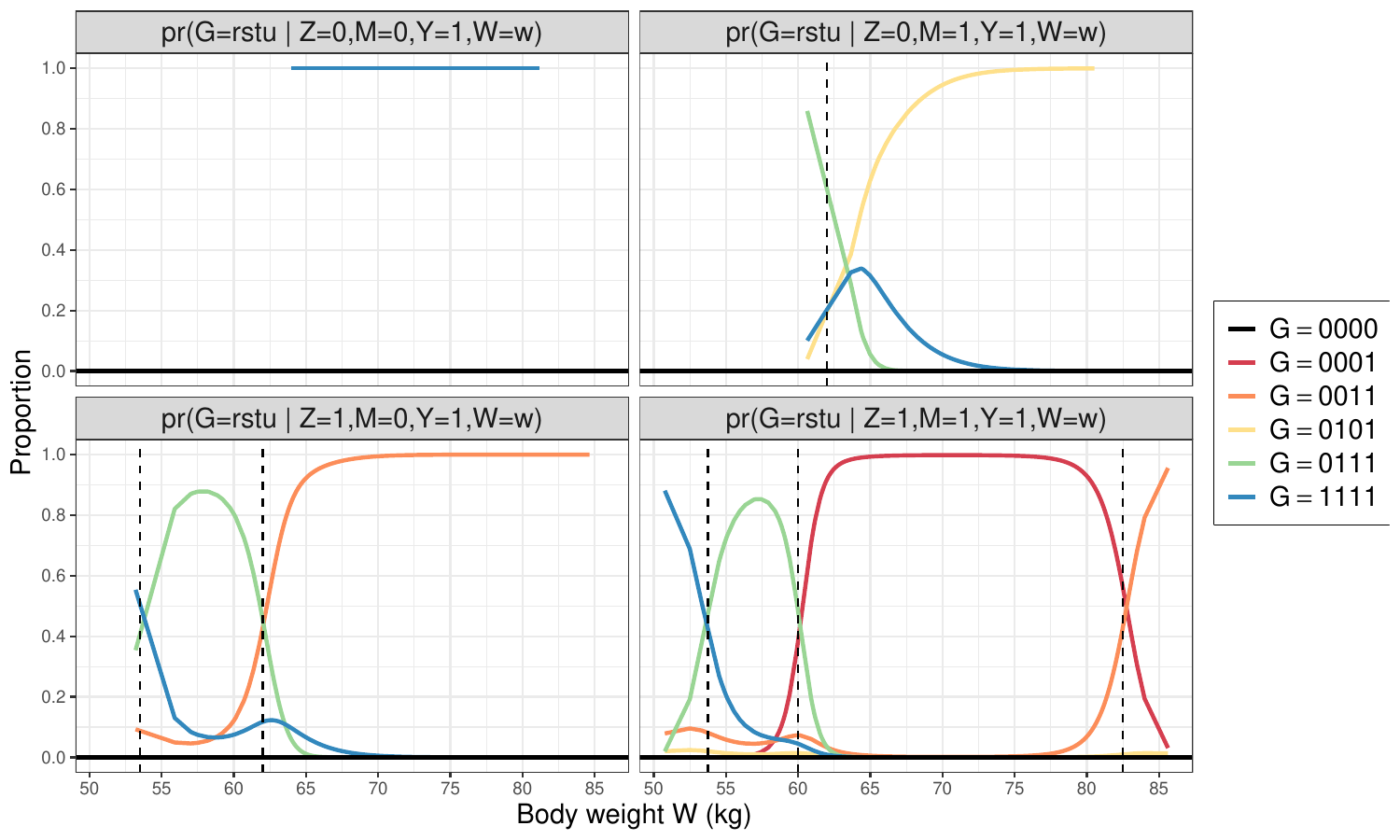}
\caption{The posterior probabilities \( \pr(G = rstu \mid Z = z, M = m, Y = 1, W) \) are plotted along the horizontal axis \( W \) for four observed scenarios. The vertical dashed lines mark the points where the two curves intersect. The thick black line represents the zero baseline on the vertical axis; for example, in the upper left panel, it covers all cases except the posterior probability \( \pr(G = 1111 \mid Z =0, M = 0, Y = 1, W) \).}
    \label{fig:enter-label-2}
\end{figure}

In addition to the primary outcome of lung cancer development, body weight $W$ is an important secondary outcome reflecting an individual's physical condition. {\black We therefore include body weight $W$ to capture heterogeneity and incorporate it in the evidence set for retrospective attribution analysis. }For clarity, we show the conditional probability $\pr(G={rstu} \mid Z=z, M=m, Y=1, W)$  plots  for different $z, m, r, s, t, u \in \{0, 1\}$ in Figure \ref{fig:enter-label-2}, where the horizontal axis represents body weight $W$.

In the upper left panel of Figure \ref{fig:enter-label-2}, for the evidence $\mathcal{O} = (Z = 0, M = 0, Y = 1)$,   the posterior probability $\text{pr}(G = 1111 \mid Z = 0, M = 0, Y = 1, W) = 1$ holds for any $W$. This indicates that all lung cancer patients in this scenario belong to  $G = 1111$.  In the upper right panel of Figure \ref{fig:enter-label-2}, for the evidence $\mathcal{O} = (Z = 0, M = 1, Y = 1)$, we find that at lower body weights (60-62 kg), most individuals belong to   $G = 0111$. This suggests that individuals with lower body weights have weaker constitutions, and either smoking or asbestos exposure alone could lead to lung cancer. As body weight increases ($\geq$ 62 kg), most individuals belong to $G = 0101$, indicating that lung cancer is primarily due to asbestos exposure.

In the lower left panel of Figure \ref{fig:enter-label-2}, for individuals who smoked but were not exposed to asbestos $\mathcal{O} = (Z = 1, M = 0, Y = 1)$, those with lower body weights (54-62  kg) are mostly in the $G = 0111$ class. With increasing body weight ($\geq$ 62 kg) and improving physical condition, $G = 0011$ has the highest posterior probability, indicating that lung cancer is primarily attributable to smoking.  In the lower right panel of Figure \ref{fig:enter-label-2}, for individuals who both smoked and were exposed to asbestos $\mathcal{O} = (Z = 1, M = 1, Y = 1)$, those with lower body weights (54-60 kg) primarily belong to the $G = 0111$ class, indicating that either asbestos or smoking alone could cause lung cancer. As body weight improves (60-82.5 kg), most individuals belong to the $G = 0001$ class, suggesting that lung cancer is primarily due to the synergistic effect of asbestos and smoking.  Finally, for individuals with stronger constitutions (body weights $\geq$ 82.5 kg), most belong to the $G = 0011$ class, with lung cancer caused solely by smoking and not by asbestos. 

\section{Discussion}\label{sec:discussion}
In this paper, we propose using the posterior probabilities to characterize causal interactions with different evidence sets. Identifying the parameters of interest primarily depends on identifying the joint distribution of four potential outcomes. We discuss the identifiability of the proposed estimands using a secondary outcome variable and Gaussian mixture distributions. We provide two likelihood-based estimation methods and conduct several numerical studies in Supplementary Material. Additionally, we apply the proposed method to analyze the classic lung cancer data related to smoking and asbestos exposure. All methods, including the nonparametric bounds \citep{miettinen1982causal}, maximum entropy method \citep{darroch1994synergism}, and the approach we propose based on the secondary outcome, provide strong evidence that the primary cause of lung cancer in individuals exposed to both smoking and asbestos is the synergistic effect of these two factors.

{\black When $Z$ and $M$ are jointly considered as a four-level exposure, our problem becomes analogous to causal analysis of ordinal treatments with a binary intermediate variable, as studied by \citet{wang2017causal} and \citet{luo2023causal}. These studies typically rely on strong monotonicity assumptions, such as (1) $Y_{0,0} \leq Y_{0,1} \leq Y_{1,0} \leq Y_{1,1}$ or (2) $Y_{0,0} \leq Y_{1,0} \leq Y_{0,1} \leq Y_{1,1}$, to identify quantities like $\mathbb{E}(W \mid Z, M, G)$. While these assumptions ensure identification of $\pi_{rstu}$ and $\pi_{rstu}(\mathcal{O})$, they are incompatible with attribution analysis because they automatically exclude meaningful causal types. For instance, monotonicity (1) rules out asbestos-only sensitive individuals ($G=0101$), while (2) rules out smoking-only sensitive individuals ($G=0011$). Since attribution analysis aims to assess all possible causal mechanisms, such strong monotonicity assumptions would exclude key causes of interest.}

{\black Note that in our real-data analysis in the Supplementary Material, the  AIC  is used for selecting between models with and without the monotonicity assumption. Since both types of models employ the same parametric specification for the secondary outcome $W$ (assuming normality), AIC provides an appropriate criterion for comparison within a unified modeling framework. AIC balances model fit and complexity by penalizing overly complex models while rewarding goodness of fit, thus offering a likelihood-based, fair, and practical approach to model comparison. It is important to emphasize that for models without the monotonicity assumption, identification relies on Assumption \ref{assump:equal-prop}, which should be carefully considered when interpreting the estimation results in practice.}

The proposed methods can be improved or extended in several directions. First, nonparametric partial identification techniques \citep{dawid2022bounding} may complement our parametric approach. Second, extending the framework beyond two exposures to handle multiple exposures and outcomes \citep{li2024Biometrika} would enhance practical applicability. Finally, developing identification strategies that relax the no unmeasured confounding assumption (Assumption \ref{assump:no-confounding}) would broaden the scope of applications.  

\appendix
\section*{Supplementary Material}

The Supplementary Material includes the proofs of the theorems and corollaries, additional graphic illustrations, estimation procedures, and details on simulation studies. It also provides an additional real data analysis based on the Job Search Intervention Study   and an extended example with covariates within the smoking and asbestos context, along with further discussion of interactive causal attribution. 

 \bibliography{mybib}
\newpage
\begin{center}
\Large\bf Supplementary Material    
\end{center}
The Supplementary Material includes the proofs of the theorems and corollaries, additional graphic illustrations, estimation procedures, and details on simulation studies. It also provides an additional real data analysis based on the Job Search Intervention Study   and an extended example with covariates within the smoking and asbestos context, along with further discussion of interactive causal attribution. 

\addtocontents{toc}{\protect\setcounter{tocdepth}{2}}
\tableofcontents
\section{Additional graphic illustration}
\label{sec:graph-ill}
 In this section, we provide visual representations of how the observed evidence $(Z=z, M=m, Y=y)$ links to the latent types $G$, both with and without monotonicity assumption \ref{ass:monotonicity}.  Table \ref{tab:no-mono} shows the identification structure without monotonicity assumptions, where all 16 latent types can appear in the observed data. Each row represents an observable situation defined by the two binary exposures $(Z,M)$, and the binary outcome $(Y)$. The colored cells show which latent types $G$ are possible under each observable situation. In contrast, Table \ref{tab:with-mono} shows the simplified structure under monotonicity assumptions (Assumption \ref{ass:monotonicity}), where only 6 latent types remain: immune ($G=0000$), synergistic ($G=0001$), smoking ($G=0011$), asbestos ($G=0101$), parallel ($G=0111$), and doomed ($G=1111$). The reduction from 16 to 6 types demonstrates how monotonicity substantially simplifies the causal structure.

To achieve identification, we rely on Assumption \ref{assumption:normal}, Assumption \ref{assump:equal-prop}(ii) and Assumption \ref{assump:equal-prop}(iii), which ensure the identifiability of Gaussian mixture distributions \citep{Titterington1985}. Assumption  \ref{assumption:normal} requires that the secondary outcome $W$ follows a normal distribution within each colored cell of Tables \ref{tab:no-mono} and \ref{tab:with-mono}. Each cell represents a specific combination of observed evidence $\mathcal{O}=(Z=z, M=m, Y=y)$ (shown in the rows) and latent type $G=rstu$ (shown in the columns). Within each cell, we assume $W \mid (Z=z, M=m, G=rstu) \sim N(\mu_{z,m,rstu}, \sigma^2_{z,m,rstu})$. Note that since $(z,m,rstu)$ determines $y$ within each cell, we omit $y$ from the conditioning set for notational simplicity. Consequently, given the observed evidence $\mathcal{O}=(Z=z, M=m, Y=y)$, the secondary outcome $W$ follows a Gaussian mixture distribution across the latent types in each row.

Specifically, Assumption \ref{assump:equal-prop}(ii) requires that the mean of $W$ depends only on the latent type $G=rstu$, not on exposure status $(Z, M)$ or outcome $Y$. That is, $\mu_{z,m,rstu} = \mu_{rstu}$ is constant across all cells with the same color (same latent type) in Tables \ref{tab:no-mono} and \ref{tab:with-mono}, but varies across different colors (different types): $\mu_{rstu} \neq \mu_{r's't'u'}$ for $rstu \neq r's't'u'$.  

Similarly, Assumption \ref{assump:equal-prop}(iii) requires that the variance $\sigma^2_{z,m,rstu} = \sigma^2_{rstu}$ depends only on latent type, remaining constant within each color but differing across different colors.   

Either of these two conditions is sufficient to achieve identification.

\begin{table}[h]
\centering
\caption{Identification structure without monotonicity: all 16 latent types.  Cells sharing the same color correspond to the same latent type across different observable combinations.}
\label{tab:no-mono}
\resizebox{0.95\textwidth}{!}{%
\renewcommand{\arraystretch}{1.5}
\begin{tabular}{|c|c|c|c|c|c|c|c|c|}
\hline
$(Z=0, M=0, Y=0)$ & \cellcolor{c1}${G=0000}$ & \cellcolor{c2}${G=0001}$ & \cellcolor{c3}${G=0010}$ & \cellcolor{c4}${G=0011}$ & \cellcolor{c5}${G=0100}$ & \cellcolor{c6}${G=0101}$ & \cellcolor{c7}${G=0110}$ & \cellcolor{c8}${G=0111}$ \\
\hline
$(Z=0, M=0, Y=1)$ & \cellcolor{c9}${G=1000}$ & \cellcolor{c10}${G=1001}$ & \cellcolor{c11}${G=1010}$ & \cellcolor{c12}${G=1011}$ & \cellcolor{c13}${G=1100}$ & \cellcolor{c14}${G=1101}$ & \cellcolor{c15}${G=1110}$ & \cellcolor{c16}${G=1111}$ \\
\hline
$(Z=0, M=1, Y=0)$ & \cellcolor{c1}${G=0000}$ & \cellcolor{c2}${G=0001}$ & \cellcolor{c3}${G=0010}$ & \cellcolor{c4}${G=0011}$ & \cellcolor{c9}${G=1000}$ & \cellcolor{c10}${G=1001}$ & \cellcolor{c11}${G=1010}$ & \cellcolor{c12}${G=1011}$ \\
\hline
$(Z=0, M=1, Y=1)$ & \cellcolor{c5}${G=0100}$ & \cellcolor{c6}${G=0101}$ & \cellcolor{c7}${G=0110}$ & \cellcolor{c8}${G=0111}$ & \cellcolor{c13}${G=1100}$ & \cellcolor{c14}${G=1101}$ & \cellcolor{c15}${G=1110}$ & \cellcolor{c16}${G=1111}$ \\
\hline
$(Z=1, M=0, Y=0)$ & \cellcolor{c1}${G=0000}$ & \cellcolor{c2}${G=0001}$ & \cellcolor{c5}${G=0100}$ & \cellcolor{c6}${G=0101}$ & \cellcolor{c9}${G=1000}$ & \cellcolor{c10}${G=1001}$ & \cellcolor{c13}${G=1100}$ & \cellcolor{c14}${G=1101}$ \\
\hline
$(Z=1, M=0, Y=1)$ & \cellcolor{c3}${G=0010}$ & \cellcolor{c4}${G=0011}$ & \cellcolor{c7}${G=0110}$ & \cellcolor{c8}${G=0111}$ & \cellcolor{c11}${G=1010}$ & \cellcolor{c12}${G=1011}$ & \cellcolor{c15}${G=1110}$ & \cellcolor{c16}${G=1111}$ \\

\hline
$(Z=1, M=1, Y=0)$ & \cellcolor{c1}${G=0000}$ & \cellcolor{c5}${G=0100}$ & \cellcolor{c9}${G=1000}$ & \cellcolor{c13}${G=1100}$ & \cellcolor{c3}${G=0010}$ & \cellcolor{c7}${G=0110}$ & \cellcolor{c11}${G=1010}$ & \cellcolor{c15}${G=1110}$ \\
\hline
$(Z=1, M=1, Y=1)$ & \cellcolor{c2}${G=0001}$ & \cellcolor{c6}${G=0101}$ & \cellcolor{c10}${G=1001}$ & \cellcolor{c14}${G=1101}$ & \cellcolor{c4}${G=0011}$ & \cellcolor{c8}${G=0111}$ & \cellcolor{c12}${G=1011}$ & \cellcolor{c16}${G=1111}$ \\
\hline
\end{tabular}%
}
\end{table}

\begin{table}[h]
\centering
\caption{Identification structure under monotonicity: only 6 latent types.  Cells sharing the same color correspond to the same latent type across different observable combinations.}
\label{tab:with-mono}
\resizebox{0.69\textwidth}{!}{%
\renewcommand{\arraystretch}{1.8}
\begin{tabular}{|l|c|c|c|c|c|}
\hline
$(Z=0, M=0, Y=0)$ & \cellcolor{c1}${G=0000}$ & \cellcolor{c2}${G=0001}$ & \cellcolor{c4}${G=0011}$ & \cellcolor{c6}${G=0101}$ & \cellcolor{c8}${G=0111}$ \\\hline
$(Z=0, M=0, Y=1)$ & \cellcolor{c16}${G=1111}$ & & & & \\
\hline
$(Z=0, M=1, Y=0)$ & \cellcolor{c1}${G=0000}$ & \cellcolor{c2}${G=0001}$ & \cellcolor{c4}${G=0011}$ & & \\
\hline
$(Z=0, M=1, Y=1)$ & \cellcolor{c6}${G=0101}$ & \cellcolor{c8}${G=0111}$ & \cellcolor{c16}${G=1111}$ & & \\
\hline
$(Z=1, M=0, Y=0)$ & \cellcolor{c1}${G=0000}$ & \cellcolor{c2}${G=0001}$ & \cellcolor{c6}${G=0101}$ & & \\
\hline
$(Z=1, M=0, Y=1)$ & \cellcolor{c4}${G=0011}$ & \cellcolor{c8}${G=0111}$ & \cellcolor{c16}${G=1111}$ & & \\
\hline
$(Z=1, M=1, Y=0)$ & \cellcolor{c1}${G=0000}$ & & & & \\
\hline
$(Z=1, M=1, Y=1)$ & \cellcolor{c2}${G=0001}$ & \cellcolor{c4}${G=0011}$ & \cellcolor{c6}${G=0101}$ & \cellcolor{c8}${G=0111}$ & \cellcolor{c16}${G=1111}$ \\
\hline
\end{tabular}%
}
\end{table}

  \section{Estimation with monotonicity assumption \ref{ass:monotonicity}}
\label{ssec: estimation}

To accommodate the practical observational studies, for each unit \(i\), in addition to the binary exposure variables \(Z_i\) and \(M_i\), the binary outcome variable \(Y_i\), and the secondary outcome variable \(W_i\), we also introduce the background covariates \(\boldsymbol{X}_i\), and discuss the estimation method in the Supplementary Material. When incorporating the covariates, we consistently assume that the more general ignorability assumption holds, i.e., \((Z, M) \indep Y_{z,m } \mid \boldsymbol{X}\), which naturally satisfies Assumption~\ref{assump:no-confounding} in the main text. Furthermore, when the covariates \(\boldsymbol{X}\) are introduced, Assumption \ref{assumption:normal} is similarly extended to hold conditional on the covariates \(\boldsymbol{X}\). 

\subsection{EM algorithms }
Under the monotonicity assumption  \ref{ass:monotonicity}, 
we can model  $\pr(G=rstu\mid \boldsymbol{X} )$ as a six level multinomial logistic model: $$\pi_{rstu}( {\boldsymbol X} ;  {\boldsymbol \theta
}):=\pr(G=rstu\mid {\boldsymbol X};  {\boldsymbol \theta
} )=\dfrac{\mathrm{exp}(  {\boldsymbol \theta
}_{rstu}^\T  {\boldsymbol X})}{\sum_{r's't'u'}\mathrm{exp}(  {\boldsymbol\theta
}_{r's't'u'}^\T  {\boldsymbol X})}, $$
    where $rstu\in \{{0000},{0001}, {0011}, {0101}, {0111}, {1111}\}$, $\boldsymbol\theta= (\boldsymbol\theta_{0000}^\T,\boldsymbol\theta_{0001}^\T,\boldsymbol\theta_{0011}^\T,\boldsymbol\theta_{0101}^\T, \boldsymbol\theta_{0111}^\T, \boldsymbol\theta_{1111}^\T)^\T,$ and ${\boldsymbol\theta
}_{0000}=\boldsymbol 0$ for identiﬁcation.

We also model  the conditional probability density ${f}\left(W\mid Z=z,M=m, G=rstu, \boldsymbol X\right)$ as the following normal distribution:  
\begin{align*} 
    {f}(W\mid Z=z,M=m, G=rstu, \boldsymbol{X};\boldsymbol\mu_{z,m,rstu},\sigma _{z,m,rstu}^2 ) = \frac{\exp\left\{- {(W - \boldsymbol\mu_{z,m,rstu}^\T\boldsymbol{X})^2}/{2\sigma^2_{z,m,rstu}}\right\}}{\sqrt{2\pi\sigma^2_{z,m,rstu}}} , 
\end{align*}
for $z,m\in\{0,1\}$ and $rstu\in \{{0000},{0001}, {0011}, {0101}, {0111}, {1111}\}$. Let $\boldsymbol{\beta}$ denote the parameter vector consisting of all $\boldsymbol{\mu}_{z,m,rstu}$ and $\sigma_{z,m,rstu}^2$ for $z,m,r,s,t,u\in\{0,1\}$.

We then use the EM algorithm to obtain a maximum likelihood estimate, where we consider $G$ as incompletely observed data. We introduce the shorthand notation $\pi_{i, rstu} = \pr(G_i = rstu \mid \boldsymbol{X}_i ;  {\boldsymbol \theta
})$ and $f_{i,z,m, rstu} = f(W_i \mid Z_i=z,M_i=m, G_i = rstu, \boldsymbol{X}_i;\boldsymbol\mu_{z,m,rstu},\sigma _{z,m,rstu}^2 )$.

In the E-step, we first compute the conditional probabilities of latent classes given the observed data. Let $\pi_{i, rstu}^{[j]}$ and $f_{i,z,m, rstu}^{[j]}$ denote the estimates at the $j$-th iteration for unit $i$: 
 $$\pi_{i,rstu}^{[j]} =\dfrac{\mathrm{exp}(  {\boldsymbol \theta
}_{rstu}^{[j]\T}  {\boldsymbol X}_i )}{\sum_{r's't'u'}\mathrm{exp}(  {\boldsymbol\theta
}_{r's't'u'}^{[j]\T} {\boldsymbol X}_i )},~~
   f_{i,z,m,rstu}^{[j]}= \frac{\exp\left\{- {(W_i  - \boldsymbol\mu_{z,m,rstu}^{[j]\T}\boldsymbol{X}_i )^2}/{2\sigma^{2[j]}_{z,m,rstu}}\right\}}{\sqrt{2\pi\sigma^{2[j]}_{z,m,rstu}}} . $$

 For  $(Z_i = 1, M_i = 1, Y_i = 1,\boldsymbol{X}_i, W_i)$, we have
\begin{align*} 
      \begin{aligned}
\operatorname{pr} ^{[j]}\left(G_i=rst1\mid -\right)=\frac{\pi_{i, rst1}^{[j]} f_{i, 1,1, rst1}^{[j]}}{\begin{pmatrix}
     \pi_{i, 0001}^{[j]} f_{i,  1,1,0001}^{[j]}+ \pi_{i, 0011}^{[j]} f_{i, 1,1, 0011}^{[j]}+ \pi_{i, 0101}^{[j]} f_{i, 1,1, 0101}^{[j]}\\\addlinespace[1.5mm]+ \pi_{i, 0111}^{[j]} f_{i,  1,1,0111}^{[j]}+ \pi_{i, 1111}^{[j]} f_{i, 1,1, 1111}^{[j]}
\end{pmatrix}},
\end{aligned} 
\end{align*}
where $rst1\in \{ {0001}, {0011}, {0101}, {0111}, {1111}\}$ and 
we use ``$-$" to simply represent the observed variables. 

For $(Z_i=1,M_i=1,Y_i=0,\boldsymbol{X}_i,W_i)$,  we have $ 
\operatorname{pr}^{[j]} \left(G_i=0000\mid -\right)=1.$  

For  $(Z_i = 1, M_i = 0, Y_i = 1,\boldsymbol{X}_i, W_i)$, we have
\begin{align*} 
      \begin{aligned}
\operatorname{pr} ^{[j]}\left(G_i=rs1u\mid -\right)=\frac{\pi_{i, rs1u}^{[j]} f_{i, 1,0, rs1u}^{[j]}}{  \pi_{i, 0011}^{[j]} f_{i,  1,0, 0011}^{[j]} + \pi_{i, 0111}^{[j]} f_{i, 1,0,  0111}^{[j]}+ \pi_{i, 1111}^{[j]} f_{i, 1,0,  1111}^{[j]}},  
\end{aligned} 
\end{align*}
where $rs1u\in \{  {0011},  {0111}, {1111}\}$. 

For $(Z_i=1,M_i=0,Y_i=0,\boldsymbol{X}_i,W_i)$,  we have 
\begin{align*} 
      \begin{aligned}
\operatorname{pr} \left(G_i=rs0u\mid -\right)=\frac{\pi_{i, rs0u}^{[j]} f_{i, 1,0,  rs0u}^{[j]}}{  \pi_{i, 0000}^{[j]} f_{i,  1,0, 0000}^{[j]} + \pi_{i, 0001}^{[j]} f_{i, 1,0,  0001}^{[j]}+ \pi_{i, 0101}^{[j]} f_{i,  1,0, 0101}^{[j]}},
\end{aligned} 
\end{align*}
where $rs0u\in \{  {0000},  {0001}, {0101}\}$. 

For  $(Z_i = 0, M_i = 1, Y_i = 1,\boldsymbol{X}_i, W_i)$, we have \begin{align*} 
      \begin{aligned}
\operatorname{pr} ^{[j]}\left(G_i=r1tu\mid -\right)=\frac{\pi_{i, r1tu}^{[j]} f_{i,0,1,  r1tu}^{[j]}}{  \pi_{i, 0101}^{[j]} f_{i,  0,1,0101}^{[j]} + \pi_{i, 0111}^{[j]} f_{i, 0,1, 0111}^{[j]}+ \pi_{i, 1111}^{[j]} f_{i, 0,1, 1111}^{[j]}},
\end{aligned} 
\end{align*}
where $r1tu\in \{  {0101},  {0111}, {1111}\}$.

For $(Z_i=0,M_i=1,Y_i=0,\boldsymbol{X}_i,W_i)$, we have 
\begin{align*} 
      \begin{aligned}
\operatorname{pr} ^{[j]}\left(G_i=r0tu\mid -\right)=\frac{\pi_{i, r0tu}^{[j]} f_{i, 0,1, r0tu}^{[j]}}{  \pi_{i, 0000}^{[j]} f_{i,  0,1,0000}^{[j]} + \pi_{i, 0001}^{[j]} f_{i, 0,1, 0001}^{[j]}+ \pi_{i, 0011}^{[j]} f_{i, 0,1, 0011}^{[j]}},
\end{aligned} 
\end{align*}
where $r0tu\in \{  {0000},  {0001}, {0011}\}$.  

For  $(Z_i = 0, M_i = 0, Y_i = 1,\boldsymbol{X}_i, W_i)$, we have $ 
\operatorname{pr}^{[j]} \left(G_i=1111\mid -\right) =1 . $ 

For $(Z_i=0,M_i=0,Y_i=0,\boldsymbol{X}_i,W_i)$, we have
\begin{align*} 
      \begin{aligned}
\operatorname{pr}^{[j]} \left(G_i=0stu\mid -\right)=\frac{\pi_{i, 0stu}^{[j]} f_{i, 0,0, 0stu}^{[j]}}{  \begin{pmatrix}
    \pi_{i, 0000}^{[j]} f_{i,  0,0, 0000}^{[j]}+\pi_{i, 0001}^{[j]} f_{i,  0,0, 0001}^{[j]}+ \pi_{i, 0011}^{[j]} f_{i,  0011}^{[j]}\\\addlinespace[1mm]+ \pi_{i, 0101}^{[j]} f_{i,  0,0, 0101}^{[j]}+ \pi_{i, 0111}^{[j]} f_{i,  0,0, 0111}^{[j]}
\end{pmatrix}},
\end{aligned} 
\end{align*}
where $0stu\in \{ {0000}, {0001}, {0011}, {0101}, {0111}\}$.

In the M-step, $\boldsymbol{\theta}^{[j+1]}$ can be obtained by maximizing the following summation expression using weighted multinomial logistic regression,
    \begin{gather*} 
\sum _{i}\begin{bmatrix}\mathbb{I}(Z_i=1,M_i=1,Y_i=1)
\sum_{rst1\in \{ {0001}, {0011}, {0101}, {0111}, {1111}\}}\operatorname{pr} ^{[j]}\left(G_i=rst1\mid-\right)\log\left\{\pi_{rst1}( {\boldsymbol X}_i ;  {\boldsymbol \theta
})\right\}\\\addlinespace[2.5mm]+
  \mathbb{I}(Z_i=1,M_i=1,Y_i=0)\operatorname{pr} ^{[j]}\left(G_i=0000\mid-\right)\log\left\{\pi_{0000}( {\boldsymbol X}_i ;  {\boldsymbol \theta
  })\right\}\\\addlinespace[2.5mm]+
  \mathbb{I}(Z_i=1,M_i=0,Y_i=1)\sum_{rs1u\in \{  {0011},  {0111}, {1111}\}}\operatorname{pr} ^{[j]}\left(G_i=rs1u\mid-\right)\log\left\{\pi_{rs1u}( {\boldsymbol X}_i ;  {\boldsymbol \theta
  })\right\}\\\addlinespace[2.5mm]+\mathbb{I}(Z_i=1,M_i=0,Y_i=0)\sum_{rs0u\in \{  {0000},  {0001}, {0101}\}}\operatorname{pr} ^{[j]}\left(G_i=rs0u\mid-\right)\log\left\{\pi_{rs0u}( {\boldsymbol X}_i ;  {\boldsymbol \theta
  })\right\}\\\addlinespace[2.5mm]+
  \mathbb{I}(Z_i=0,M_i=1,Y_i=1)\sum_{r1tu\in \{  {0101},  {0111}, {1111}\}}\operatorname{pr} ^{[j]}\left(G_i=r1tu\mid-\right)\log\left\{\pi_{r1tu}( {\boldsymbol X}_i ;  {\boldsymbol \theta
  })\right\}\\\addlinespace[2.5mm]+\mathbb{I}(Z_i=0,M_i=1,Y_i=0)\sum_{r0tu\in \{  {0000},  {0001}, {0011}\}}\operatorname{pr} ^{[j]}\left(G_i=r0tu\mid-\right)\log\left\{\pi_{r0tu}( {\boldsymbol X}_i ;  {\boldsymbol \theta
  })\right\}\\\addlinespace[2.5mm]+ 
  \mathbb{I}(Z_i=0,M_i=0,Y_i=1)\operatorname{pr} ^{[j]}\left(G_i=1111\mid-\right)\log\left\{\pi_{1stu}( {\boldsymbol X}_i ;  {\boldsymbol \theta
  })\right\}\\\addlinespace[2.5mm]+\mathbb{I}(Z_i=0,M_i=0,Y_i=0)\sum_{0stu\in \{ {0000}, {0001}, {0011}, {0101}, {0111}\}}\operatorname{pr} ^{[j]}\left(G_i=0stu\mid-\right)\log\left\{\pi_{0stu}( {\boldsymbol X}_i ;  {\boldsymbol \theta
  })\right\}
\end{bmatrix}. 
\end{gather*} 
For example, for unit \(i\) with \( (Z_i = 1, M_i = 0, Y_i = 1, \boldsymbol{X}_i,W_i) \), we can create three observations: \( (Z_i = 1, M_i = 0, Y_i = 1, \boldsymbol{X}_i,W_i, G_i = 0011) \), \( (Z_i = 1, M_i = 0, Y_i = 1, \boldsymbol{X}_i,W_i, G_i = 0111) \), and \( (Z_i = 1, M_i = 0, Y_i = 1, \boldsymbol{X}_i,W_i, G_i = 1111) \), with corresponding weights \( \operatorname{pr}^{[j]}(G_i = 0011 \mid -) \), \( \operatorname{pr}^{[j]}(G_i = 0111 \mid -) \), and \( \operatorname{pr}^{[j]}(G_i = 1111 \mid -) \) used in the weighted multinomial logistic regression. Similarly, for unit \(i\) with \( (Z_i = 1, M_i = 0, Y_i = 0, \boldsymbol{X}_i,W_i) \), we create three observations: \( (Z_i = 1, M_i = 0, Y_i = 0, \boldsymbol{X}_i,W_i, G_i = 0000) \), \( (Z_i = 1, M_i = 0, Y_i = 0, \boldsymbol{X}_i,W_i, G_i = 0001) \), and \( (Z_i = 1, M_i = 0, Y_i = 0, \boldsymbol{X}_i,W_i, G_i = 0101) \), with corresponding weights \( \operatorname{pr}^{[j]}(G_i = 0000 \mid -) \), \( \operatorname{pr}^{[j]}(G_i = 0001 \mid -) \), and \( \operatorname{pr}^{[j]}(G_i = 0101 \mid -) \) used in the weighted multinomial logistic regression. 

 Also, $\boldsymbol\beta^{[j+1]}$ can be obtained by minimizing the following summation expression using weighted least squares,
    \begin{gather*} 
\sum _{i}\begin{bmatrix}\mathbb{I}(Z_i=1,M_i=1,Y_i=1)
\sum_{rst1\in \{ {0001}, {0011}, {0101} \}}{\operatorname{pr} ^{[j]}\left(G_i=rst1\mid-\right)}{\left(W_i - \boldsymbol\mu_{1,1,rst1}^\T\boldsymbol{X}_i\right)^2}/{2\sigma_{1,1,rst1}^2}\\\addlinespace[2.5mm]+\mathbb{I}(Z_i=1,M_i=1,Y_i=1)
\sum_{rst1\in \{  {0111}, {1111}\}}{\operatorname{pr} ^{[j]}\left(G_i=rst1\mid-\right)}{\left(W_i - \boldsymbol\mu_{1,1,rst1}^\T\boldsymbol{X}_i\right)^2}/{2\sigma_{1,1,rst1}^2}\\\addlinespace[2.5mm]+
  \mathbb{I}(Z_i=1,M_i=1,Y_i=0)\operatorname{pr} ^{[j]}\left(G_i=0000\mid-\right)   {\left(W_i - \boldsymbol\mu_{1,1,0000}^\T\boldsymbol{X}_i\right)^2}/{2\sigma_{1,1,0000}^2}\\\addlinespace[2.5mm]+ 
  \mathbb{I}(Z_i=1,M_i=0,Y_i=1)\sum_{rs1u\in \{  {0011},  {0111}, {1111}\}}\operatorname{pr} ^{[j]}\left(G_i=rs1u\mid-\right){\left(W_i - \boldsymbol\mu_{1,0,rs1u}^\T\boldsymbol{X}_i\right)^2}/{2\sigma_{1,0,rs1u}^2}\\\addlinespace[2.5mm]+\mathbb{I}(Z_i=1,M_i=0,Y_i=0)\sum_{rs0u\in \{  {0000},  {0001}, {0101}\}}\operatorname{pr} ^{[j]}\left(G_i=rs0u\mid-\right){\left(W_i - \boldsymbol\mu_{1,0,rs0u}^\T\boldsymbol{X}_i\right)^2}/{2\sigma_{1,0,rs0u}^2}\\\addlinespace[2.5mm]+
  \mathbb{I}(Z_i=0,M_i=1,Y_i=1)\sum_{r1tu\in \{  {0101},  {0111}, {1111}\}}\operatorname{pr} ^{[j]}\left(G_i=r1tu\mid-\right){\left(W_i - \boldsymbol\mu_{0,1,r1tu}^\T\boldsymbol{X}_i\right)^2}/{2\sigma_{0,1,r1tu}^2}\\\addlinespace[2.5mm]+\mathbb{I}(Z_i=0,M_i=1,Y_i=0)\sum_{r0tu\in \{  {0000},  {0001}, {0011}\}}\operatorname{pr} ^{[j]}\left(G_i=r0tu\mid-\right){\left(W_i - \boldsymbol\mu_{0,1,r0tu}^\T\boldsymbol{X}_i\right)^2}/{2\sigma_{0,1,r0tu}^2}\\\addlinespace[2.5mm]+ 
  \mathbb{I}(Z_i=0,M_i=0,Y_i=1)\operatorname{pr} ^{[j]}\left(G_i=1111\mid-\right){\left(W_i - \boldsymbol\mu_{0,0,1111}^\T\boldsymbol{X}_i\right)^2}/{2\sigma_{0,0,1111}^2}\\\addlinespace[2.5mm]+\mathbb{I}(Z_i=0,M_i=0,Y_i=0)\sum_{0stu\in \{ {0000}, {0001}, {0011} \}}\operatorname{pr} ^{[j]}\left(G_i=0stu\mid-\right){\left(W_i - \boldsymbol\mu_{0,0,0stu}^\T\boldsymbol{X}_i\right)^2}/{2\sigma_{0,0,0stu}^2}\\\addlinespace[2.5mm]+\mathbb{I}(Z_i=0,M_i=0,Y_i=0)\sum_{0stu\in \{  {0101}, {0111}\}}\operatorname{pr} ^{[j]}\left(G_i=0stu\mid-\right){\left(W_i - \boldsymbol\mu_{0,0,0stu}^\T\boldsymbol{X}_i\right)^2}/{2\sigma_{0,0,0stu}^2}
\end{bmatrix}. 
\end{gather*} 
For example, for unit \(i\) with \( (Z_i = 1, M_i = 0, Y_i = 1, \boldsymbol{X}_i,W_i) \), we can apply the weight  \( \operatorname{pr}^{[j]}(G_i = 0011 \mid -) \) in the weighted linear regression to estimate the parameters \( (\boldsymbol{\mu}^{[j+1]}_{1,0,0011}, \sigma^{2[j+1]}_{1,0,0011}) \); for unit \(i\) with \( (Z_i = 1, M_i = 0, Y_i = 1, \boldsymbol{X}_i,W_i) \), we can apply the weight   \( \operatorname{pr}^{[j]}(G_i = 0111 \mid -) \)  in the weighted linear regression to estimate  the parameters \( (\boldsymbol{\mu}^{[j+1]}_{1,0,0111}, \sigma^{2[j+1]}_{1,0,0111}) \); for unit \(i\) with \( (Z_i = 1, M_i = 0, Y_i = 1, \boldsymbol{X}_i,W_i) \), we can apply the weight \( \operatorname{pr}^{[j]}(G_i = 1111 \mid -) \) in the weighted linear regression to estimate   the parameters \( (\boldsymbol{\mu}^{[j+1]}_{1,0,1111}, \sigma^{2[j+1]}_{1,0,1111}) \).
 
\subsection{Estimating posterior probabilities}
\label{ssec:est-prop}
Let $\hat{\boldsymbol\theta} $ and $\hat{\boldsymbol\beta} $ denote the MLEs obtained by the EM algorithm in the previous subsection.
After obtaining $\hat{\boldsymbol\theta} $ and $\hat{\boldsymbol\beta} $, it will be straightforward for us to estimate the causal quantities of interest $\pi_{rstu}({\mathcal{O}})$. For example, for ${\mathcal{O}}=(Z=1,M=1,Y=1)$ and $\tilde {\mathcal{O}}=({\mathcal{O}},W=w)$, we can estimate the posterior probabilities $\pi_{rstu}( {\mathcal{O}} )$  and $\pi_{rstu}( \tilde{\mathcal{O}} )$ through the following steps:
\begin{itemize}
    \item[] {\it Step 1}. For each unit $i$, we can estimate $\pr(G_i=rst1\mid  Z_i= M_i= Y_i=1,\boldsymbol{X}_i;\hat{\boldsymbol\theta})$ as follows:
 $$\hat{\pr} (G_i=rst1\mid Z_i= M_i= Y_i=1, \boldsymbol{X}_i;\hat{\boldsymbol\theta})= \dfrac{  \pr(G_i=rst1\mid \boldsymbol X_i;\hat{ \boldsymbol\theta})} {\sum_{rst1\in \{ {0001}, {0011}, {0101}, {0111}, {1111}\}} \pr(G_i=rst1\mid\boldsymbol X_i;\hat{\boldsymbol\theta})} .$$ 

   \item[] {\it Step 2}. For each unit $i$, we can estimate $\pi_{rst1}( {\mathcal{O}} )$ using the following expression: $$\hat\pi_{rst1}( {\mathcal{O}} )= \dfrac{\sum_{i=1}^n {\mathbb{I}(Z_i=M_i=Y_i=1)\pr(G_i=rst1\mid  Z_i= M_i= Y_i=1,\boldsymbol{X}_i;\hat{\boldsymbol\theta} )}}{\sum_{i=1}^n {\mathbb{I}(Z_i=M_i=Y_i=1) }} .$$       \item[] {\it Step 3}. For each unit $i$, we can estimate $\pr(W_i=w,G_i=rst1\mid  Z_i= M_i= Y_i=1,\boldsymbol{X}_i;\hat{\boldsymbol\theta},\hat{\boldsymbol\beta})$ as follows:\begin{align*}
     \hat{\pr}& (W_i=w,G_i =rst1\mid Z_i =M_i =Y_i =1,\boldsymbol X_i;\hat{\boldsymbol\theta},\hat{\boldsymbol\beta} )\\&= { {f}(W_i=w\mid  Z_i =M_i =Y_i =1, G_i=rst1, \boldsymbol X_i;\hat{\boldsymbol\beta})\pr(G_i=rst1\mid  Z_i =M_i =Y_i =1,  \boldsymbol   X_i;\hat{ \boldsymbol\theta})}  .
 \end{align*}
   \item[] {\it Step 4}. We then estimate $\pr(W=w,G=rstu\mid  {\mathcal{O}} )$ using the following expression: \begin{align*}
       \hat{\pr}& (W=w,G =rst1\mid  {\mathcal{O}})\\&=\dfrac{\sum_{i=1}^n {\mathbb{I}(Z_i=M_i=Y_i=1)\pr(W_i=w,G_i=rst1\mid  Z_i= M_i= Y_i=1,\boldsymbol{X}_i;\hat{\boldsymbol\theta} ,\hat{\boldsymbol\beta})}}{\sum_{i=1}^n {\mathbb{I}(Z_i=M_i=Y_i=1)}} .
   \end{align*}    \item[] {\it Step 5}. We then estimate $\pi_{rst1}( \tilde {\mathcal{O}} )$ using the following expression: \begin{align*}
  \hat\pi_{rst1}( \tilde {\mathcal{O}} )= \dfrac{ \hat{\pr}  \left(W=w,G =rst1\mid  {\mathcal{O}}\right)}{\sum_{rst1\in \{ {0001}, {0011}, {0101}, {0111}, {1111}\}} \hat{\pr}  \left(W=w,G =rst1\mid  {\mathcal{O}}\right)} .
   \end{align*}
\end{itemize}
 \section{Estimation without monotonicity assumption  \ref{ass:monotonicity}}
\label{ssec: estimation-nomo}
In this section, parallel to the estimating procedure of Section \ref{ssec: estimation}, we also introduce the baseline  covariates \(\boldsymbol X\). Without the monotonicity assumption  \ref{ass:monotonicity}, 
we can model  $\pr(G=rstu\mid \boldsymbol{X} )$ as a sixteen-level multinomial logistic model: $$\pi_{rstu}( {\boldsymbol X} ;  {\boldsymbol \theta
}):=\pr(G=rstu\mid {\boldsymbol X};  {\boldsymbol \theta
} )=\dfrac{\mathrm{exp}(  {\boldsymbol \theta
}_{rstu}^\T  {\boldsymbol X})}{\sum_{r's't'u'}\mathrm{exp}(  {\boldsymbol\theta
}_{r's't'u'}^\T  {\boldsymbol X})}, $$
    where $r,s,t,u\in \{0,1\}$ and ${\boldsymbol\theta
}_{0000}=\boldsymbol 0$ for identiﬁcation.

To satisfy identification assumption \ref{assump:equal-prop}, without the monotonicity assumption, we need to impose additional modeling restrictions. For example, we can model the conditional probability density function $ f(W \mid Z=z, M=m, G=rstu, \boldsymbol{X}) $ as:
\begin{align*} 
    {f} (W\mid Z=z,M=m, G=rstu, \boldsymbol{X};\boldsymbol\mu_{z,m,rstu},\sigma _{z,m, rstu}^2  ) = \frac{\exp\left\{- {(W - \boldsymbol\mu_{z,m, rstu}^\T\boldsymbol{X})^2}/{2\sigma^2_{z,m, rstu}}\right\}}{\sqrt{2\pi\sigma^2_{z,m, rstu}}} . 
\end{align*} Let $\boldsymbol{\beta}$ denote the parameter vector consisting of all $\boldsymbol{\mu}_{z,m,rstu}$ and $\sigma_{z,m,z,m, rstu}^2$ for $ z,m,r,s,t,u\in\{0,1\}$.  We also introduce the shorthand notation  $\pi_{i, rstu} = \pr(G_i = rstu \mid \boldsymbol{X}_i ;  {\boldsymbol \theta
})$ and $f_{i,z,m, rstu} = f(W_i \mid Z_i=z,M_i=m, G_i = rstu, \boldsymbol{X}_i;\boldsymbol\mu_{z,m, rstu},\sigma _{z,m, rstu}^2 )$ to simplify the exposition.

In the E-step, we first calculate the conditional probabilities  of latent classes given the observed data. Let $\pi_{i, rstu}^{[j]}$ and $f_{i,z,m, rstu}^{[j]}$ denote the estimates at the $j$-th iteration: 
 $$\pi_{i,rstu}^{[j]} =\dfrac{\mathrm{exp}(  {\boldsymbol \theta
}_{rstu}^{[j]\T}  {\boldsymbol X}_i )}{\sum_{r's't'u'}\mathrm{exp}(  {\boldsymbol\theta
}_{r's't'u'}^{[j]\T} {\boldsymbol X}_i )},~~
   f_{i,z,m, rstu}^{[j]}= \frac{\exp\left\{- {(W_i  - \boldsymbol\mu_{z,m,rstu}^{[j]\T}\boldsymbol{X}_i )^2}/{2\sigma^{2[j]}_{z,m, rstu}}\right\}}{\sqrt{2\pi\sigma^{2[j]}_{z,m, rstu}}} . $$  For  $(Z_i = 1, M_i = 1, Y_i = 1,\boldsymbol{X}_i, W_i)$, we have
\begin{align*} 
      \begin{aligned}
\operatorname{pr} ^{[j]}\left(G_i=rst1\mid -\right)=\frac{\pi_{i, rst1}^{[j]} f_{i,1,1,  rst1}^{[j]}}{\sum_{r,s,t=0}^1\pi_{i, rst1}^{[j]} f_{i,1,1,  rst1}^{[j]}},~~~~\text{for}~~r,s,t\in\{0,1\},
\end{aligned} 
\end{align*} 
For $(Z_i=1,M_i=1,Y_i=0,\boldsymbol{X}_i,W_i)$,  we have \begin{align*} 
      \begin{aligned}
\operatorname{pr} ^{[j]}\left(G_i=rst0\mid -\right)=\frac{\pi_{i, rst0}^{[j]} f_{i,1,1,  rst0}^{[j]}}{\sum_{r,s,t=0}^1\pi_{i, rst0}^{[j]} f_{i,1,1,  rst0}^{[j]}},~~~~\text{for}~~r,s,t\in\{0,1\},
\end{aligned} 
\end{align*}
  For  $(Z_i = 1, M_i = 0, Y_i = 1,\boldsymbol{X}_i, W_i)$, we have
\begin{align*} 
      \begin{aligned}
\operatorname{pr} ^{[j]}\left(G_i=rs1u\mid -\right)=\frac{\pi_{i, rs1u}^{[j]} f_{i,1,0,  rs1u}^{[j]}}{\sum_{r,s,u=0}^1\pi_{i, rs1u}^{[j]} f_{i,1,0,  rs1u}^{[j]}},~~~~\text{for}~~r,s,u\in\{0,1\},
\end{aligned}  
\end{align*} 
For $(Z_i=1,M_i=0,Y_i=0,\boldsymbol{X}_i,W_i)$,  we have 
\begin{align*} 
      \begin{aligned} 
\operatorname{pr} ^{[j]}\left(G_i=rs0u\mid -\right)=\frac{\pi_{i, rs0u}^{[j]} f_{i,1,0,  rs0u}^{[j]}}{\sum_{r,s,u=0}^1\pi_{i, rs0u}^{[j]} f_{i,1,0,  rs0u}^{[j]}},~~~~\text{for}~~r,s,u\in\{0,1\},
\end{aligned} 
\end{align*} 
For  $(Z_i = 0, M_i = 1, Y_i = 1,\boldsymbol{X}_i, W_i)$, we have\begin{align*} 
      \begin{aligned} 
\operatorname{pr} ^{[j]}\left(G_i=r1tu\mid -\right)=\frac{\pi_{i, r1tu}^{[j]} f_{i,0,1,  r1tu}^{[j]}}{\sum_{r,t,u=0}^1\pi_{i, r1tu}^{[j]} f_{i,0,1,  r1tu}^{[j]}},~~~~\text{for}~~r,t,u\in\{0,1\},
\end{aligned} 
\end{align*}  
For $(Z_i=0,M_i=1,Y_i=0,\boldsymbol{X}_i,W_i)$,  we have 
\begin{align*} 
      \begin{aligned} 
\operatorname{pr} ^{[j]}\left(G_i=r0tu\mid -\right)=\frac{\pi_{i,r0tu}^{[j]} f_{i,0,1,  r0tu}^{[j]}}{\sum_{r,t,u=0}^1\pi_{i, r0tu}^{[j]} f_{i,0,1,  r0tu}^{[j]}},~~~~\text{for}~~r,t,u\in\{0,1\},
\end{aligned} 
\end{align*}  For  $(Z_i = 0, M_i = 0, Y_i = 1,\boldsymbol{X}_i, W_i)$, we have\begin{align*} 
      \begin{aligned}
\operatorname{pr}^{[j]} \left(G_i=1stu\mid -\right)=\frac{\pi_{i,1stu}^{[j]} f_{i,0, 0,  1stu}^{[j]}}{\sum_{s,t,u=0}^1\pi_{i, 1stu}^{[j]} f_{i,0, 0,  1stu}^{[j]}},~~~~\text{for}~~s,t,u\in\{0,1\}.
\end{aligned} 
\end{align*} 
For $(Z_i=0,M_i=0,Y_i=0,\boldsymbol{X}_i,W_i)$,  we have
\begin{align*} 
      \begin{aligned}
\operatorname{pr}^{[j]} \left(G_i=0stu\mid -\right)=\frac{\pi_{i,0stu}^{[j]} f_{i,0, 0,  0stu}^{[j]}}{\sum_{s,t,u=0}^1\pi_{i, 0stu}^{[j]} f_{i,0, 0,  0stu}^{[j]}},~~~~\text{for}~~s,t,u\in\{0,1\}.
\end{aligned} 
\end{align*}

In the M-step, $\boldsymbol{\theta}^{[j+1]}$ can be obtained by maximizing the following summation expression using weighted multinomial logistic regression,
\begin{gather*} 
\sum _{i}\begin{bmatrix}\mathbb{I}(Z_i=1,M_i=1,Y_i=1)
\sum_{r,s,t=0}^1\operatorname{pr} ^{[j]}\left(G_i=rst1\mid-\right)\log\left\{\pi_{rst1}( {\boldsymbol X}_i ;  {\boldsymbol \theta
})\right\}\\\addlinespace[2.5mm]+
  \mathbb{I}(Z_i=1,M_i=1,Y_i=0)  \sum_{r,s,t=0}^1\operatorname{pr} ^{[j]}\left(G_i=rst0\mid-\right)\log\left\{\pi_{rst0}( {\boldsymbol X}_i ;  {\boldsymbol \theta
  })\right\}\\\addlinespace[2.5mm]+
  \mathbb{I}(Z_i=1,M_i=0,Y_i=1)\sum_{r,s,u=0}^1\operatorname{pr} ^{[j]}\left(G_i=rs1u\mid-\right)\log\left\{\pi_{rs1u}( {\boldsymbol X}_i ;  {\boldsymbol \theta
  })\right\}\\\addlinespace[2.5mm]+\mathbb{I}(Z_i=1,M_i=0,Y_i=0)\sum_{r,s,u=0}^1\operatorname{pr} ^{[j]}\left(G_i=rs0u\mid-\right)\log\left\{\pi_{rs0u}( {\boldsymbol X}_i ;  {\boldsymbol \theta
  })\right\}\\\addlinespace[2.5mm]+
  \mathbb{I}(Z_i=0,M_i=1,Y_i=1)\sum_{r,t,u=0}^1\operatorname{pr} ^{[j]}\left(G_i=r1tu\mid-\right)\log\left\{\pi_{r1tu}( {\boldsymbol X}_i ;  {\boldsymbol \theta
  })\right\}\\\addlinespace[2.5mm]+\mathbb{I}(Z_i=0,M_i=1,Y_i=0)\sum_{r,t,u=0}^1 \operatorname{pr} ^{[j]}\left(G_i=r0tu\mid-\right)\log\left\{\pi_{r0tu}( {\boldsymbol X}_i ;  {\boldsymbol \theta
  })\right\}\\\addlinespace[2.5mm]+ 
  \mathbb{I}(Z_i=0,M_i=0,Y_i=1)\sum_{s,t,u=0}^1 \operatorname{pr} ^{[j]}\left(G_i=1stu\mid-\right)\log\left\{\pi_{1stu}( {\boldsymbol X}_i ;  {\boldsymbol \theta
  })\right\}\\\addlinespace[2.5mm]+\mathbb{I}(Z_i=0,M_i=0,Y_i=0)\sum_{s,t,u=0}^1\operatorname{pr} ^{[j]}\left(G_i=0stu\mid-\right)\log\left\{\pi_{0stu}( {\boldsymbol X}_i ;  {\boldsymbol \theta
  })\right\}
\end{bmatrix}. 
\end{gather*} 
For example, for unit $i$ with $(Z_i=1, M_i=1, Y_i=1, \boldsymbol{X}_i)$, we can create eight observations: $(Z_i=1, M_i=1, Y_i=1, \boldsymbol{X}_i, G_i=0001), \dots, (Z_i=1, M_i=1, Y_i=1, \boldsymbol{X}_i, G_i=1111)$, with corresponding weights $\operatorname{pr}^{[j]}(G_i=0001\mid-), \dots, \operatorname{pr}^{[j]}(G_i=1111\mid-)$ used in the weighted multinomial logistic regression. Similarly, for unit $i$ with $(Z_i=0, M_i=0, Y_i=1, \boldsymbol{X}_i)$, we create eight observations: $(Z_i=0, M_i=0, Y_i=1, \boldsymbol{X}_i, G_i=1000), \dots, (Z_i=0, M_i=0, Y_i=1, \boldsymbol{X}_i, G_i=1111)$, with corresponding weights $\operatorname{pr}^{[j]}(G_i=1000\mid-), \dots,  \operatorname{pr}^{[j]}(G_i=1111\mid-)$ used in the weighted multinomial logistic regression. Other units can be similarly created with eight observations and corresponding weights for the weighted multinomial logistic regression.

 Also, $\boldsymbol\beta^{[j+1]}$ can be obtained by minimizing the following summation expression using weighted least squares,
    \begin{gather*} 
\sum _{i}\begin{bmatrix}\mathbb{I}(Z_i=1,M_i=1,Y_i=1)
\sum_{r,s,t=0}^1{\operatorname{pr} ^{[j]}\left(G_i=rst1\mid-\right)}{\left(W_i - \boldsymbol\mu_{1,1,rst1}^\T\boldsymbol{X}_i\right)^2}/{2\sigma_{1,1,rst1}^2}\\\addlinespace[2.5mm] +
  \mathbb{I}(Z_i=1,M_i=1,Y_i=0)
\sum_{r,s,t=0}^1{\operatorname{pr} ^{[j]}\left(G_i=rst0\mid-\right)}{\left(W_i - \boldsymbol\mu_{1,1,rst0}^\T\boldsymbol{X}_i\right)^2}/{2\sigma_{1,1,rst0}^2}\\\addlinespace[2.5mm]+ 
  \mathbb{I}(Z_i=1,M_i=0,Y_i=1)
\sum_{r,s,u=0}^1\operatorname{pr} ^{[j]}\left(G_i=rs1u\mid-\right){\left(W_i - \boldsymbol\mu_{1,0,rs1u}^\T\boldsymbol{X}_i\right)^2}/{2\sigma_{1,0,rs1u}^2}\\\addlinespace[2.5mm]+\mathbb{I}(Z_i=1,M_i=0,Y_i=0)\sum_{r,s,u=0}^1\operatorname{pr} ^{[j]}\left(G_i=rs0u\mid-\right){\left(W_i - \boldsymbol\mu_{1,0,rs0u}^\T\boldsymbol{X}_i\right)^2}/{2\sigma_{1,0,rs0u}^2}\\\addlinespace[2.5mm]+
  \mathbb{I}(Z_i=0,M_i=1,Y_i=1)\sum_{r,t,u=0}^1\operatorname{pr} ^{[j]}\left(G_i=r1tu\mid-\right){\left(W_i - \boldsymbol\mu_{0,1,r1tu}^\T\boldsymbol{X}_i\right)^2}/{2\sigma_{0,1,r1tu}^2}\\\addlinespace[2.5mm]+\mathbb{I}(Z_i=0,M_i=1,Y_i=0)\sum_{r,t,u=0}^1\operatorname{pr} ^{[j]}\left(G_i=r0tu\mid-\right){\left(W_i - \boldsymbol\mu_{0,1,r0tu}^\T\boldsymbol{X}_i\right)^2}/{2\sigma_{0,1,r0tu}^2}\\\addlinespace[2.5mm]+ 
  \mathbb{I}(Z_i=0,M_i=0,Y_i=1)\sum_{s,t,u=0}^1\operatorname{pr} ^{[j]}\left(G_i=1stu\mid-\right){\left(W_i - \boldsymbol\mu_{0,0,1stu}^\T\boldsymbol{X}_i\right)^2}/{2\sigma_{0,0,1stu}^2}\\\addlinespace[2.5mm]+\mathbb{I}(Z_i=0,M_i=0,Y_i=0)\sum_{s,t,u=0}^1\operatorname{pr} ^{[j]}\left(G_i=0stu\mid-\right){\left(W_i - \boldsymbol\mu_{0,0,0stu}^\T\boldsymbol{X}_i\right)^2}/{2\sigma_{0,0,0stu}^2} 
\end{bmatrix}. 
\end{gather*} 
To estimate the parameters $\boldsymbol{\mu}_{1,1,1111}^{[j+1]}$ and $\sigma_{1,1,1111}^{2[j+1]}$, we can use individuals from the subpopulations $(Z_i=1, M_i=1, Y_i=1, \boldsymbol{X}_i, W_i)$, $(Z_i=0, M_i=1, Y_i=1, \boldsymbol{X}_i, W_i)$, $(Z_i=1, M_i=0, Y_i=1, \boldsymbol{X}_i, W_i)$, and $(Z_i=0, M_i=0, Y_i=1, \boldsymbol{X}_i, W_i)$, respectively. The estimation process is performed using four $j$-th iteration  weights $\operatorname{pr}^{[j]}\left(G_i=1111 \mid Z_i=1, M_i=1, Y_i=1, \boldsymbol{X}_i, W_i\right)$, $ \operatorname{pr}^{[j]}(G_i=1111 \mid Z_i=0, M_i=1, Y_i=1, \boldsymbol{X}_i, W_i)$,  $ \operatorname{pr}^{[j]}\left(G_i=1111 \mid Z_i=1, M_i=0, Y_i=1, \boldsymbol{X}_i, W_i\right)$, and $\operatorname{pr}^{[j]}\left(G_i=1111 \mid Z_i=0, M_i=0, Y_i=1, \boldsymbol{X}_i, W_i\right)$, to perform weighted least squares estimation. 

After obtaining  $\hat{\boldsymbol\theta} $ and $\hat{\boldsymbol\beta} $, we can similarly estimate the posterior probabilities as in Section \ref{ssec:est-prop}.
\section{Numerical experiments}
\label{sec:sim-detail}
\subsection{Simulation settings}

\label{ssec:simulation-settings}
In this section, we perform simulation studies to assess the finite sample performance of the proposed method described in the previous sections. We consider two sample sizes: 500 and 1000, with 500 simulations conducted for each size. Specifically, we use the following data-generating mechanism, with the detailed parameter settings provided in Section \ref{SEC:sim-detial} of the Supplementary Material.
\begin{enumerate}
\item  We generate the covariates $X_{ 1},X_{2}\sim N(0,1)$, and let ${\boldsymbol X}=(1,X_1,X_2)^\T$.
\item  We generate the binary  exposures $Z$ and   $M$ from a  multinomial logistic model:
  $$ \pr(Z=z,M=m\mid {\boldsymbol X})={\mathrm{exp}\left({\boldsymbol \alpha}_{z,m }^\T{\boldsymbol X} \right)}\Big/{\textstyle\sum_{z{},m{}=0}^1 \mathrm{exp}\left({\boldsymbol \alpha}_{z,m }^\T{\boldsymbol X} \right)},$$     where  $ \boldsymbol \alpha_{00 } =(0,0,0)^\T $ is used for identification.
  \item We generate the latent variable $G={rstu} $ with $ r, s, t, u \in \{0,1\} $  from a multinomial logistic model  under the monotonicity assumption  \ref{ass:monotonicity}:
    \begin{equation*} 
  \pr(G=rstu\mid {\boldsymbol X})={\mathrm{exp}\left({\boldsymbol  \theta}_{rstu }^\T{\boldsymbol X} \right)}\Big/{\textstyle\sum_{r{},s{},t{},u{}=0}^1 \mathrm{exp}\left({\boldsymbol  \theta}_{r{} s{} t{} u{}}^\T{\boldsymbol  X} \right)},
  \end{equation*}
    where  $ \boldsymbol  \theta_{0000}=(0,0,0)^\T$ is used for identification.
    We then generate the binary outcomes $Y $  according to Table \ref{tab: probabilities}.
    \item We generate the    secondary outcome $W$ from  the  normal   distribution: \begin{align}
    \label{eq:second-distribution}
    W\mid Z=z,M=m,G=rstu, {\boldsymbol X}\sim {{\boldsymbol \mu}}_{z,m,rstu}^\T {\boldsymbol X}+\epsilon, ~~~\epsilon\sim  N( 0,\sigma_{z,m,rstu}^2).
    \end{align} 
    \end{enumerate}

\subsection{Simulation results for $\pi(\mathcal{O})$}
\label{ssec:results-pi-O}

\begin{table}[h]
\centering \centering \caption{The estimation results for the proportion $ \pi_{rstu} $ with $ r, s, t, u \in \{0,1\} $, where the error distribution is generated from five distributions. The sample sizes for the first and second tables are 500 and 1000, respectively.}
\label{fig:enter-label}\resizebox{0.9480\textwidth}{!}{
\begin{tabular}{lcccccc}
  \toprule\addlinespace[1mm]
 & $\pi_{0000}$ &  $\pi_{0001}$& $\pi_{0011}$ &  $\pi_{0101}$ &  $\pi_{0111}$ & $\pi_{1111}$\\ \addlinespace[1mm]
  \cline{2-7}\addlinespace[1mm]
(1)  Normal distribution & 0.44 (1.63) & 1.81 (2.64) & -2.02 (2.71) & -0.25 (2.40) & -0.37 (2.99) & 0.40 (2.42) \\
(2)  $t$ distribution   & 0.56 (1.74) & 1.82 (2.78) & -1.68 (2.80) & -0.53 (2.44) & -0.65 (2.90) & 0.49 (2.32) \\
(3) Uniform distribution & 0.06 (1.68) & 1.41 (2.14) & -0.60 (2.25) & -0.32 (1.82) & -0.74 (2.21) & 0.18 (1.80) \\
(4) Bernoulli distribution  & 0.40 (1.69) & 1.88 (2.71) & -1.83 (2.68) & -0.67 (2.29) & -0.04 (2.82) & 0.28 (2.14) \\
 (5) Gamma distribution  & 2.29 (2.11) & 0.93 (3.29) & -1.91 (3.02) & -0.66 (2.70) & -1.85 (3.84) & 1.21 (2.68)\\ \addlinespace[0.5mm]
  \toprule\addlinespace[1mm]
 & $\pi_{0000}$ &  $\pi_{0001}$& $\pi_{0011}$ &  $\pi_{0101}$ &  $\pi_{0111}$ & $\pi_{1111}$\\ \addlinespace[1mm]
  \cline{2-7}\addlinespace[1mm]
  (1)  Normal distribution & 0.34 (1.17) & 1.69 (1.89) & -1.67 (1.93) & -0.26 (1.60) & -0.35 (1.92) & 0.25 (1.56) \\
  (2)  $t$ distribution  & 0.62 (1.30) & 1.66 (2.08) & -1.39 (1.94) & -0.65 (1.58) & -0.51 (1.93) & 0.27 (1.53) \\
  (3) Uniform distribution  & 0.05 (1.16) & 1.29 (1.57) & -0.30 (1.51) & -0.44 (1.33) & -0.65 (1.64) & 0.05 (1.28) \\
  (4) Bernoulli distribution  & 0.38 (1.20) & 1.76 (1.93) & -1.54 (1.78) & -0.71 (1.61) & 0.01 (1.95) & 0.09 (1.49) \\
  (5) Gamma distribution   & 2.72 (1.50) & 0.46 (2.22) & -1.67 (2.01) & -0.60 (1.86) & -2.33 (2.71) & 1.42 (1.91) \\ \addlinespace[0.5mm]
   \bottomrule\addlinespace[1mm]
\end{tabular}
}
\end{table}In this section, we first consider {simulation results for $\pi(\mathcal{O})$}. 
When the normality assumption is violated, it affects the validity of the proposed method discussed in Section \ref{ssec: estimation}, which is based solely on Gaussian mixture models.  In this case, to assess the sensitivity of the proposed estimation method, we will examine the different settings of the error distribution in \eqref{eq:second-distribution}, which represents a deviation from normality while preserving the other data-generating mechanisms. Specifically, we considered the following five scenarios:
(1) the noise $\epsilon_{rstu}$  follows a $t$-distribution with 5 degrees of freedom;
(2) the noise  $\epsilon_{rstu}$  is generated from a uniform distribution over the interval $[-1, 1]$;
(3) the noise  $\epsilon_{rstu}$ follows a Bernoulli distribution with equal probabilities: $\pr(\epsilon_{rstu} = 1) = \pr(\epsilon_{rstu} = -1) = 0.5$.
(4) the noise  $\epsilon_{rstu}$  follows a Gamma distribution with a shape parameter of 2 and a rate parameter of 0.5.


 We present the bias and standard error  (in parentheses)  of $\pi_{rstu}$ with $r,s,t,u\in\{0,1\}$ in Table  \ref{fig:enter-label}, with all estimation results scaled by  100.  We also present the estimation results of the posterior probabilities   $\pi_{rstu}( {\mathcal{O}})$, specifically for    ${\mathcal{O}}=(Z=1,M=1,Y=1)$ and ${\mathcal{O}}=(Z=0,M=0,Y=0)$. Posterior probabilities for other evidence sets can be similarly  estimated using the equations in Section \ref{sec:secondary-outcome} of the main text. 
The estimation results of    $\pr(G=rst1\mid Z=1,M=1,Y=1)$ and $\pr(G=0stu\mid Z=0,M=0,Y=0)$   are presented in Tables \ref{tab:like-comp-GN} and \ref{tab:like-comp-GS}. All results are based on 500 bootstrap runs. We find that all the estimates are very close to the true values and exhibit small estimation bias regardless of whether the noise follows a normal distribution or not. In addition, the results for non-normal noise are comparable to those for normal noise regarding the standard error. Moreover, bias and standard errors become smaller as the sample size increases. These simulation results indicate that, in most cases, our proposed estimation method for $\pi(\mathcal{O})$ is not highly sensitive to violations of Assumption \ref{assumption:normal}.

\begin{table}[t]
\centering
\centering \centering
\caption{The estimation results of  $\pr(G=rst1\mid Z=1,M=1,Y=1)$ for $r,s,t\in\{0,1\}$, where the error distribution is generated from five distributions. The sample sizes for the first and second tables are 500 and 1000, respectively.}
\label{tab:like-comp-GN}\resizebox{1\textwidth}{!}{
  \begin{tabular}{lccccc}
  \toprule\addlinespace[1mm]
  & $(r,s,t)=(0,0,1)$ &  $(r,s,t)=(0,1,1)$  &  $(r,s,t)=(1,0,1)$  &  $(r,s,t)=(1,1,1)$  &  $(r,s,t)=(1,1,1)$ \\ \addlinespace[1mm]
  \cline{2-6}\addlinespace[1mm]
  (1)  Normal distribution  & 2.24 (3.09) & -2.30 (3.16) & -0.22 (2.82) & -0.29 (3.57) & 0.57 (2.88) \\
  (2)  $t$ distribution  & 2.26 (3.21) & -1.87 (3.27) & -0.52 (2.88) & -0.58 (3.48) & 0.71 (2.82) \\
  (3) Uniform distribution & 1.68 (2.51) & -0.70 (2.60) & -0.37 (2.12) & -0.85 (2.60) & 0.23 (2.11) \\
  (4) Bernoulli distribution  & 2.30 (3.13) & -2.08 (3.13) & -0.72 (2.70) & 0.09 (3.36) & 0.42 (2.57) \\
  (5) Gamma distribution & 1.56 (3.76) & -1.73 (3.60) & -0.37 (3.21) & -1.43 (4.76) & 1.98 (3.41) \\  \addlinespace[0.5mm]
  \toprule\addlinespace[1mm]
  & $(r,s,t)=(0,0,1)$ &  $(r,s,t)=(0,1,1)$  &  $(r,s,t)=(1,0,1)$  &  $(r,s,t)=(1,1,1)$  &  $(r,s,t)=(1,1,1)$ \\ \addlinespace[1mm]
  \cline{2-6}\addlinespace[1mm]
  (1)  Normal distribution & 2.07 (2.18) & -1.90 (2.27) & -0.25 (1.88) & -0.30 (2.27) & 0.37 (1.88) \\
  (2)  $t$ distribution & 2.09 (2.37) & -1.50 (2.29) & -0.65 (1.87) & -0.39 (2.30) & 0.45 (1.87) \\
  (3) Uniform distribution & 1.53 (1.81) & -0.34 (1.77) & -0.51 (1.56) & -0.75 (1.91) & 0.07 (1.52) \\
  (4) Bernoulli distribution & 2.16 (2.20) & -1.73 (2.11) & -0.77 (1.90) & 0.15 (2.30) & 0.19 (1.81) \\
  (5) Gamma distribution & 1.11 (2.54) & -1.33 (2.40) & -0.21 (2.23) & -1.90 (3.32) & 2.32 (2.47) \\
  \bottomrule
  \end{tabular}}
\end{table}

\begin{table}[h]
\centering
\centering \centering \caption{The estimation results of   $\pr(G=0stu\mid Z=0,M=0,Y=0)$ for $ s,t,u\in\{0,1\}$,  where the error distribution is generated from five distributions. The sample sizes for the first and second tables are 500 and 1000, respectively.}
\label{tab:like-comp-GS}\resizebox{1\textwidth}{!}{
  \begin{tabular}{lccccc}
  \toprule\addlinespace[1mm]
  & $(s,t,u)=(0,0,0)$ &  $(s,t,u)=(0,0,1)$  &  $(s,t,u)=(0,1,1)$  &  $(s,t,u)=(1,0,1)$  &  $(s,t,u)=(1,1,1)$ \\ \addlinespace[1mm]
  \cline{2-6}\addlinespace[1mm]
  (1)  Normal distribution  & 0.61 (1.98) & 2.22 (3.05) & -2.27 (3.19) & -0.22 (2.80) & -0.33 (3.27) \\
  (2)  $t$ distribution & 0.78 (2.15) & 2.23 (3.21) & -1.86 (3.28) & -0.53 (2.86) & -0.62 (3.20) \\
  (3) Uniform distribution  & 0.12 (1.95) & 1.69 (2.45) & -0.65 (2.64) & -0.34 (2.13) & -0.81 (2.49) \\
  (4) Bernoulli distribution & 0.54 (2.04) & 2.26 (3.13) & -2.08 (3.13) & -0.73 (2.69) & 0.02 (3.11) \\
  (5) Gamma distribution & 3.02 (2.74) & 1.33 (3.79) & -1.97 (3.56) & -0.56 (3.18) & -1.82 (4.30) \\ \addlinespace[0.5mm]
  \toprule\addlinespace[1mm]
  & $(s,t,u)=(0,0,0)$ &  $(s,t,u)=(0,0,1)$  &  $(s,t,u)=(0,1,1)$  &  $(s,t,u)=(1,0,1)$  &  $(s,t,u)=(1,1,1)$ \\ \addlinespace[1mm]
  \cline{2-6}\addlinespace[1mm]
  (1)  Normal distribution  & 0.46 (1.42) & 2.04 (2.20) & -1.90 (2.23) & -0.26 (1.87) & -0.34 (2.07) \\
  (2)  $t$ distribution  & 0.79 (1.60) & 2.00 (2.41) & -1.57 (2.25) & -0.71 (1.85) & -0.52 (2.11) \\
  (3) Uniform distribution & 0.07 (1.37) & 1.52 (1.83) & -0.34 (1.76) & -0.50 (1.54) & -0.75 (1.80) \\
  (4) Bernoulli distribution & 0.47 (1.47) & 2.08 (2.23) & -1.78 (2.08) & -0.82 (1.88) & 0.04 (2.10) \\
  (5) Gamma distribution & 3.57 (2.00) & 0.83 (2.57) & -1.63 (2.37) & -0.45 (2.20) & -2.33 (2.97) \\
  \bottomrule
  \end{tabular}}

\end{table}
\subsection{Simulation results for $\pi(\tilde{\mathcal{O}})$}
\label{ssec:results-pi-tildeO}
{\black In this section, we present the estimation results for $\pi_{rstu}(\tilde{\mathcal{O}})$ stratified by tertiles of the secondary outcome $W$. The estimation results of $\pr(G=rst1\mid Z=1,M=1,Y=1, w_{\min} \leq W \leq w_{1/3})$, $\pr(G=rst1\mid Z=1,M=1,Y=1, w_{1/3} < W \leq w_{2/3})$, and $\pr(G=rst1\mid Z=1,M=1,Y=1, w_{2/3} < W \leq w_{\max})$ are presented in Tables \ref{tab:S4_Q1}, \ref{tab:S4_Q2}, and \ref{tab:S4_Q3}, respectively, where $w_{1/3}$ and $w_{2/3}$ denote the first and second tertiles of $W$ among individuals with $(Z=1, M=1, Y=1)$. Similarly, the estimation results of $\pr(G=0stu\mid Z=0,M=0,Y=0, W)$ stratified by tertiles of $W$ among individuals with $(Z=0, M=0, Y=0)$ are presented in Tables \ref{tab:S5_Q1}, \ref{tab:S5_Q2}, and \ref{tab:S5_Q3}.

All estimation results are based on 500 simulation runs. In each run, we generate five datasets under five different error distributions, including normal, $t$, uniform, Bernoulli, and gamma distributions, respectively. We then compute the true posterior probability values for each distribution and estimate them using our proposed method, which assumes normality. The reported bias is defined as the difference between these estimates and the true values under the actual data-generating distribution.

As expected, when the data are generated under the normal distribution (which aligns with our modeling assumption), the estimates exhibit very small bias (close to zero) across all tertiles of $W$ and all latent types, with standard errors generally below 5\%. When the data are generated under non-normal error distributions but the estimation procedure assumes normality (i.e., model misspecification), we observe varying degrees of bias depending on the true error distribution and the tertile of $W$. The standard errors remain relatively small (mostly below 13\%) across all scenarios and consistently decrease as the sample size increases from 500 to 1000.

Comparing these results with those in Tables \ref{tab:like-comp-GN} and \ref{tab:like-comp-GS}, we observe an important distinction: the estimation of $\pi_{rstu}(\mathcal{O})$ (without conditioning on $W$) demonstrates strong robustness to distributional misspecification. In contrast, incorporating the secondary outcome $W$ through $\pi_{rstu}(\tilde{\mathcal{O}})$ introduces greater sensitivity to the normality assumption in Assumption \ref{assumption:normal}. This suggests that practitioners should assess the normality assumption more carefully when stratifying attribution by secondary outcomes.}

\begin{table}[h]
\centering
\centering \centering
\caption{The estimation results of  $\pr(G=rst1\mid Z=1,M=1,Y=1, w_{min} \leq W \leq w_{1/3})$ for $r,s,t\in\{0,1\}$, where $w_{min}$ and $w_{1/3}$ are the minimum value and first tertile of $W$, respectively, and the sample sizes for the first and second tables are 500 and 1000, respectively.}
\label{tab:S4_Q1}\resizebox{1\textwidth}{!}{
  \begin{tabular}{lccccc}
  \toprule\addlinespace[1mm]
  & $(r,s,t)=(0,0,1)$ &  $(r,s,t)=(0,1,1)$  &  $(r,s,t)=(1,0,1)$  &  $(r,s,t)=(1,1,1)$  &  $(r,s,t)=(1,1,1)$ \\ \addlinespace[1mm]
  \cline{2-6}\addlinespace[1mm]
(1) Normal distribution & 0.26 (4.46) & -0.38 (4.74) & 0.11 (1.75) & 0.02 (0.38) & 0.00 (0.01) \\
(2) $t$ distribution & 27.04 (9.86) & 16.67 (12.69) & -2.76 (7.68) & -24.54 (2.18) & -16.42 (1.60) \\
(3) Uniform distribution & -5.69 (3.83) & 16.05 (5.60) & 7.46 (5.15) & -15.04 (4.92) & -2.77 (4.56) \\
(4) Bernoulli distribution & 21.48 (8.36) & 6.70 (5.85) & -3.04 (4.09) & -15.33 (2.49) & -9.81 (1.62) \\
(5) Gamma distribution & 27.63 (9.14) & 9.02 (7.95) & -3.10 (5.54) & -18.29 (4.57) & -15.26 (1.68) \\
\addlinespace[0.5mm]
  \toprule\addlinespace[1mm]
  & $(r,s,t)=(0,0,1)$ &  $(r,s,t)=(0,1,1)$  &  $(r,s,t)=(1,0,1)$  &  $(r,s,t)=(1,1,1)$  &  $(r,s,t)=(1,1,1)$ \\ \addlinespace[1mm]
  \cline{2-6}\addlinespace[1mm]
(1) Normal distribution & 0.07 (3.04) & -0.10 (3.30) & -0.01 (1.08) & 0.03 (0.23) & 0.00 (0.00) \\
(2) $t$ distribution & 27.33 (9.20) & 16.21 (11.83) & -2.64 (7.14) & -24.57 (1.89) & -16.34 (1.41) \\
(3) Uniform distribution & -5.83 (2.54) & 16.29 (4.07) & 7.59 (3.63) & -15.27 (3.32) & -2.78 (3.08) \\
(4) Bernoulli distribution & 20.03 (5.97) & 7.96 (4.04) & -3.00 (2.82) & -15.29 (1.72) & -9.70 (1.15) \\
(5) Gamma distribution & 27.81 (6.28) & 9.11 (5.59) & -2.86 (3.58) & -18.81 (2.71) & -15.26 (1.15) \\
  \bottomrule
  \end{tabular}}
\end{table}

\begin{table}[h]
\centering
\centering \centering
\caption{The estimation results of  $\pr(G=rst1\mid Z=1,M=1,Y=1, w_{1/3} < W \leq w_{2/3})$ for $r,s,t\in\{0,1\}$, where $w_{1/3}$ and $w_{2/3}$ are the first and second tertiles of $W$, respectively, and the sample sizes for the first and second tables are 500 and 1000, respectively.}
\label{tab:S4_Q2}\resizebox{1\textwidth}{!}{
  \begin{tabular}{lccccc}
  \toprule\addlinespace[1mm]
  & $(r,s,t)=(0,0,1)$ &  $(r,s,t)=(0,1,1)$  &  $(r,s,t)=(1,0,1)$  &  $(r,s,t)=(1,1,1)$  &  $(r,s,t)=(1,1,1)$ \\ \addlinespace[1mm]
  \cline{2-6}\addlinespace[1mm]
(1) Normal distribution & -0.05 (1.38) & -0.21 (2.50) & 0.19 (2.75) & 0.13 (2.38) & -0.06 (1.16) \\
(2) $t$ distribution & 10.33 (6.76) & 9.05 (5.77) & 2.92 (4.22) & -10.05 (4.70) & -12.25 (1.57) \\
(3) Uniform distribution & -10.25 (1.95) & 14.67 (3.55) & 10.20 (3.47) & -11.74 (3.80) & -2.88 (2.98) \\
(4) Bernoulli distribution & -3.89 (4.04) & 20.21 (5.26) & -7.56 (6.63) & -7.75 (6.21) & -1.02 (2.04) \\
(5) Gamma distribution & -10.69 (4.72) & 4.42 (6.59) & 7.08 (5.24) & 3.93 (7.07) & -4.73 (3.54) \\
\addlinespace[0.5mm]
  \toprule\addlinespace[1mm]
  & $(r,s,t)=(0,0,1)$ &  $(r,s,t)=(0,1,1)$  &  $(r,s,t)=(1,0,1)$  &  $(r,s,t)=(1,1,1)$  &  $(r,s,t)=(1,1,1)$ \\ \addlinespace[1mm]
  \cline{2-6}\addlinespace[1mm]
(1) Normal distribution & -0.02 (0.82) & -0.09 (1.56) & -0.01 (1.80) & 0.14 (1.47) & -0.02 (0.71) \\
(2) $t$ distribution & 10.00 (5.64) & 9.23 (5.70) & 2.75 (3.66) & -9.71 (3.95) & -12.27 (1.39) \\
(3) Uniform distribution & -10.37 (1.22) & 14.81 (2.57) & 10.27 (2.52) & -11.69 (2.59) & -3.02 (1.88) \\
(4) Bernoulli distribution & -4.04 (2.64) & 20.81 (3.51) & -7.76 (4.53) & -8.19 (4.07) & -0.83 (1.48) \\
(5) Gamma distribution & -11.86 (3.19) & 5.23 (4.82) & 6.72 (4.06) & 4.31 (4.78) & -4.40 (2.44) \\
  \bottomrule
  \end{tabular}}
\end{table}

\begin{table}[h]
\centering
\centering \centering
\caption{The estimation results of  $\pr(G=rst1\mid Z=1,M=1,Y=1, w_{2/3} < W \leq w_{max})$ for $r,s,t\in\{0,1\}$, where $w_{2/3}$ and $w_{max}$ are the second tertile and maximum value of $W$, respectively, and the sample sizes for the first and second tables are 500 and 1000, respectively.}
\label{tab:S4_Q3}\resizebox{1\textwidth}{!}{
  \begin{tabular}{lccccc}
  \toprule\addlinespace[1mm]
  & $(r,s,t)=(0,0,1)$ &  $(r,s,t)=(0,1,1)$  &  $(r,s,t)=(1,0,1)$  &  $(r,s,t)=(1,1,1)$  &  $(r,s,t)=(1,1,1)$ \\ \addlinespace[1mm]
  \cline{2-6}\addlinespace[1mm]
(1) Normal distribution & 0.00 (0.16) & -0.11 (1.04) & 0.02 (2.09) & 0.06 (2.65) & 0.02 (3.19) \\
(2) $t$ distribution & -11.76 (3.91) & 4.80 (4.34) & 7.71 (4.12) & 2.90 (4.75) & -3.66 (2.61) \\
(3) Uniform distribution & -13.67 (2.33) & 12.89 (5.31) & 12.74 (5.30) & -8.64 (4.78) & -3.31 (4.33) \\
(4) Bernoulli distribution & -1.19 (0.77) & 1.07 (2.14) & 1.68 (4.19) & 2.97 (6.22) & -4.52 (7.09) \\
(5) Gamma distribution & -16.97 (2.02) & -4.86 (7.03) & 1.95 (6.01) & -1.22 (6.69) & 21.09 (7.03) \\
\addlinespace[0.5mm]
  \toprule\addlinespace[1mm]
  & $(r,s,t)=(0,0,1)$ &  $(r,s,t)=(0,1,1)$  &  $(r,s,t)=(1,0,1)$  &  $(r,s,t)=(1,1,1)$  &  $(r,s,t)=(1,1,1)$ \\ \addlinespace[1mm]
  \cline{2-6}\addlinespace[1mm]
(1) Normal distribution & 0.00 (0.07) & 0.04 (0.76) & 0.02 (1.21) & -0.02 (1.72) & -0.05 (2.07) \\
(2) $t$ distribution & -11.91 (3.62) & 4.70 (4.19) & 7.67 (3.87) & 2.99 (4.50) & -3.45 (2.44) \\
(3) Uniform distribution & -13.92 (1.54) & 13.22 (3.73) & 12.67 (3.49) & -8.54 (3.41) & -3.43 (2.75) \\
(4) Bernoulli distribution & -1.13 (0.57) & 1.13 (1.52) & 1.37 (2.71) & 3.17 (4.73) & -4.54 (5.08) \\
(5) Gamma distribution & -17.33 (1.21) & -4.92 (5.36) & 3.17 (4.63) & -1.93 (4.76) & 21.01 (4.93) \\
  \bottomrule
  \end{tabular}}
\end{table}

\begin{table}[h]
\centering
\centering \centering
\caption{The estimation results of  $\pr(G=0stu\mid Z=0,M=0,Y=0, w_{min} \leq W \leq w_{1/3})$ for $s,t,u\in\{0,1\}$, where $w_{min}$ and $w_{1/3}$ are the minimum value and first tertile of $W$, respectively, and the sample sizes for the first and second tables are 500 and 1000, respectively.}
\label{tab:S5_Q1}\resizebox{1\textwidth}{!}{
  \begin{tabular}{lccccc}
  \toprule\addlinespace[1mm]
  & $(s,t,u)=(0,0,0)$ &  $(s,t,u)=(0,0,1)$  &  $(s,t,u)=(0,1,1)$  &  $(s,t,u)=(1,0,1)$  &  $(s,t,u)=(1,1,1)$ \\ \addlinespace[1mm]
  \cline{2-6}\addlinespace[1mm]
(1) Normal distribution & 0.02 (2.32) & 0.26 (2.56) & -0.35 (2.08) & 0.06 (0.59) & 0.01 (0.25) \\
(2) $t$ distribution & 45.28 (6.90) & 1.16 (6.29) & -6.13 (4.65) & -13.72 (2.08) & -26.59 (1.41) \\
(3) Uniform distribution & -4.62 (4.22) & -5.16 (4.44) & 17.88 (5.41) & 8.31 (5.12) & -16.42 (4.57) \\
(4) Bernoulli distribution & -0.83 (5.92) & 3.35 (6.24) & 1.10 (4.22) & -0.91 (1.78) & -2.70 (1.64) \\
(5) Gamma distribution & 24.63 (7.99) & 12.05 (7.36) & -2.52 (4.85) & -10.65 (2.93) & -23.51 (2.88) \\
\addlinespace[0.5mm]
  \toprule\addlinespace[1mm]
  & $(s,t,u)=(0,0,0)$ &  $(s,t,u)=(0,0,1)$  &  $(s,t,u)=(0,1,1)$  &  $(s,t,u)=(1,0,1)$  &  $(s,t,u)=(1,1,1)$ \\ \addlinespace[1mm]
  \cline{2-6}\addlinespace[1mm]
(1) Normal distribution & -0.09 (1.61) & 0.19 (1.84) & -0.11 (1.51) & 0.00 (0.34) & 0.01 (0.13) \\
(2) $t$ distribution & 45.09 (6.28) & 1.35 (5.79) & -6.17 (4.63) & -13.77 (1.90) & -26.50 (1.03) \\
(3) Uniform distribution & -4.48 (2.99) & -5.72 (3.04) & 17.94 (3.88) & 8.33 (3.85) & -16.08 (3.43) \\
(4) Bernoulli distribution & -1.43 (4.37) & 3.61 (4.90) & 1.65 (2.84) & -1.12 (1.15) & -2.71 (1.09) \\
(5) Gamma distribution & 24.52 (5.20) & 12.00 (5.54) & -2.18 (3.62) & -10.84 (1.97) & -23.51 (2.08) \\
  \bottomrule
  \end{tabular}}
\end{table}

\begin{table}[h]
\centering
\centering \centering
\caption{The estimation results of  $\pr(G=0stu\mid Z=0,M=0,Y=0, w_{1/3} < W \leq w_{2/3})$ for $s,t,u\in\{0,1\}$, where $w_{1/3}$ and $w_{2/3}$ are the first and second tertiles of $W$, and the sample sizes for the first and second tables are 500 and 1000, respectively.}
\label{tab:S5_Q2}\resizebox{1\textwidth}{!}{
  \begin{tabular}{lccccc}
  \toprule\addlinespace[1mm]
  & $(s,t,u)=(0,0,0)$ &  $(s,t,u)=(0,0,1)$  &  $(s,t,u)=(0,1,1)$  &  $(s,t,u)=(1,0,1)$  &  $(s,t,u)=(1,1,1)$ \\ \addlinespace[1mm]
  \cline{2-6}\addlinespace[1mm]
(1) Normal distribution & -0.05 (1.01) & -0.06 (2.04) & -0.10 (2.93) & 0.25 (2.45) & -0.04 (1.65) \\
(2) $t$ distribution & 21.22 (5.76) & 9.73 (4.97) & -2.81 (4.04) & -6.51 (2.80) & -21.64 (3.08) \\
(3) Uniform distribution & -4.01 (2.91) & -10.22 (2.44) & 15.73 (3.53) & 10.99 (3.46) & -12.48 (3.94) \\
(4) Bernoulli distribution & -2.05 (3.43) & 1.27 (5.45) & 1.90 (6.97) & -0.45 (5.47) & -0.67 (4.38) \\
(5) Gamma distribution & -5.53 (5.02) & -4.70 (6.21) & 6.05 (6.81) & 5.96 (5.69) & -1.78 (7.24) \\
\addlinespace[0.5mm]
  \toprule\addlinespace[1mm]
  & $(s,t,u)=(0,0,0)$ &  $(s,t,u)=(0,0,1)$  &  $(s,t,u)=(0,1,1)$  &  $(s,t,u)=(1,0,1)$  &  $(s,t,u)=(1,1,1)$ \\ \addlinespace[1mm]
  \cline{2-6}\addlinespace[1mm]
(1) Normal distribution & 0.03 (0.63) & -0.09 (1.26) & -0.02 (1.81) & -0.00 (1.60) & 0.08 (1.14) \\
(2) $t$ distribution & 20.91 (5.10) & 9.80 (4.39) & -2.75 (3.69) & -6.51 (2.55) & -21.45 (2.58) \\
(3) Uniform distribution & -3.91 (2.01) & -10.52 (1.60) & 15.86 (2.51) & 11.14 (2.50) & -12.57 (2.77) \\
(4) Bernoulli distribution & -2.02 (2.26) & 1.69 (3.63) & 1.38 (4.90) & -0.43 (4.05) & -0.62 (2.80) \\
(5) Gamma distribution & -5.32 (3.42) & -5.61 (4.37) & 6.68 (5.09) & 5.45 (4.14) & -1.20 (5.05) \\
  \bottomrule
  \end{tabular}}
\end{table}

\begin{table}[h]
\centering
\centering \centering
\caption{The estimation results of $\pr(G=0stu\mid Z=0,M=0,Y=0, w_{2/3} < W \leq w_{max})$ for $s,t,u\in{0,1}$, where $w_{2/3}$ and $w_{max}$ are the second tertile and maximum value of $W$, respectively, and the sample sizes for the first and second tables are 500 and 1000, respectively.}
\label{tab:S5_Q3}\resizebox{1\textwidth}{!}{
  \begin{tabular}{lccccc}
  \toprule\addlinespace[1mm]
  & $(s,t,u)=(0,0,0)$ &  $(s,t,u)=(0,0,1)$  &  $(s,t,u)=(0,1,1)$  &  $(s,t,u)=(1,0,1)$  &  $(s,t,u)=(1,1,1)$ \\ \addlinespace[1mm]
  \cline{2-6}\addlinespace[1mm]
(1) Normal distribution & 0.00 (0.08) & 0.01 (0.39) & -0.19 (1.41) & 0.11 (1.82) & 0.06 (2.13) \\
(2) $t$ distribution & -7.11 (3.55) & -7.11 (5.33) & 6.71 (4.75) & 6.29 (4.14) & 1.21 (5.13) \\
(3) Uniform distribution & -3.87 (4.29) & -14.08 (2.88) & 13.55 (5.10) & 13.47 (5.12) & -9.07 (4.80) \\
(4) Bernoulli distribution & -2.61 (1.19) & -2.32 (1.50) & 0.39 (3.05) & 1.38 (4.85) & 3.17 (5.31) \\
(5) Gamma distribution & -16.62 (1.63) & -16.33 (3.50) & 4.35 (9.16) & 12.11 (7.62) & 16.49 (8.88) \\
\addlinespace[0.5mm]
  \toprule\addlinespace[1mm]
  & $(s,t,u)=(0,0,0)$ &  $(s,t,u)=(0,0,1)$  &  $(s,t,u)=(0,1,1)$  &  $(s,t,u)=(1,0,1)$  &  $(s,t,u)=(1,1,1)$ \\ \addlinespace[1mm]
  \cline{2-6}\addlinespace[1mm]
(1) Normal distribution & 0.00 (0.05) & 0.00 (0.24) & -0.01 (0.97) & -0.04 (1.27) & 0.04 (1.32) \\
(2) $t$ distribution & -7.71 (2.90) & -7.39 (5.06) & 6.88 (4.55) & 6.54 (3.75) & 1.69 (4.72) \\
(3) Uniform distribution & -3.66 (3.04) & -14.37 (1.94) & 13.76 (3.56) & 13.70 (3.75) & -9.43 (3.38) \\
(4) Bernoulli distribution & -2.68 (0.77) & -2.44 (1.04) & 0.55 (1.97) & 1.21 (3.32) & 3.36 (3.89) \\
(5) Gamma distribution & -16.76 (1.04) & -17.14 (1.95) & 4.57 (6.72) & 13.43 (6.08) & 15.89 (6.57) \\
  \bottomrule
  \end{tabular}}
\end{table}

\subsection{Simulation details for Section \ref{ssec:simulation-settings}}
\label{SEC:sim-detial}
In this section, we present the detailed parameter settings used in Section \ref{ssec:simulation-settings}. Specifically:

\begin{itemize} \item[(1)] $\boldsymbol{\alpha}_{0,0}=(0.00,0.00,0.00)^\T$, $\boldsymbol{\alpha}_{0,1}=(-0.36,-0.37,-0.26)^\T$, $\boldsymbol{\alpha}_{1,0}=(-0.28,-0.19,0.29)^\T$, $\boldsymbol{\alpha}_{1,1}=(-0.28,-0.19,0.29)^\T$;

\item[(2)] For any $z,m\in\{0,1\}$: $\boldsymbol{\theta}_{z,m,0000}=(0.00,0.00,0.00)^\T$, $\boldsymbol{\theta}_{z,m,0001}=(-0.07,0.36,0.06)^\T$, $\boldsymbol{\theta}_{z,m,0011}=(0.03,-0.38,-0.37)^\T$, $\boldsymbol{\theta}_{z,m,0101}=(-0.19,0.08,-0.27)^\T$, $\boldsymbol{\theta}_{z,m,0111}=(0.38,-0.03,0.25)^\T$, and $\boldsymbol{\theta}_{z,m,1111}=(-0.10,0.27,0.30)^\T$;

\item[(3)] For any $z,m\in\{0,1\}$: $\boldsymbol{\mu}_{z,m,0000}=(-5.00,-1.77,-1.39)^\T$, $\boldsymbol{\mu}_{z,m,0001}=(-3.00,-1.46,1.40)^\T$, $\boldsymbol{\mu}_{z,m,0011}=(-1.00,-1.72,-0.41)^\T$, $\boldsymbol{\mu}_{z,m,0101}=(1.00,1.62,-1.29)^\T$, $\boldsymbol{\mu}_{z,m,0111}=(3.00,-1.08,\\1.595)^\T$, and $\boldsymbol{\mu}_{z,m,1111}=(5.00,0.35,-0.06)^\T$;

\item[(4)] For any $z,m\in\{0,1\}$: $\sigma_{z,m,0000}=0.65$, $\sigma_{z,m,0001}=1.38$, $\sigma_{z,m,0011}=1.81$, $\sigma_{z,m,0101}=1.17$, $\sigma_{z,m,0111}=1.16$, and $\sigma_{z,m,1111}=1.39$. \end{itemize}
\section{Simulation details for the smoking and asbestos example}
 \label{ssec:bodyweight-second}
In this section, we provide the sampling details and secondary outcome variable generation details for the smoking and asbestos exposure example, based on Table 4.

\begin{table}[h]
    \centering
    \caption{Summary data on lung cancer under smoking habits and asbestos exposure, where the raw data comes from Table 1 in \citet{vanderweele2014tutorial} and Figure 2 in \citet{hilt1986previous}.}
    \label{tab:summary-stat}
    \resizebox{0.84905\linewidth}{!}{%
   \begin{tabular}{ccc}
\addlinespace[2mm]
  \toprule
                           & No asbestos exposure $(M=0)$  & Asbestos   exposure $(M=1)$    \\\addlinespace[1mm] \hline\addlinespace[1mm] 
Non-smokers $(Z=0)$      & $ \delta_{0,0}=6 / 5 057 \approx 0.12\%$    & $ \delta_{0,1}= 5 / 749 \approx 0.67\%$     \\\addlinespace[1mm]  \hline\addlinespace[1mm] 
Smokers $(Z=1)$ & $  \delta_{1,0}=118/ 12383\approx 0. 95\%$ & $ \delta_{1,1}= 141 / 3130 \approx 4.51\%$ \\  \toprule
\end{tabular}
}
\end{table}

\begin{figure}[h]
     \centering
     \includegraphics[width=0.45\linewidth]{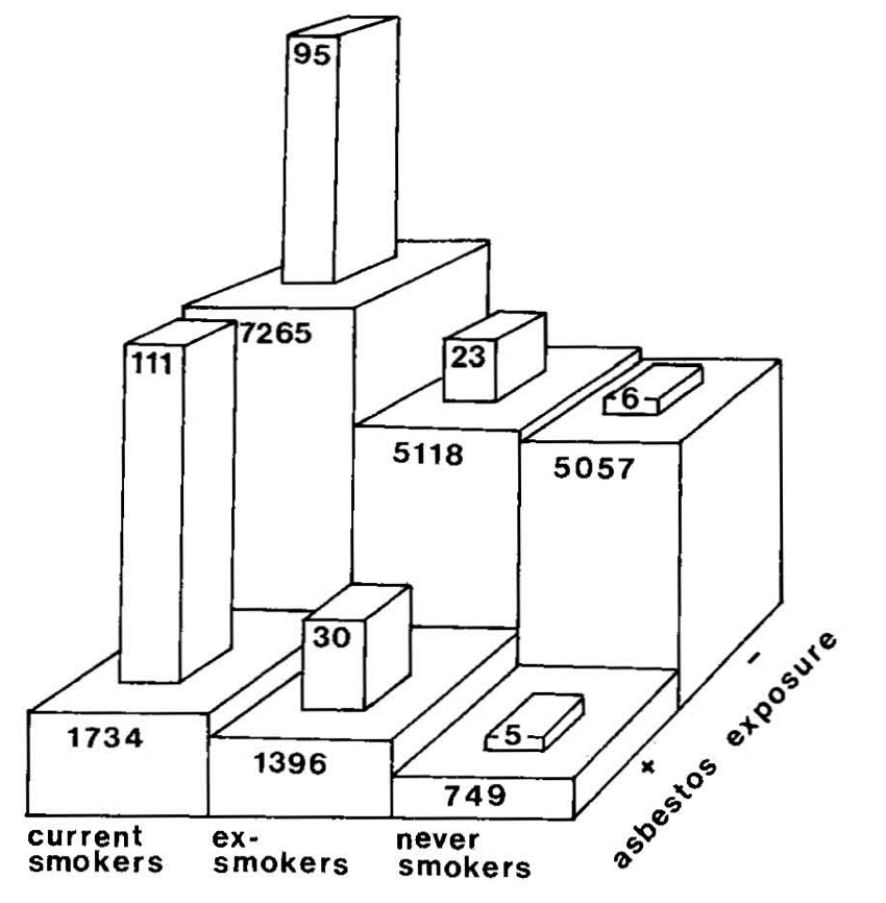}
     \caption{Study subjects (21319) grouped according to questionnaire information about asbestos exposure and smoking habits (shown as boxes), with the number of lung cancers expected to occur in each group during the next decade (shown as columns). Following \citet{vanderweele2014tutorial}, we classify ex-smokers and current smokers as $Z = 1$ and never smokers as $Z = 0$. We categorize those who have been exposed to asbestos as $M = 1$ and those who have not been exposed as $M = 0$. Here, $Y = 1$ denotes individuals with lung cancer, and $Y = 0$ denotes individuals without lung cancer.}
     \label{fig:summary-fig}
 \end{figure}

Figure \ref{fig:summary-fig} is taken from Figure 2 of \citet{hilt1986previous} and shows study subjects (\( 21319\)) grouped according to questionnaire information about asbestos exposure and smoking habits (boxes represent the groups). Figure \ref{fig:summary-fig} contains the raw calculation information for Table \ref{tab:summary-stat} in the main text, and we will use the specific values in Figure \ref{fig:summary-fig} for the sampling.
 \begin{enumerate}
     \item We sample  $21 319$ individuals from the   data provided in Figure \ref{fig:summary-fig}. Among them, there are $6$ individuals with $(Z=0, M=0, Y=1)$, $5057-6=5051$ individuals with $(Z=0, M=0, Y=0)$,  5 individuals with $(Z=0, M=1, Y=1)$, $749-5=744$ individuals with $(Z=0, M=1, Y=0)$, $95+23=118$ individuals with $(Z=1, M=0, Y=1)$, $7265+5118-95-23=12265$ individuals with $(Z=1, M=0, Y=0)$, $111+30=141$ individuals with $(Z=1, M=1, Y=1)$, and $1734+1396-111-30=2989$ individuals with $(Z=1, M=1, Y=0)$. 

\item  For the individual $i$ with $(Z_i=0, M_i=0, Y_i=1)$, the secondary outcome $W_i$ is sampled from the following Gaussian mixture model,
     \begin{align*}
W_i \mid Z_i=0, M_i=0, Y_i=1 \sim N(55,5^2). 
     \end{align*}

  For the individual $i$ with $(Z_i=0, M_i=0, Y_i=0)$, the secondary outcome $W_i$ is sampled from the following Gaussian mixture model,
     \begin{align*}
W_i \mid Z_i=0,  M_i=0, Y_i=0 \sim \begin{Bmatrix}0.9561  N(72,4^2) + 0.0320 N(70,3^2) \\+ 0.0064 N(68,6.5^2)+0.0035 N(65,6^2) +0.0020 N(58,2^2)
\end{Bmatrix}. 
     \end{align*}
     
       For the individual $i$ with $(Z_i=0, M_i=1, Y_i=1)$, the secondary outcome $W_i$ is sampled from the following Gaussian mixture model,
     \begin{align*}
W_i \mid Z_i=0, M_i=1, Y_i=1\sim 0.5289  N(65,6 ^2) +0.2934 N(58,2^2)  +0.1777N(55,5^2) . 
     \end{align*}

            For the individual $i$ with $(Z_i=0, M_i=1, Y_i=0)$, the secondary outcome $W_i$ is sampled from the following Gaussian mixture model,
     \begin{align*}
W_i \mid Z_i=0, M_i=1, Y_i=0\sim 0.9614  N(72,4^2) +0.0322 N(70,3^2)  +0.0064N(68,6.5^2) . 
     \end{align*}   
     
             For the individual $i$ with $(Z_i=1, M_i=0, Y_i=1)$, the secondary outcome $W_i$ is sampled from the following Gaussian mixture model,
     \begin{align*}
W_i \mid Z_i=1, M_i=0, Y_i=1\sim 0.6700  N(68,6.5 ^2) +0.2055 N(58,2^2)  +0.1245N(55,5^2) . 
     \end{align*}
     
             For the individual $i$ with $(Z_i=1, M_i=0, Y_i=0)$, the secondary outcome $W_i$ is sampled from the following Gaussian mixture model,
     \begin{align*}
W_i \mid Z_i=1, M_i=0, Y_i=0\sim 0.9641  N(72,4 ^2) +0.0323 N(70,3^2)  +0.0036N(65,6^2) . 
     \end{align*}

  For the individual $i$ with $(Z_i=1, M_i=1, Y_i=1)$, the secondary outcome $W_i$ is sampled from the following Gaussian mixture model,
     \begin{align*}
W_i \mid Z_i=1,  M_i=1, Y_i=1 \sim \begin{Bmatrix} 0.7101 N(70,3^2)+ 0.1417 N(68,6.5^2) \\+0.0784 N(65,6^2) +0.0435 N(58,2^2)+ 0.0263  N(55,5^2) 
\end{Bmatrix}. 
     \end{align*}
      For the individual $i$ with $(Z_i=1, M_i=1, Y_i=0)$, the secondary outcome $W_i$ is sampled from the following Gaussian mixture model,      \begin{align*}
W_i \mid Z_i=1, M_i=1, Y_i=0 \sim N(72,4^2). 
     \end{align*}   \end{enumerate}
\section{The proof of  Theorem \ref{thm:identification}}
\label{sec:proof-bound}
\begin{proof}
First of all, we can identify the proportion $\pi_{1111}$ by the conditional probability $\pi_{1111}=\pr(Y=1 \mid Z=0, M=0)$, and identify the proportion $\pi_{0000}$ by the conditional probability $\pi_{0000}=\pr(Y=0 \mid Z=1, M=1)$. By the Law of Iterated Expectations, we have:
\begin{equation*} 
\begin{aligned}
f(W \mid & Z=0, M=1, Y=1) \\& = f\left(W \mid {Z=0, M=1}, G=0101\right) \frac{\pi_{0101}}{\delta_{0,1}} + f\left(W \mid {Z=0, M=1}, G=0111\right) \frac{\pi_{0111}}{\delta_{0,1}} \\
&~~+ f\left(W \mid {Z=0, M=1}, G=1111\right) \frac{\pi_{1111}}{\delta_{0,1}},\\
f(W \mid&  Z=1, M=0, Y=1)\\ & = f\left(W \mid {Z=1, M=0}, G=0011\right) \frac{\pi_{0011}}{\delta_{1,0}}+ f\left(W \mid {Z=1, M=0}, G=0111\right) \frac{\pi_{0111}}{\delta_{1,0}} \\
&~~+ f\left(W \mid {Z=1, M=0}, G=1111\right) \frac{\pi_{1111}}{\delta_{1,0}},\\
\end{aligned}
\end{equation*}
\begin{equation*} 
\begin{aligned} 
f(W \mid & Z=1, M=1, Y=1) \\& = f\left(W \mid Z=1, M=1, G=0001\right) \frac{\pi_{0001}}{{\delta_{1,1}}}  + f\left(W \mid Z=1, M=1, G=0011\right) \frac{\pi_{0011}}{{\delta_{1,1}}} \\
&~~+ f\left(W \mid Z=1, M=1, G=0101\right) \frac{\pi_{0101}}{{\delta_{1,1}}}  + f\left(W \mid Z=1, M=1, G=0111\right) \frac{\pi_{0111}}{{\delta_{1,1}}} \\
&~~+ f\left(W \mid Z=1, M=1, G=1111\right) \frac{\pi_{1111}}{{\delta_{1,1}}}.
\end{aligned}
\end{equation*}
Given Assumption \ref{assumption:normal} (Example 3.1.4 and Theorem 3.1.2 in \citet{Titterington1985}), we can identify the following three sets:
\begin{gather*}
  \left\{ \frac{\pi_{0101}}{\delta_{0,1}}, \frac{\pi_{0111}}{\delta_{0,1}}, \frac{\pi_{1111}}{\delta_{0,1}}\right\}, ~~(\text{by~using~Assumption~3~to~}  f(W\mid Z=0,M=1,Y=1)),\\
 \left\{ \frac{\pi_{0011}}{\delta_{1,0}}, \frac{\pi_{0111}}{\delta_{1,0}}, \frac{\pi_{1111}}{\delta_{1,0}}\right\},  ~~(\text{by~using~Assumption~3~to~}  f(W\mid Z=1,M=0,Y=1)),\\
 \left\{\frac{\pi_{0001}}{{\delta_{1,1}}},\dfrac{\pi_{0011}}{{\delta_{1,1}}}, \dfrac{\pi_{0101}}{{\delta_{1,1}}}, \dfrac{\pi_{0111}}{{\delta_{1,1}}},\dfrac{\pi_{1111}}{{\delta_{1,1}}}\right\} ,~~(\text{by~using~Assumption~3~to~}  f(W\mid Z=1,M=1,Y=1)).
\end{gather*}Since the denominator is identifiable, we can also identify the following three sets,\begin{gather*}
     \mathcal{S}_{011}^\circ= \left\{ {\pi_{0101}}, {\pi_{0111}}, {\pi_{1111}}\right\}, \\
     \mathcal{S}_{101} ^\circ= \left\{ {\pi_{0011}}, {\pi_{0111}}, {\pi_{1111}}\right\}, \\
     \mathcal{S}_{111}^\circ = \left\{ {\pi_{0001}}, {\pi_{0011}},  {\pi_{0101}},  {\pi_{0111}}, {\pi_{1111}}\right\}.
\end{gather*}
Since $\pi_{1111}$ is identified by $\pi_{1111}=\pr(Y=1\mid Z=0,M=0)$. We hence can identify the sets  $\mathcal{C}_{011}^\circ=\mathcal{S}_{011}^\circ\backslash \pi_{1111}= \left\{ {\pi_{0101}}, {\pi_{0111}} \right\}$, $\mathcal{C}_{101}^\circ=\mathcal{S}_{101}^\circ\backslash \pi_{1111}= \left\{ {\pi_{0011}}, {\pi_{0111}} \right\}$, and $\mathcal{C}_{111}^\circ=\mathcal{S}_{111}^\circ\backslash \pi_{1111}= \left\{ {\pi_{0001}}, {\pi_{0011}},  {\pi_{0101}},  {\pi_{0111}} \right\}$.
Through the union set $\mathcal{C}_{011}^\circ \cup \mathcal{C}_{101}^\circ$, we can use the set $\mathcal{C}^\circ_{111}$ to identify the proportion $\pi_{0001}$.   We also note that under the monotonicity assumption, we have
\begin{align*}
    \pr(Y=0 \mid Z=1, M=0) = \pi_{0000} + \pi_{0001} + \pi_{0101},
\end{align*}
then $\pi_{0101}$ can be identified through $\pi_{0101} = \pr(Y=0 \mid Z=1, M=0) - \pi_{0000} - \pi_{0001}$.    We also note that under the monotonicity assumption, we have
\begin{align*}
    \pr(Y=0 \mid Z=0, M=1) = \pi_{0000} + \pi_{0001} + \pi_{0011},
\end{align*}
then $\pi_{0011}$ can be identified through $\pi_{0011} = \pr(Y=0 \mid Z=0, M=1) - \pi_{0000} - \pi_{0001}$.   Therefore, all the posterior probabilities are identifiable.

\end{proof}

\section{The proof of Corollary \ref{coro:equal-probs}}
\begin{proof}
By using  Assumption \ref{assumption:normal} to $f(W\mid Z=1,M=1,Y=1)$, we can identify the following sets, 
\begin{align*}
  \mathcal{C}_{111}^\circ&= \begin{Bmatrix}
       & (\pi_{0001},\mu_{1,1,0001},\sigma^2_{1,1,0001}),&~(\pi_{0011},\mu_{1,1,0011},\sigma^2_{1,1,0011}),& \\& (\pi_{0101},\mu_{1,1,0101},\sigma^2_{1,1,0101}),&~(\pi_{0111},\mu_{1,1,0111},\sigma^2_{1,1,0111} ),& \\&(\pi_{1111},\mu_{1,1,1111},\sigma^2_{1,1,1111})&
    \end{Bmatrix}.\\
      \mathcal{S}_{111}^\circ&= \{\pi_{0001}, \pi_{0011}, \pi_{0101}, \pi_{0111}, \pi_{1111}\}.
\end{align*}Here, we may have some slight abuse of notation, as we will still use $  \mathcal{C}_{111}^\circ$ to denote the parameters identified in the  Gaussian mixture model. Of course, the identification here still refers to identification up to label swapping. 
By using  Assumption \ref{assumption:normal} to $f(W\mid Z=0,M=1,Y=1)$, we can identify the following sets, 
\begin{align*}
  \mathcal{C}_{011}^\circ&= \begin{Bmatrix} 
       & (\pi_{0101},\mu_{0,1,0101},\sigma^2_{0,1,0101}),&~(\pi_{0111},\mu_{0,1,0111},\sigma^2_{0,1,0111} ),& \\&(\pi_{1111},\mu_{0,1,1111},\sigma^2_{0,1,1111})&
    \end{Bmatrix}.\\
      \mathcal{S}_{011}^\circ&= \{\pi_{0101}, \pi_{0111}, \pi_{1111} \}.
\end{align*}By using  Assumption \ref{assumption:normal} to $f(W\mid Z=1,M=0,Y=1)$, we can identify the following sets, 
\begin{align*}
  \mathcal{C}_{101}^\circ&= \begin{Bmatrix} 
       & (\pi_{0011},\mu_{1,0,0011},\sigma^2_{1,0,0011}),&~(\pi_{0111},\mu_{1,0,0111},\sigma^2_{1,0,0111} ),& \\&(\pi_{1111},\mu_{1,0,1111},\sigma^2_{1,0,1111})&
    \end{Bmatrix}.\\
      \mathcal{S}_{101}^\circ&= \{\pi_{0011}, \pi_{0111}, \pi_{1111} \}.
\end{align*}By using  Assumption \ref{assumption:normal} to $f(W\mid Z=0,M=0,Y=1)$, we can identify the following sets, 
\begin{align*}
  \mathcal{C}_{001}^\circ&= \begin{Bmatrix} & (\pi_{1111},\mu_{0,0,1111},\sigma^2_{0,0,1111})&
    \end{Bmatrix}.\\  \mathcal{S}_{001}^\circ&=\{\pi_{1111}\}.
\end{align*}By using  Assumption \ref{assumption:normal} to $f(W\mid Z=0,M=0,Y=0)$, we can identify the following sets, 
\begin{align*}  \mathcal{C}_{000}^\circ&= \begin{Bmatrix}
       & (\pi_{0001},\mu_{0,0,0001},\sigma^2_{0,0,0001}),&~(\pi_{0011},\mu_{0,0,0011},\sigma^2_{0,0,0011}),& \\& (\pi_{0101},\mu_{0,0,0101},\sigma^2_{0,0,0101}),&~(\pi_{0111},\mu_{0,0,0111},\sigma^2_{0,0,0111} ),& \\&(\pi_{0000},\mu_{0,0,0000},\sigma^2_{0,0,0000})&
    \end{Bmatrix}.\\  \Sset_{000}^\circ&= \{\pi_{0001}, \pi_{0011}, \pi_{0101},  \pi_{0111},\pi_{0000}\} .
\end{align*}By using  Assumption \ref{assumption:normal} to $f(W\mid Z=0,M=1,Y=0)$, we can identify the following sets, 
\begin{align*}  \mathcal{C}_{010}^\circ&= \begin{Bmatrix}
       & (\pi_{0001},\mu_{0,1,0001},\sigma^2_{0,1,0001}),&~(\pi_{0011},\mu_{0,1,0011},\sigma^2_{0,1,0011}),& \\& (\pi_{0000},\mu_{0,1,0000},\sigma^2_{0,1,0000})&
    \end{Bmatrix}.\\\Sset_{010}^\circ&= \{\pi_{0001}, \pi_{0011},\pi_{0000}
    \}
\end{align*}By using  Assumption \ref{assumption:normal} to $f(W\mid Z=1,M=0,Y=0)$, we can identify the following sets, 
\begin{align*}  \mathcal{C}_{100}^\circ& = \begin{Bmatrix}&  (\pi_{0101},\mu_{1,0,0101},\sigma^2_{1,0,0101}),  &(\pi_{0001},\mu_{1,0,0111},\sigma^2_{1,0,0111} ),  \\&(\pi_{0000},\mu_{1,0,0000},\sigma^2_{1,0,0000}) ~&  
    \end{Bmatrix}. \\\Sset_{100}^\circ&= \{\pi_{0101},\pi_{0001}, \pi_{0000}\}.
\end{align*}By using  Assumption \ref{assumption:normal} to $f(W\mid Z=1,M=1,Y=0)$, we can identify the following sets, 
\begin{align*}
  \mathcal{C}_{110}^\circ& = \begin{Bmatrix}
       & (\pi_{0000},\mu_{1,1,0000},\sigma^2_{1,1,0000})&
    \end{Bmatrix}.\\\Sset_{110}& = \{\pi_{0000} \}
\end{align*}
  Theorem \ref{thm:identification} shows that we can identify $\pi_{rstu}$. If we additionally assume that $\pi_{rstu} \neq \pi_{r^\prime s^\prime t^\prime u^\prime}$ for any $rstu \neq r^\prime s^\prime t^\prime u^\prime$, we can further distinguish the labels of each element in $\mathcal{C}_{zmy}^\circ$. Once the labels are determined, we can identify the corresponding mean and variance parameters. Consequently, the conditional probability of $f(W\mid Z=z,M=m,Y=y,G=rstu)$ can also be identified. For example, by using the sets $\mathcal{S}^\circ_{001}$, $\mathcal{S}^\circ_{011}$, $\mathcal{S}^\circ_{101}$, and $\mathcal{S}^\circ_{111}$, the intersection of the four sets is given by $\mathcal{S}^\circ_{001}\cap\mathcal{S}^\circ_{011}\cap\mathcal{S}^\circ_{101}\cap\mathcal{S}^\circ_{111}$. Since we assume that each $\pi_{rstu}$ is unique, we can determine the label of $\pi_{1111}$ in each set. This allows us to identify the parameters $(\mu_{0,0,1111}, \sigma^2_{0,0,1111})$, $(\mu_{0,1,1111}, \sigma^2_{0,1,1111})$, $(\mu_{1,0,1111}, \sigma^2_{1,0,1111})$, and $(\mu_{1,1,1111}, \sigma^2_{1,1,1111})$.  Consequently, $f(W\mid Z=0,M=0, G=1111)$, $f(W\mid Z=0,M=1,G=1111)$, $f(W\mid Z=1,M=0,G=1111)$ and $f(W\mid Z=1,M=1,G=1111)$ can be identified.  
  
Finally, once $\pi_{rstu}({\mathcal{O}})$ and $f(W\mid Z=z,M=m,Y=y)$ are identified, the conditional probability $\pi_{rstu}(\tilde{{\mathcal{O}}})$ can be identified for ${\mathcal{O}}=(Z=z, M=m, Y=y)$ and $\tilde{{\mathcal{O}}}=({\mathcal{O}}, W=w)$. For example, suppose ${\mathcal{O}}=(Z=1,M=1,Y=1)$ and $\tilde{\mathcal{O}}=({\mathcal{O}},W )$, we then have, 
\begin{equation*}
   \begin{aligned}
\pi_{rst1}(\tilde{\mathcal{O}}) =\dfrac{f(W\mid Z=1,M=1,G=rst1)\pi_{rst1}( {\mathcal{O}})}{\sum_{r,s,t=0}^1 f(W\mid Z=1,M=1,G=rst1)\pi_{rst1}( {\mathcal{O}})}.
   \end{aligned}
\end{equation*} 
\end{proof}

 \section{The proof of  Theorem 2}
\label{sec:proof-prop1}
\begin{proof}
Under Assumption~\ref{assump:no-confounding}, by the Law of Iterated Expectations, we have,
\begin{equation*} 
\begin{aligned}
f(W \mid& Z=0,M=0, Y=1)  ={\textstyle  \sum_{ s,t,u=0}^1}~ f\left(W  \mid Z=0,M=0, G=1stu\right) \frac{\pr\left(G=  1stu \right)}{\operatorname{pr}(Y=1 \mid Z=0,M=0)}   ,\\
f(W \mid &Z=0,M=1, Y=1)  ={\textstyle  \sum_{r ,t,u=0}^1}~ f\left(W \mid {Z=0,M=1}, G=r1tu\right) \frac{\pr\left(G=  r1tu \right)}{\operatorname{pr}(Y=1 \mid Z=0,M=1)}   ,\\
f(W \mid &Z=1,M=0, Y=1)   ={\textstyle  \sum_{r,s,u=0}^1}~ f\left(W  \mid Z=1,M=0, G=rs1u\right) \frac{\pr\left(G=  rs1u  \right)}{\operatorname{pr}(Y=1 \mid Z=1,M=0)}   ,\\
f(W \mid & Z=1,M=1, Y=1)   ={\textstyle  \sum_{r, s,t=0}^1}~ f\left(W  \mid Z=1,M=1, G=rst1\right) \frac{\pr\left(G=  rst1 \right)}{\operatorname{pr}(Y=1 \mid Z=1,M=1)},\\f(W \mid &Z=0,M=0, Y=0)  ={\textstyle  \sum_{s ,t,u=0}^1}~ f\left(W  \mid Z=0,M=0, G=0stu\right) \frac{\pr\left(G=  0stu \right)}{\operatorname{pr}(Y=0 \mid Z=0,M=0)}   ,\\
f(W \mid &Z=0,M=1, Y=0)  ={\textstyle  \sum_{r,t,u=0}^1}~ f\left(W \mid {Z=0,M=1}, G=r0tu\right) \frac{\pr\left(G=  r0tu \right)}{\operatorname{pr}(Y=0 \mid Z=0,M=1)}   ,\\ 
f(W \mid& Z=1,M=0, Y=0)  ={\textstyle  \sum_{r,s,u=0}^1}~ f\left(W  \mid Z=1,M=0, G=rs0u\right) \frac{\pr\left(G=  rs0u  \right)}{\operatorname{pr}(Y=0 \mid Z=1,M=0)}   ,\\
f(W \mid& Z=1,M=1, Y=0)  ={\textstyle  \sum_{r,s,t=0}^1}~ f\left(W  \mid Z=1,M=1, G=rst0\right) \frac{\pr\left(G=  rst0 \right)}{\operatorname{pr}(Y=0 \mid Z=1,M=1)}.
\end{aligned}
\end{equation*}
Given Assumption \ref{assumption:normal}    (Example 3.1.4 and Theorem 3.1.2 in \citet{Titterington1985}), we can identify the following  sets:
\begin{equation}
    \label{eq:sets-iden}
    \begin{aligned} 
    \mathcal{C}_{001} =\{(\pi_{1stu},\mu_{0,0,1stu},\sigma_{0,0,1stu}^2):s,t,u=0,1\},\\ 
    \mathcal{C}_{011} =\{(\pi_{r1tu},\mu_{0,1,r1tu},\sigma_{0,1,r1tu}^2):r,t,u=0,1\},\\ 
    \mathcal{C}_{101} =\{(\pi_{rs1u},\mu_{1,0,rs1u},\sigma_{1,0,rs1u}^2):r,s, u=0,1\},\\ 
    \mathcal{C}_{111} =\{(\pi_{rst1},\mu_{1,1,rst1},\sigma_{1,1,rst1}^2):r,s,t =0,1\},\\ 
    \mathcal{C}_{000} =\{(\pi_{0stu},\mu_{0,0,0stu},\sigma_{0,0,0stu}^2):s,t,u=0,1\},\\ 
    \mathcal{C}_{010} =\{(\pi_{r0tu},\mu_{0,1,r0tu},\sigma_{0,1,r0tu}^2):r,t,u=0,1\},\\ 
    \mathcal{C}_{100} =\{(\pi_{rs0u},\mu_{1,0,rs0u},\sigma_{1,0,rs0u}^2):r,s,u=0,1\},\\ 
    \mathcal{C}_{110} =\{(\pi_{rst0},\mu_{1,1,rst0},\sigma_{1,1,rst0}^2):r,s,t =0,1\}.\\
\end{aligned}
\end{equation}
The identification here is still up to label swapping. Next, we will determine the labels under Assumption \ref{assump:equal-prop}.
\begin{itemize}
    \item[(1)] We first discuss the identifiability of $\pi_{rstu}({\mathcal{O}})$ and $\pi_{rstu}(\tilde{\mathcal{O}})$ under Assumption \ref{assump:equal-prop}(i). We   consider a simpler structure of \eqref{eq:sets-iden} as follows:
\begin{equation*} 
    \begin{aligned}
    \Sset_{001}&=\{\pi_{1stu}:s,t,u=0,1\}, ~~
    \Sset_{011} =\{\pi_{r1tu}:r,t,u=0,1\}, \\
    \Sset_{101}&=\{\pi_{rs1u}:r,s,u=0,1\},  ~~
    \Sset_{111} =\{\pi_{rst1}:r,s,t=0,1\}, \\
    \Sset_{000}&=\{\pi_{0stu}:s,t,u=0,1\}, ~~
    \Sset_{010}  =\{\pi_{r0tu}:r,t,u=0,1\}, \\
    \Sset_{100}&=\{\pi_{rs0u}:r,s,u=0,1\},~~
    \Sset_{110} =\{\pi_{rst0}:r,s,t=0,1\} .\\
\end{aligned}
\end{equation*}Without loss of generality, we assume that the labels of $  \Sset_{zmy}$ and $  \mathcal{C}_{zmy}$ are aligned with respect to $\pi_{rstu}$. We will determine the labels and other parameters in $  \mathcal{C}_{zmy}$ in  \eqref{eq:sets-iden}  through $  \Sset_{zmy}$.  For any type $G=rstu$,  let $\mathcal{T}_{rstu}=\Sset_{00r}\cap \Sset_{01s}\cap \Sset_{10t} \cap\Sset_{11u}$.
We must have the unique element in $\mathcal{T}_{rstu}$, which is $\pi_{rstu}$, since we assume that $\pi_{rstu}\neq \pi_{g'}$ in Assumption \ref{assump:equal-prop}(i). Once the labels are determined, we can identify the corresponding mean and variance parameters.   Consequently, the conditional probability of $f(W\mid Z=z,M=m,  G=rstu)$ can also be identified for $z,m, r,s,t,u\in\{0,1\}$.  
 For example, from  Table \ref{tab: probabilities}, it is observed that the class $G=1111$ appears in all four subpopulations, indicating that $\pi_{1111} \in \mathcal{S}_{001}\cap\mathcal{S}_{011}\cap\mathcal{S}_{101}\cap\mathcal{S}_{111}$. Assumption \ref{assump:equal-prop}(i) implies that the intersection $\mathcal{S}_{001}\cap\mathcal{S}_{011}\cap\mathcal{S}_{101}\cap\mathcal{S}_{111}$ must contain only one element, which allows us to identify the proportion $\pi_{1111}$ and determine the label of $G=1111$. This allows us to identify the parameters $(\mu_{0,0,1111}, \sigma^2_{0,0,1111})$, $(\mu_{0,1,1111}, \sigma^2_{0,1,1111})$, $(\mu_{1,0,1111}, \sigma^2_{1,0,1111})$, and $(\mu_{1,1,1111}, \sigma^2_{1,1,1111})$.  Consequently, $f(W\mid Z=0,M=0,  G=1111)$, $f(W\mid Z=0,M=1, G=1111)$, $f(W\mid Z=1,M=0, G=1111)$ and $f(W\mid Z=1,M=1, G=1111)$ can be identified.  

Finally, once $\pi_{rstu}({\mathcal{O}})$ and $f(W\mid Z=z,M=m,Y=y)$ are identified, the conditional probability $\pi_{rstu}(\tilde{{\mathcal{O}}})$ can be identified for ${\mathcal{O}}=(Z=z, M=m, Y=y)$ and $\tilde{{\mathcal{O}}}=({\mathcal{O}}, W=w)$. For example, suppose ${\mathcal{O}}=(Z=1,M=1,Y=1)$ and $\tilde{\mathcal{O}}=({\mathcal{O}},W )$, we then have, 
\begin{equation*}
   \begin{aligned}
\pi_{rst1}(\tilde{\mathcal{O}}) =\dfrac{f(W\mid Z=1,M =1,G=rst1)\pi_{rst1}( {\mathcal{O}})}{\sum_{r,s,t=0}^1 f(W\mid Z=1,M =1,G=rst1)\pi_{rst1}( {\mathcal{O}})}.
   \end{aligned}
\end{equation*} 
  \item[(2)] We next discuss the identifiability of $\pi_{rstu}({\mathcal{O}})$ and $\pi_{rstu}(\tilde{\mathcal{O}})$ under Assumption \ref{assump:equal-prop}(ii). We   consider a simpler structure of \eqref{eq:sets-iden} as follows: \begin{equation*} 
    \begin{aligned}
    \Sset_{001}&=\{\mu_{1stu}:s,t,u=0,1\}, ~~
    \Sset_{011} =\{\mu_{r1tu}:r,t,u=0,1\}, \\
    \Sset_{101}&=\{\mu_{rs1u}:r,s,u=0,1\},  ~~
    \Sset_{111} =\{\mu_{rst1}:r,s,t=0,1\}, \\
    \Sset_{000}&=\{\mu_{0stu}:s,t,u=0,1\}, ~~
    \Sset_{010}  =\{\mu_{r0tu}:r,t,u=0,1\}, \\
    \Sset_{100}&=\{\mu_{rs0u}:r,s,u=0,1\},~~
    \Sset_{110} =\{\mu_{rst0}:r,s,t=0,1\} .\\
\end{aligned}
\end{equation*}Without loss of generality, we assume that the labels of $  \Sset_{zmy}$ and $  \mathcal{C}_{zmy}$ are aligned with respect to $\mu_{rstu}$. We will determine the labels and other parameters in $  \mathcal{C}_{zmy}$ in  \eqref{eq:sets-iden}  through $  \Sset_{zmy}$.  For any type $G=rstu$,  let $\mathcal{T}_{rstu}=\Sset_{00r}\cap \Sset_{01s}\cap \Sset_{10t} \cap\Sset_{11u}$.
We must have the unique element in $\mathcal{T}_{rstu}$, which is $\mu_{rstu}$, since we assume that $\mu_{rstu}\neq \mu_{g'}$ in Assumption \ref{assump:equal-prop}(ii). Once the labels are determined, we can identify the corresponding proportion and variance parameters. Consequently, the conditional probabilities $\pi_{rstu}$ and $f(W\mid Z=z,M=m, G=rstu)$ can also be identified for $z,m, r,s,t,u\in\{0,1\}$.    For example, from  Table \ref{tab: probabilities}, it is observed that the class $G=1111$ appears in all four subpopulations, indicating that $\mu_{1111} \in \mathcal{S}_{001}\cap\mathcal{S}_{011}\cap\mathcal{S}_{101}\cap\mathcal{S}_{111}$. Assumption \ref{assump:equal-prop}(ii) implies that the intersection $\mathcal{S}_{001}\cap\mathcal{S}_{011}\cap\mathcal{S}_{101}\cap\mathcal{S}_{111}$ must contain only one element, which allows us to identify the proportion $\mu_{1111}$ and determine the label of $G=1111$. This allows us to identify the parameters $(\pi_{ 1111}, \sigma^2_{0,0,1111})$, $(\pi_{ 1111},  \sigma^2_{0,1,1111})$, $(\pi_{ 1111},  \sigma^2_{1,0,1111})$, and $(\pi_{ 1111}, \sigma^2_{1,1,1111})$.    Finally, once $\pi_{rstu} $ and $f(W\mid Z=z,M=m,G=rstu)$ are identified, the conditional probability $\pi_{rstu}(\tilde{{\mathcal{O}}})$ can be identified for ${\mathcal{O}}=(Z=z, M=m, Y=y)$ and $\tilde{{\mathcal{O}}}=({\mathcal{O}}, W=w)$. For example, suppose ${\mathcal{O}}=(Z=1,M=1,Y=1)$ and $\tilde{\mathcal{O}}=({\mathcal{O}},W )$, we then have, 
\begin{equation*}
   \begin{aligned}
\pi_{rst1}(\tilde{\mathcal{O}}) =\dfrac{f(W\mid Z=1,M=1,G=rst1)\pi_{rst1}( {\mathcal{O}})}{\sum_{r,s,t=0}^1 f(W\mid Z=1,M=1,G=rst1)\pi_{rst1}( {\mathcal{O}})}.
   \end{aligned}
\end{equation*}  
\item[(3)] The proof under Assumption \ref{assump:equal-prop}(iii) follows similarly to the proof under Assumption \ref{assump:equal-prop}(ii); hence, we omit the details for simplicity. 
\end{itemize}

\end{proof}

 \section{Application to the Job Search Intervention Study}
\label{ssec:app-job}
In this section, we apply the methods proposed in previous sections to the Job Search Intervention Study (JOBS II) \citep{vinokur1997mastery}.  The study aims to improve both the re-employment rates and the psychological well-being of unemployed job seekers. In the JOBS II field experiment, unemployed individuals completed a prescreening questionnaire and were randomly assigned to either a treated group or a control group. Participants in the treated group \((Z=1)\) attended a job search skills workshop to learn job search techniques and strategies for coping with job search challenges, while participants in the control group \((Z=0)\) received a booklet on job search skills. The variable \(M\) represents education level, where \(M=1\) indicates having a bachelor's degree or higher, and \(M=0\) indicates less than a bachelor's degree. In this analysis, the randomized experiment ensures that \(Z\) and \(M\) are independent, allowing us to adopt the proposed approach. The outcome variable \(Y\) is participants' self-assessment of their confidence and ability to succeed in finding a job, where \(Y=1\) indicates high job search self-efficacy, and \(Y=0\) indicates low job search self-efficacy. Baseline covariates \(X\) include economic hardship, pre-treatment depressive symptoms, gender, age, occupation type, marital status, and ethnicity, totaling eight variables. Post-treatment measures of depressive symptoms serve as a secondary outcome variable \(W\). After excluding some missing data, we obtained 899 observations.


 We aim to investigate whether self-efficacy in the JOBS II program is caused by vocational training alone, by education alone, or by two synergistic effects of $Z$ and $M$. We consider two different types of monotonicity restrictions: the monotonicity assumption \ref{ass:monotonicity} holds and the absence of any monotonicity assumption.  We use a multinomial logistic model for $\pr(G = rstu\mid X)$ and a normal model for \( f(W\mid Z, M, G = rstu, X)\).   We use the EM algorithm in Section \ref{ssec: estimation} to obtain MLEs of the parameters.  

{\black   {To evaluate model fit, we compare the estimated log-likelihoods under the two specifications and find that the model without monotonicity yields a higher likelihood and a lower AIC. These results suggest that Assumption \ref{ass:monotonicity} (monotonicity) may not be supported by this dataset. 
Accordingly, we conduct the subsequent analysis under Assumption \ref{assump:equal-prop}, which does not impose monotonicity. It is important to note that Assumption \ref{assump:equal-prop} imposes structural constraints that are not directly testable from the observed data. While the selected model provides better fit according to AIC, the validity of the attribution results depends critically on whether Assumption \ref{assump:equal-prop} holds in practice. Violations of this assumption may lead to biased estimates of the posterior probabilities. We therefore recommend that readers interpret these results with due caution and consider the plausibility of Assumption \ref{assump:equal-prop} in the context of   specific application.}

 To assess which condition in Assumption \ref{assump:equal-prop} holds empirically, we examine the estimated parameters $\pi_{rstu}$, $\mu_{z,m,rstu}$, and $\sigma^2_{z,m,rstu}$ for $z,m,r,s,t,u\in\{0,1\}$ with 200 Bootstrap. We find that the mean parameters vary not only across latent classes but also across exposure combinations within each class, indicating that Assumption \ref{assump:equal-prop}(ii) does not hold empirically. Similarly, the variance parameters also depend on the exposure patterns rather than solely on the latent class, suggesting that Assumption \ref{assump:equal-prop}(iii) does not hold. However, the estimated class proportions presented in the first part of Table \ref{SM-tab:est-res} show clear separation across all 16 latent classes, with no two classes having equal   proportions. This empirical evidence indicates that Assumption \ref{assump:equal-prop}(i) holds in this dataset, allowing identification through the distinct mixture weights. Consequently, $W$ (depressive symptoms) can be affected by the exposures while still providing valid identification of the latent class structure.

 }

\subsection{Point estimation of posterior probabilities}

\begin{table*}[t!]
\centering
\caption{Estimation results without any monotonicity, where ${\mathcal{O}}_1=(Z=M=Y=1) $ and  ${\mathcal{O}}_2=(Z=M=Y=0).$}
\label{SM-tab:est-res}
\resizebox{0.849049790\textwidth}{!}{  
  \begin{threeparttable}
\begin{tabular}{cccccc}
\toprule
$\pi_{rstu} $        & Point estimates and 95\% CIs &  &  & $\pi_{rstu} $        & Point estimates and 95\% CIs \\\addlinespace[0.25mm]  \cline{1-2} \cline{5-6}\addlinespace[0.5mm] 
$\pi_{0001} $            & 4.45\% (4.08\%, 4.89\%)    &  &  & $\pi_{0000}$             & 3.71\% (3.24\%, 4.46\%)    \\\addlinespace[0.5mm]
$\pi_{1001} $            & 7.39\% (6.60\%, 7.94\%)     &  &  & $\pi_{1000}$             & 4.83\% (4.12\%, 4.92\%)    \\\addlinespace[0.5mm]
$\pi_{0101} $            & 4.86\% (4.25\%, 5.12\%)    &  &  & $\pi_{0100}$             & 3.45\% (3.09\%, 3.65\%)    \\\addlinespace[0.5mm]
$\pi_{1101} $            & 8.76\% (7.68\%, 8.92\%)    &  &  & $\pi_{1100}$             & 5.35\% (4.52\%, 5.32\%)    \\\addlinespace[0.5mm]
$\pi_{0011} $            & 4.85\% (4.31\%, 5.23\%)    &  &  & $\pi_{0010}$             & 3.40\% (2.99\%, 3.58\%)     \\\addlinespace[0.5mm]
$\pi_{1011} $            & 8.72\% (7.71\%, 8.80\%)     &  &  & $\pi_{1010}$             & 5.17\% (4.41\%, 5.12\%)    \\\addlinespace[0.5mm]
$\pi_{0111} $            & 6.26\% (5.44\%, 6.61\%)    &  &  & $\pi_{0110}$             & 3.87\% (3.33\%, 3.96\%)    \\\addlinespace[0.5mm]
$\pi_{1111} $            & \bf  17.97\% (19.65\%, 23.23\%) &  &  & $\pi_{1110}$             & 6.96\% (5.95\%, 6.88\%)    \\\addlinespace[0.25mm] \toprule \addlinespace[0.5mm] 
$\pi_{rstu}( {\mathcal{O}}_1)$  & Point estimates and 95\% CIs &  &  & $\pi_{rstu}( {\mathcal{O}}_2)$  & Point estimates and 95\% CIs \\ \addlinespace[0.25mm]\cline{1-2} \cline{5-6} \addlinespace[0.5mm]
$\pi_{0 0 0 1} ({\mathcal{O}}_{1})$ & 7.58\% (7.04\%, 8.49\%)    &  &  & $\pi_{0 0 0 0}({\mathcal{O}}_{2}) $ & 7.63\% (6.87\%, 8.78\%)    \\\addlinespace[0.5mm]
$\pi_{1 0 0 1} ({\mathcal{O}}_{1})$ & 12.21\% (10.53\%, 12.53\%) &  &  & $\pi_{0 1 0 0}({\mathcal{O}}_{2}) $ & 9.54\% (8.81\%, 10.10\%)    \\\addlinespace[0.5mm]
$\pi_{0 1 0 1} ({\mathcal{O}}_{1})$ & 7.08\% (6.97\%, 8.32\%)    &  &  & $\pi_{0 0 1 0}({\mathcal{O}}_{2}) $ & 9.54\% (8.67\%, 9.81\%)    \\\addlinespace[0.5mm]
$\pi_{1 1 0 1} ({\mathcal{O}}_{1})$ & 12.26\% (11.78\%, 13.59\%) &  &  & $\pi_{0 1 1 0}({\mathcal{O}}_{2}) $ & 11.22\% (10.18\%, 11.55\%) \\\addlinespace[0.5mm]
$\pi_{0 0 1 1} ({\mathcal{O}}_{1})$ & 7.78\% (7.15\%, 8.67\%)    &  &  & $\pi_{0 0 0 1}({\mathcal{O}}_{2}) $ & 13.77\% (12.16\%, 13.76\%) \\\addlinespace[0.5mm]
$\pi_{1 0 1 1} ({\mathcal{O}}_{1})$ & 12.35\% (11.79\%, 13.43\%) &  &  & $\pi_{0 1 0 1}({\mathcal{O}}_{2}) $ & 13.89\% (13.75\%, 15.72\%) \\\addlinespace[0.5mm]
$\pi_{0 1 1 1} ({\mathcal{O}}_{1})$ & 9.33\% (8.60\%, 10.28\%)    &  &  & $\pi_{0 0 1 1}({\mathcal{O}}_{2}) $ & 15.05\% (13.71\%, 15.42\%) \\\addlinespace[0.5mm]
$\pi_{1 1 1 1} ({\mathcal{O}}_{1})$ &\bf 31.42\% (28.2\%, 32.65\%)  &  &  & $\pi_{0 1 1 1}({\mathcal{O}}_{2}) $ &\bf 19.36\% (19.17\%, 21.55\%) \\ \addlinespace[0.5mm]\bottomrule
\end{tabular}
  \end{threeparttable}}
\end{table*}
Without any monotonicity assumption, we present the point estimates and 95\% confidence intervals (95\% CI) of the proportion \(\pr(Y_{0,0}=r,Y_{0,1}=s,Y_{1,0}=t,Y_{1,1}=u)\) where \(r, s, t, u \in \{0,1\}\) in Table \ref{SM-tab:est-res}. Firstly, we find that the subgroup $G=1111$ with the largest proportion, approximately \( 17.97\%\), consists of individuals who can achieve higher job search self-efficacy regardless of whether they participate in job training or their educational status. In addition, the second largest subgroup consisted of individuals who had low job search self-efficacy only when they had both vocational training and low education \((\hat{\pi}_{1101} \approx 8.76\%\)), suggesting that the most significant synergistic effect was due to \((Z=1, M=0)\).

We also consider the attribution analysis with the evidence \({\mathcal{O}}_1 = (Z = M = Y = 1)\) in the second part of Table \ref{SM-tab:est-res}. We find that the units with these characteristics are most likely to belong to the subgroup \(G = 1111\)   because \(\pi_{1111}({\mathcal{O}}_1) \approx 31.42\%\), which indicates that their high self-efficacy is more likely to be caused by factors other than education or job training, as these units will have self-efficacy under any combination of exposures.   Secondly, we consider some counterfactual probabilities applied to future career plan recommendations. We find that \(\pr(Y_{0,0}=0\mid {\mathcal{O}}_1)=\pi_{0001}({\mathcal{O}}_1)+\pi_{0101}({\mathcal{O}}_1) +\pi_{0011}({\mathcal{O}}_1) +\pi_{0111}({\mathcal{O}}_1) \approx 31.77\%\); similarly, \(\pr(Y_{0,1}=0\mid {\mathcal{O}}_1)\approx 39.92\%\) and \(\pr(Y_{1,0}=0\mid {\mathcal{O}}_1)\approx 39.12\%\). These quantities can be interpreted as certain post-intervention probability of necessity \citep{zhao2023conditional}. For instance, $\pr(Y_{0,1}=0\mid {\mathcal{O}}_1)$ represents the probability that job training is a necessary cause for high self-efficacy among individuals controlled for \(M=1\) (higher education group), while $\pr(Y_{1,0}=0\mid {\mathcal{O}}_1)$ represents the probability that education is a necessary cause for high self-efficacy among individuals controlled for \(Z=1\) (job training group). We note that \(\pr(Y_{0,1}=0 \mid {\mathcal{O}}_1) \approx 39.92\%\) is the largest among the three counterfactual probabilities. This indicates that for future similar training programs, practitioners should continue providing job training for these unemployed individuals with higher education (\(M=1\)). This is because, with job training, they are likely to avoid low self-satisfaction as much as possible. 


We also conduct the attribution analysis with the evidence \({\mathcal{O}}_2 = (Z = M = Y = 0)\), as shown in the second part of Table \ref{SM-tab:est-res}. We find that the subgroup \( G = 0111 \) has the highest proportion among individuals with low self-efficacy, with \( \pi_{0111}({\mathcal{O}}_2) \approx 19.36\% \). This indicates that low job search self-efficacy occurs only when individuals receive no job training \( (Z=0) \) and are not highly educated \( (M=0) \). Thus, the most likely cause of low self-efficacy in this context is the absence of this synergistic factor \( (Z=0, M=0) \).  Furthermore, we observe that \(\pr(Y_{0,1}=1\mid {\mathcal{O}}_2) \approx 54.01\%\), \(\pr(Y_{1,0}=1\mid {\mathcal{O}}_2) \approx 55.17\%\), and \(\pr(Y_{1,1}=1\mid {\mathcal{O}}_2) \approx 62.07\%\). Given that the conditional probability $\pr(Y_{1,1}=1\mid {\mathcal{O}}_2) \approx 62.07\%$ is the highest, this indicates that in future training programs, practitioners should focus on providing this subpopulation with job-related skills and instruction similar to higher education to maximize their self-efficacy.

 In conclusion, for the subgroup \( (Z=1, M=1, Y=1) \), we find that the main reason for the high self-efficacy may be due to other factors. Conversely, for the subgroup \( (Z=0, M=0, Y=0) \), we find that the main reason for the low self-efficacy may be due to the lack of job training and high education simultaneously. 

 \subsection{Further analysis with secondary outcome}
In labor economics and social psychology, the secondary outcome variable, depressive symptoms \(W\), is crucial for evaluating policy effectiveness.  We now incorporate \(W\) into the evidence set for retrospective attribution analysis. The empirical range of depressive symptoms \(W\) in the observed data is \([1, 4.5]\), with higher values indicating more severe depression. We classify \(W\) into three categories: \(1 \leq W < 2\) as a mild depressive state, \(2 \leq W < 3.5\) as a moderate depressive state, and \(3.5 \leq W < 4.5\) as a severe depressive state. Our goal is to explore whether employment self-efficacy in different depressive states is caused by job training alone \(Z\), education alone \(M\), or various synergistic effects of \(Z\) and \(M\).

\begin{table*}[t!]
\centering
\caption{Estimation results with secondary outcome.}
\label{tab:est-res-2}
\resizebox{0.9949790\textwidth}{!}{  
  \begin{threeparttable}
\begin{tabular}{ccccccc}\toprule
$\pi_{rstu} (\tilde{\mathcal{O}}_{1k})$      &  & $\tilde{\mathcal{O}}_{1,1}=(Z=M=Y=1,W\leq 2)$ &  & $\tilde{\mathcal{O}}_{12}=(Z=M=Y=1,2 <W\leq 3.5)$ &  & $\tilde{\mathcal{O}}_{13}=(Z=M=Y=1,W>3.5)$ \\\addlinespace[0.25mm] \cline{1-1} \cline{3-3} \cline{5-5} \cline{7-7} \addlinespace[0.75mm]
$\pi_{0 0 0 1} (\tilde{\mathcal{O}}_{1k})$    &  & 5.87\% (5.38\%, 6.34\%)           &  & 7.61\% (7.02\%, 8.21\%)                &  & 9.99\% (7.87\%, 12.14\%)        \\\addlinespace[0.25mm]
$\pi_{1 0 0 1} (\tilde{\mathcal{O}}_{1k})$    &  & 8.86\% (8.49\%, 9.24\%)           &  & 10.42\% (9.97\%, 10.89\%)              &  & 12.24\% (10.26\%, 14.34\%)      \\\addlinespace[0.25mm]
$\pi_{0 1 0 1} (\tilde{\mathcal{O}}_{1k})$    &  & 6.3\% (6.03\%, 6.57\%)            &  & 8.62\% (8.29\%, 8.94\%)                &  & 9.14\% (8.12\%, 10.23\%)        \\\addlinespace[0.25mm]
$\pi_{1 1 0 1} (\tilde{\mathcal{O}}_{1k})$    &  & 12.18\% (11.78\%, 12.56\%)        &  & 14.21\% (13.93\%, 14.47\%)             &  & 13.31\% (12.52\%, 14.16\%)      \\\addlinespace[0.25mm]
$\pi_{0 0 1 1} (\tilde{\mathcal{O}}_{1k})$    &  & 6.62\% (6.23\%, 6.99\%)           &  & 8.39\% (8.02\%, 8.75\%)                &  & 10.28\% (9.17\%, 11.39\%)       \\\addlinespace[0.25mm]
$\pi_{1 0 1 1} (\tilde{\mathcal{O}}_{1k})$    &  & 13.36\% (12.99\%, 13.75\%)        &  & 13.6\% (13.35\%, 13.83\%)              &  & 13.41\% (12.54\%, 14.26\%)      \\\addlinespace[0.25mm]
$\pi_{0 1 1 1} (\tilde{\mathcal{O}}_{1k})$    &  & 10.39\% (10.09\%, 10.7\%)         &  & 11.39\% (11.10\%, 11.66\%)              &  & 11.06\% (9.92\%, 12.2\%)        \\\addlinespace[0.25mm]
$\pi_{1 1 1 1} (\tilde{\mathcal{O}}_{1k})$    &  & \bf 36.41\% (35.05\%, 37.83\%)        &  & \bf 25.76\% (24.43\%, 27.12\%)             &  &\bf  20.58\% (16.04\%, 24.85\%)      \\\addlinespace[0.25mm]
$\pr(Y_{0,0}=1\mid \tilde{\mathcal{O}}_{1k})$ &  & 29.19\% (28.04\%, 30.29\%)        &  & 36.01\% (34.77\%, 37.23\%)             &  & 40.47\% (36.51\%, 44.53\%)      \\\addlinespace[0.25mm]
$\pr(Y_{0,1}=1\mid \tilde{\mathcal{O}}_{1k})$ &  & \bf 34.71\% (33.48\%, 35.92\%)        &  & 40.02\% (38.84\%, 41.21\%)             &  &\bf  45.92\% (42.21\%, 49.76\%)      \\\addlinespace[0.25mm]
$\pr(Y_{1,0}=1\mid \tilde{\mathcal{O}}_{1k})$ &  & 33.22\% (32.19\%, 34.18\%)        &  & \bf 40.86\% (39.73\%, 41.99\%)             &  & 44.68\% (40.48\%, 49.15\%)      \\\addlinespace[0.75mm]\toprule 
$\pi_{rstu} (\tilde{\mathcal{O}}_{2k})$      &  & $\tilde{\mathcal{O}}_{21}=(Z=M=Y=0,W\leq 2)$ &  & $\tilde{\mathcal{O}}_{22}=(Z=M=Y=0,2 <W\leq 3.5)$ &  & $\tilde{\mathcal{O}}_{23}=(Z=M=Y=0,W>3.5)$ \\ \cline{1-1} \cline{3-3} \cline{5-5} \cline{7-7} \addlinespace[0.75mm] 
$\pi_{0 0 0 0} (\tilde{\mathcal{O}}_{2k})$    &  & 6.24\% (5.46\%, 7\%)              &  & 8.6\% (7.31\%, 9.9\%)                  &  & 11.02\% (4.65\%, 17.24\%)       \\\addlinespace[0.25mm]
$\pi_{0 1 0 0} (\tilde{\mathcal{O}}_{2k})$    &  & 7.67\% (7.52\%, 7.85\%)           &  & 9.2\% (8.95\%, 9.44\%)                 &  & 10.13\% (9.34\%, 10.94\%)       \\\addlinespace[0.25mm]
$\pi_{0 0 1 0} (\tilde{\mathcal{O}}_{2k})$    &  & 7.31\% (7.10\%, 7.54\%)            &  & 8.68\% (8.43\%, 8.92\%)                &  & 9.7\% (9\%, 10.42\%)            \\\addlinespace[0.25mm]
$\pi_{0 1 1 0} (\tilde{\mathcal{O}}_{2k})$    &  & 10.07\% (9.8\%, 10.41\%)          &  & 10.39\% (10.10\%, 10.65\%)              &  & 11.05\% (9.86\%, 12.27\%)       \\\addlinespace[0.25mm]
$\pi_{0 0 0 1} (\tilde{\mathcal{O}}_{2k})$    &  & 10.52\% (10.11\%, 10.91\%)        &  & 11.51\% (11.09\%, 11.94\%)             &  & 13.12\% (11.5\%, 14.77\%)       \\\addlinespace[0.25mm]
$\pi_{0 1 0 1} (\tilde{\mathcal{O}}_{2k})$    &  & 15.26\% (14.79\%, 15.69\%)        &  & 15.15\% (14.74\%, 15.58\%)             &  & 13.09\% (11.64\%, 14.64\%)      \\\addlinespace[0.25mm]
$\pi_{0 0 1 1} (\tilde{\mathcal{O}}_{2k})$    &  & 13.62\% (13.28\%, 13.96\%)        &  & 14.22\% (13.9\%, 14.55\%)              &  & 14.7\% (13.16\%, 16.24\%)       \\\addlinespace[0.25mm]
$\pi_{0 1 1 1} (\tilde{\mathcal{O}}_{2k})$    &  & \bf 29.31\% (28.24\%, 30.37\%)        &  &\bf  22.24\% (21.14\%, 23.34\%)             &  & \bf 17.18\% (13.65\%, 20.7\%)       \\\addlinespace[0.25mm]
$\pr(Y_{0,1}=1\mid \tilde{\mathcal{O}}_{2k})$ &  & 62.31\% (61.05\%, 63.61\%)        &  & 56.99\% (55.56\%, 58.39\%)             &  & 51.45\% (46.28\%, 56.75\%)      \\\addlinespace[0.25mm]
$\pr(Y_{1,0}=1\mid \tilde{\mathcal{O}}_{2k})$ &  & 60.31\% (59.42\%, 61.27\%)        &  & 55.54\% (54.29\%, 56.75\%)             &  & 52.63\% (47.27\%, 58.01\%)      \\\addlinespace[0.25mm]
$\pr(Y_{1,1}=1\mid \tilde{\mathcal{O}}_{2k})$ &  & \bf 68.71\% (67.69\%, 69.65\%)        &  &\bf  63.13\% (61.79\%, 64.5\%)              &  &\bf  58.09\% (52.3\%, 63.99\%)       \\ \toprule
\end{tabular}
\end{threeparttable}}
\end{table*}

The first part of Table \ref{tab:est-res-2} shows the estimated posterior   probabilities \(\pr(G = rst1 \mid \tilde{{\mathcal{O}}}_{1k})\) for $k\in\{1,2,3\}$, where \(\tilde{{\mathcal{O}}}_{1,1}=(Z=M=Y=1, W\leq 2)\), \(\tilde{{\mathcal{O}}}_{12}=(Z=M=Y=1, 2 <W\leq 3.5)\), and \(\tilde{{\mathcal{O}}}_{13}=(Z=M=Y=1, W>3.5)\). We observe that among the three different levels of depression, the main latent subgroup is \(G = 1111\), which consists of individuals with high job search self-efficacy in all situations. This conclusion is very similar to that of Table \ref{SM-tab:est-res}. However, in the high depression group, the proportion of \(G = 1111\) is reduced compared to the low and moderate depression groups. This suggests that depression status and employment self-efficacy are likely caused by other factors unrelated to employment and education. We also calculated a series of policy metrics to guide future employment programs, including $\pr (Y_{0,0}=0\mid\tilde{\mathcal{O}}_{1k})$, $\pr (Y_{0,1}=0\mid\tilde{\mathcal{O}}_{1k})$, and $\pr (Y_{1,1}=0\mid\tilde{\mathcal{O}}_{1k})$ in Table \ref{tab:est-res-2}. We find that for groups with low and high levels of depression, i.e., $\tilde{\mathcal{O}}_{1,1}$ and $\tilde{\mathcal{O}}_{13}$, continuous job training should be provided to enhance their employment self-efficacy. For the moderate depression group $\tilde{\mathcal{O}}_{12}$, since individuals in this group have already received higher education (\(M=1\)), it is challenging to achieve the counterfactual $ \pr(Y_{1,0}=0\mid\tilde {\mathcal{O}}_{12}) \approx 40.86\%$ by changing their state of \(M=1\) to \(M=0\). Therefore, in practice, we recommend using $ \pr(Y_{0,1}=0\mid\tilde {\mathcal{O}}_{12})\approx 40.02\% $ as guidance and still suggest offering vocational training in the future.

The second part of Table \ref{tab:est-res-2} presents the estimated posterior   probabilities \(\pr(G = 0stu \mid \tilde{{\mathcal{O}}}_{2k})\), where \(k \in \{1,2,3\}\), \(\tilde{{\mathcal{O}}}_{21}=(Z=M=Y=0, W \leq 2)\), \(\tilde{{\mathcal{O}}}_{22}=(Z=M=Y=0, 2 < W \leq 3.5)\), and \(\tilde{{\mathcal{O}}}_{23}=(Z=M=Y=0, W > 3.5)\). We observe that among the three levels of depression, the main latent subgroup is \(G = 0111\), which includes individuals with low job search self-efficacy only if they have not received job training and lack higher education. However, in the high depression group, the proportion of \(G = 0111\) is also reduced compared to the low and moderate depression groups. These results suggest that low self-efficacy in \(\tilde{{\mathcal{O}}}_2\) is most likely due to the lack of job training \((Z=0)\) and higher education \((M=0)\). Furthermore, future career policy metrics indicate that individuals with low job search self-efficacy should be provided with both job training and higher education, regardless of their depression level.

{\black  The results presented in this section rely on Assumption \ref{assump:equal-prop} rather than monotonicity assumptions. While this approach provides flexibility in incorporating secondary outcomes, practitioners should interpret these causal attribution results with appropriate caution. Violations of Assumption \ref{assump:equal-prop} may lead to biased estimates of posterior probabilities. We recommend that researchers carefully assess the plausibility of at least one of the three conditions in Assumption \ref{assump:equal-prop} in their specific application context, and consider the model selection when possible.}

\section{An additional example with additional covariates}
\label{sec:example-cov-extend}
In this section, we extend the smoking and asbestos example from the main text to a more realistic scenario by incorporating additional covariates, including age, genotype variables, and drinking history. Given the strong biological evidence \citep{lerman1999evidence}, we employ genotype variables to characterize individual susceptibility to carcinogen exposure. Moreover, in addition to BMI considered in the previous section, we include the Quality of Life Scale (QOLS), denoted as $W_2$, as a secondary outcome to capture patient-centered well-being beyond clinical endpoints.

The QOLS is a widely used metric for measuring subjective well-being across physical, psychological, and social dimensions. Higher values of $W_2$ reflect better perceived quality of life, with healthy individuals typically reporting higher scores, those with moderate chronic conditions showing intermediate values, and individuals with severe health conditions generally reporting substantially lower scores.  

\subsection{Data generation}
  {\begin{enumerate} 
\item  We generate the covariates as follows: $X_1 \sim \text{Uniform}(35, 55)$ which is then standardized, $X_2$ is a genotype variable taking values in $\{0,1,2\}$ with probabilities $\{0.6, 0.3, 0.1\}$ respectively, and $X_3$ is a drinking choice variable taking values in $\{0,1\}$ with equal probabilities of $0.5$. Let ${\boldsymbol X}=(1,X_1,X_2,X_3)^\T$.

\item  We generate the binary exposure variables $Z$ and $M$ from a multinomial logistic model:
  $$ \pr(Z=z,M=m\mid {\boldsymbol X})={\mathrm{exp}\left({\boldsymbol \alpha}_{z,m }^\T{\boldsymbol X} \right)}\Big/{\textstyle\sum_{z{},m{}=0}^1 \mathrm{exp}\left({\boldsymbol \alpha}_{z,m }^\T{\boldsymbol X} \right)},$$     
  where $\boldsymbol \alpha_{0,0} =(0.00, 0.00, 0.00, 0.00)^\T$ is used for identification. The other parameters are $\boldsymbol \alpha_{0,1} = (-0.31, 0.10, 0.09, 0.10)^\T$, $\boldsymbol \alpha_{1,0} = (0.29, 0.11, -0.39, -0.21)^\T$, and $\boldsymbol \alpha_{1,1} = (0.13, 0.01, 0.15, -0.04)^\T$.

\item The secondary outcomes  $W_{i2}$ is  generated conditional on the covariates $X_{i1}$,  $X_{i2}$, and $X_{i3}$:
\begin{enumerate}[label=(\alph*)]
    \item For $(Z_i=0, M_i=0, Y_i=1)$:
    \[
    W_{i2} \sim N(35 - 0.015 X_{i1} - 0.256 X_{i2} + 0.265 X_{i3}, 5^2).
    \]
    
    \item For $(Z_i=0, M_i=0, Y_i=0)$:
    \[
    \begin{aligned}
    W_{i2} &\sim~ \left\{\begin{gathered}
       \frac{\exp({\boldsymbol \theta}_{0000}^\T {\boldsymbol X})}{\sum_{g \in \{0000,0001,0011,0101,0111\}} \exp({\boldsymbol \theta}_{rstu}^\T {\boldsymbol X})}N(88 - 0.195 X_{i1} + 0.007 X_{i2} - 0.319 X_{i3}, 4^2) \\
       + \frac{\exp({\boldsymbol \theta}_{0001}^\T {\boldsymbol X})}{\sum_{g \in \{0000,0001,0011,0101,0111\}} \exp({\boldsymbol \theta}_{rstu}^\T {\boldsymbol X})}N(75 + 0.260 X_{i1} - 0.299 X_{i2} - 0.241 X_{i3}, 3^2) \\
     +  \frac{\exp({\boldsymbol \theta}_{0011}^\T {\boldsymbol X})}{\sum_{g \in \{0000,0001,0011,0101,0111\}} \exp({\boldsymbol \theta}_{rstu}^\T {\boldsymbol X})}N(80 + 0.492 X_{i1} + 0.307 X_{i2} + 0.053 X_{i3}, 1^2) \\
     + \frac{\exp({\boldsymbol \theta}_{0101}^\T {\boldsymbol X})}{\sum_{g \in \{0000,0001,0011,0101,0111\}} \exp({\boldsymbol \theta}_{rstu}^\T {\boldsymbol X})}N(60 + 0.146 X_{i1} - 0.188 X_{i2} + 0.122 X_{i3}, 3.5^2)\\
     +  \frac{\exp({\boldsymbol \theta}_{0111}^\T {\boldsymbol X})}{\sum_{g \in \{0000,0001,0011,0101,0111\}} \exp({\boldsymbol \theta}_{rstu}^\T {\boldsymbol X})}N(50 - 0.170 X_{i1} + 0.002 X_{i2} + 0.177 X_{i3}, 2^2)
    \end{gathered}\right\},
    \end{aligned}
    \]   where $\boldsymbol \theta_{0000}=(0.00, 0.00, 0.00, 0.00)^\T$ is used for identification. The other parameters are $\boldsymbol \theta_{0001} = (-0.92, -0.13, 0.25, 0.46)^\T$, $\boldsymbol \theta_{0011} = (-0.98, -0.12, 0.20, 0.42)^\T$, $\boldsymbol \theta_{0101} = (-0.89, -0.13, -0.14, 0.02)^\T$, and $\boldsymbol \theta_{0111} = (-0.59, -0.19, -0.16, 0.23)^\T$.
    
    \item For $(Z_i=0, M_i=1, Y_i=1)$:
    \[
    W_{i2} \sim \begin{Bmatrix}
        \begin{gathered}
            \frac{\exp({\boldsymbol \theta}_{0101}^\T {\boldsymbol X})}{\sum_{g \in \{0101,0111,1111\}} \exp({\boldsymbol \theta}_{rstu}^\T {\boldsymbol X})}N(60 + 0.146 X_{i1} - 0.188 X_{i2} + 0.122 X_{i3}, 3.5^2) \\
            + \frac{\exp({\boldsymbol \theta}_{0111}^\T {\boldsymbol X})}{\sum_{g \in \{0101,0111,1111\}} \exp({\boldsymbol \theta}_{rstu}^\T {\boldsymbol X})}N(50 - 0.170 X_{i1} + 0.002 X_{i2} + 0.177 X_{i3}, 2^2) \\
            + \frac{\exp({\boldsymbol \theta}_{1111}^\T {\boldsymbol X})}{\sum_{g \in \{0101,0111,1111\}} \exp({\boldsymbol \theta}_{rstu}^\T {\boldsymbol X})}N(35 - 0.015 X_{i1} - 0.256 X_{i2} + 0.265 X_{i3}, 5^2)
        \end{gathered}
    \end{Bmatrix},
        \] where $\boldsymbol \theta_{1111} = (-0.74, -0.11, -0.12, 0.13)^\T$.
    
    \item For $(Z_i=0, M_i=1, Y_i=0)$:
    \[
    W_{i2} \sim~ \left\{\begin{gathered} 
    \frac{\exp({\boldsymbol \theta}_{0000}^\T {\boldsymbol X})}{\sum_{g \in \{0000,0001,0011\}} \exp({\boldsymbol \theta}_{rstu}^\T {\boldsymbol X})}N(88 - 0.195 X_{i1} + 0.007 X_{i2} - 0.319 X_{i3}, 4^2) \\
    + \frac{\exp({\boldsymbol \theta}_{0001}^\T {\boldsymbol X})}{\sum_{g \in \{0000,0001,0011\}} \exp({\boldsymbol \theta}_{rstu}^\T {\boldsymbol X})}N(75 + 0.260 X_{i1} - 0.299 X_{i2} - 0.241 X_{i3}, 3^2)\\
    + \frac{\exp({\boldsymbol \theta}_{0011}^\T {\boldsymbol X})}{\sum_{g \in \{0000,0001,0011\}} \exp({\boldsymbol \theta}_{rstu}^\T {\boldsymbol X})}N(80 + 0.492 X_{i1} + 0.307 X_{i2} + 0.053 X_{i3}, 1^2)
    \end{gathered}\right\}.
    \] 
    
    \item For $(Z_i=1, M_i=0, Y_i=1)$:
    \[
    W_{i2} \sim \begin{Bmatrix}
        \begin{gathered}
            \frac{\exp({\boldsymbol \theta}_{0011}^\T {\boldsymbol X})}{\sum_{g \in \{0011,0111,1111\}} \exp({\boldsymbol \theta}_{rstu}^\T {\boldsymbol X})}N(80 + 0.492 X_{i1} + 0.307 X_{i2} + 0.053 X_{i3}, 1^2) \\
            + \frac{\exp({\boldsymbol \theta}_{0111}^\T {\boldsymbol X})}{\sum_{g \in \{0011,0111,1111\}} \exp({\boldsymbol \theta}_{rstu}^\T {\boldsymbol X})}N(50 - 0.170 X_{i1} + 0.002 X_{i2} + 0.177 X_{i3}, 2^2) \\
            + \frac{\exp({\boldsymbol \theta}_{1111}^\T {\boldsymbol X})}{\sum_{g \in \{0011,0111,1111\}} \exp({\boldsymbol \theta}_{rstu}^\T {\boldsymbol X})}N(35 - 0.015 X_{i1} - 0.256 X_{i2} + 0.265 X_{i3}, 5^2)
        \end{gathered}
    \end{Bmatrix}.
    \] 
    
    \item For $(Z_i=1, M_i=0, Y_i=0)$:
    \[
  \begin{aligned}
        W_{i2} \sim \begin{Bmatrix}
            \begin{gathered}
                \frac{\exp({\boldsymbol \theta}_{0000}^\T {\boldsymbol X})}{\sum_{g \in \{0000,0001,0101\}} \exp({\boldsymbol \theta}_{rstu}^\T {\boldsymbol X})}N(88 - 0.195 X_{i1} + 0.007 X_{i2} - 0.319 X_{i3}, 4^2) \\  
                + \frac{\exp({\boldsymbol \theta}_{0001}^\T {\boldsymbol X})}{\sum_{g \in \{0000,0001,0101\}} \exp({\boldsymbol \theta}_{rstu}^\T {\boldsymbol X})}N(75 + 0.260 X_{i1} - 0.299 X_{i2} - 0.241 X_{i3}, 3^2)\\
                + \frac{\exp({\boldsymbol \theta}_{0101}^\T {\boldsymbol X})}{\sum_{g \in \{0000,0001,0101\}} \exp({\boldsymbol \theta}_{rstu}^\T {\boldsymbol X})}N(60 + 0.146 X_{i1} - 0.188 X_{i2} + 0.122 X_{i3}, 3.5^2)
            \end{gathered}
        \end{Bmatrix}.
  \end{aligned}
    \] 
    
    \item For $(Z_i=1, M_i=1, Y_i=1)$:
    \[
    \begin{aligned}
    W_{i2} \sim~\left\{\begin{gathered}
        \frac{\exp({\boldsymbol \theta}_{0001}^\T {\boldsymbol X})}{\sum_{g \in \{0001,0011,0101,0111,1111\}} \exp({\boldsymbol \theta}_{rstu}^\T {\boldsymbol X})}N(75 + 0.260 X_{i1} - 0.299 X_{i2} - 0.241 X_{i3}, 3^2) \\
        + \frac{\exp({\boldsymbol \theta}_{0011}^\T {\boldsymbol X})}{\sum_{g \in \{0001,0011,0101,0111,1111\}} \exp({\boldsymbol \theta}_{rstu}^\T {\boldsymbol X})}N(80 + 0.492 X_{i1} + 0.307 X_{i2} + 0.053 X_{i3}, 1^2) \\
     +  \frac{\exp({\boldsymbol \theta}_{0101}^\T {\boldsymbol X})}{\sum_{g \in \{0001,0011,0101,0111,1111\}} \exp({\boldsymbol \theta}_{rstu}^\T {\boldsymbol X})}N(60 + 0.146 X_{i1} - 0.188 X_{i2} + 0.122 X_{i3}, 3.5^2) \\
     + \frac{\exp({\boldsymbol \theta}_{0111}^\T {\boldsymbol X})}{\sum_{g \in \{0001,0011,0101,0111,1111\}} \exp({\boldsymbol \theta}_{rstu}^\T {\boldsymbol X})}N(50 - 0.170 X_{i1} + 0.002 X_{i2} + 0.177 X_{i3}, 2^2)\\
     + \frac{\exp({\boldsymbol \theta}_{1111}^\T {\boldsymbol X})}{\sum_{g \in \{0001,0011,0101,0111,1111\}} \exp({\boldsymbol \theta}_{rstu}^\T {\boldsymbol X})}N(35 - 0.015 X_{i1} - 0.256 X_{i2} + 0.265 X_{i3}, 5^2)
    \end{gathered}\right\}.
    \end{aligned}
    \] 
    
    \item For $(Z_i=1, M_i=1, Y_i=0)$:
    \[
    W_{i2} \sim  N(88 - 0.195 X_{i1} + 0.007 X_{i2} - 0.319 X_{i3}, 4^2) .  
    \] 
\end{enumerate} 
\end{enumerate}}

\subsection{Casual attribution analysis}
\begin{table}[h]
\centering
\caption{The posterior probabilities $\pi_{rstu}(\mathcal{O})$ for different evidence.}
\vspace{0.25cm}
\label{tab:est-res-updated}
\resizebox{0.99949790\textwidth}{!}{  
  \begin{threeparttable}
\setlength{\arrayrulewidth}{1.25pt} 
\begin{tabular}{ccc|ccc}
\multicolumn{3}{c|}{(a) Proposed method for $\mathcal{O}=\emptyset$}                    & \multicolumn{3}{c}{(d) Proposed method for $\mathcal{O}=(Z=1,M=0,Y=1)$}                  \\\addlinespace[-2.5mm]
\multicolumn{3}{l|}{}                                                          & \multicolumn{3}{l}{}                                                            \\
Synergistic              & Smoking                  & Immune                   & Synergistic              & Smoking                  & Immune                    \\
$\pi_{0001}=$16.38$\%$     & $\pi_{0011}=$14.62$\%$     & $\pi_{0000}=$28.08$\%$     & $\pi_{0001}(\mathcal{O})=$0$\%$     & $\pi_{0011}(\mathcal{O})=$66.95$\%$ & $\pi_{0000}(\mathcal{O})=$0$\%$      \\
\multicolumn{1}{l}{}     & \multicolumn{1}{l}{}     & \multicolumn{1}{l|}{}    & \multicolumn{1}{l}{}     & \multicolumn{1}{l}{}     & \multicolumn{1}{l}{}      \\\addlinespace[-2.5mm]
Asbestos                 & Parallel                 & Doomed                   & Asbestos                 & Parallel                 & Doomed                    \\
$\pi_{0101}=$11.13$\%$     & $\pi_{0111}=$16.20$\%$     & $\pi_{1111}=$13.59$\%$     & $\pi_{0101}(\mathcal{O})=$0$\%$     & $\pi_{0111}(\mathcal{O})=$20.57$\%$ & $\pi_{1111}(\mathcal{O})=$12.49$\%$  \\
                         &                          &                          &                          &                          &                           \\\addlinespace[-1mm] 
\hline\addlinespace[-1mm]
                         &                          &                          &                          &                          &                           \\
\multicolumn{3}{c|}{(b) Proposed method for $\mathcal{O}=(Z=0,M=0,Y=1)$}                & \multicolumn{3}{c}{(e) Proposed method for $\mathcal{O}=(Z=1,M=1,Y=1)$}                  \\
\multicolumn{3}{l|}{}                                                          & \multicolumn{3}{l}{}                                                            \\\addlinespace[-2.5mm]
Synergistic              & Smoking                  & Immune                   & Synergistic              & Smoking                  & Immune                    \\
$\pi_{0001}(\mathcal{O})=$0$\%$     & $\pi_{0011}(\mathcal{O})=$0$\%$     & $\pi_{0000}(\mathcal{O})=$0$\%$     & $\pi_{0001}(\mathcal{O})=$71.01$\%$ & $\pi_{0011}(\mathcal{O})=$14.16$\%$ & $\pi_{0000}(\mathcal{O})=$0.00$\%$   \\
\multicolumn{1}{l}{}     & \multicolumn{1}{l}{}     & \multicolumn{1}{l|}{}    & \multicolumn{1}{l}{}     & \multicolumn{1}{l}{}     & \multicolumn{1}{l}{}      \\\addlinespace[-2.5mm]
Asbestos                 & Parallel                 & Doomed                   & Asbestos                 & Parallel                 & Doomed                    \\
$\pi_{0101}(\mathcal{O})=$0$\%$     & $\pi_{0111}(\mathcal{O})=$0$\%$     & $\pi_{1111}(\mathcal{O})=$100$\%$   & $\pi_{0101}(\mathcal{O})=$7.84$\%$  & $\pi_{0111}(\mathcal{O})=$4.35$\%$  & $\pi_{1111}(\mathcal{O})=$2.64$\%$   \\
                         &                          &                          &                          &                          &                           \\ \addlinespace[-1mm]
\hline\addlinespace[-1mm]
                         &                          &                          &                          &                          & \multicolumn{1}{l}{}      \\
\multicolumn{3}{c|}{(c) Proposed method for $\mathcal{O}=(Z=0,M=1,Y=1)$}                &                          &                          &                           \\\addlinespace[-2.5mm]
\multicolumn{3}{l|}{}                                                          & \multicolumn{1}{l}{}     & \multicolumn{1}{l}{}     & \multicolumn{1}{l}{}      \\
Synergistic              & Smoking                  & Immune                   &                          &                          &                           \\
$\pi_{0001}(\mathcal{O})=$0$\%$     & $\pi_{0011}(\mathcal{O})=$0$\%$     & $\pi_{0000}(\mathcal{O})=$0$\%$     &                          &                          &                           \\ \addlinespace[-2.5mm]
\multicolumn{1}{l}{}     & \multicolumn{1}{l}{}     & \multicolumn{1}{l|}{}    & \multicolumn{1}{l}{}     & \multicolumn{1}{l}{}     & \multicolumn{1}{l}{}      \\
Asbestos                 & Parallel                 & Doomed                   &                          &                          &                           \\
$\pi_{0101}(\mathcal{O})=$52.88$\%$ & $\pi_{0111}(\mathcal{O})=$29.36$\%$ & $\pi_{1111}(\mathcal{O})=$17.83$\%$ &                          &                          &                          
\end{tabular} 
  \end{threeparttable}}
\end{table}Since the estimation methods follow the same approach as in Sections  \ref{ssec:simulation-settings} and \ref{ssec:app-job}, we focus on interpreting the causal attribution probabilities using the true parameters. Table \ref{tab:est-res-updated} presents the posterior probabilities $\pi_{rstu}(\mathcal{O})$ under different evidence scenarios.

Table \ref{tab:est-res-updated}(a) shows the posterior distribution $\pi_{rstu}(\mathcal{O})$ when the evidence set is empty ($\mathcal{O}=\emptyset$).  The results reveal distinct population subgroups: 28.08\% are immune to all known exposures ($\pi_{0000}$); 16.38\% respond only to synergistic type ($\pi_{0001}$); 16.20\% are sensitive through parallel pathways ($\pi_{0111}$); 14.62\% are sensitive to smoking alone ($\pi_{0011}$); 13.59\% are doomed to develop cancer under all scenarios including the unexposed state ($\pi_{1111}$); and 11.13\% are sensitive to asbestos alone ($\pi_{0101}$).

Table \ref{tab:est-res-updated}(b) examines individuals who developed lung cancer despite no exposure to smoking or asbestos ($\mathcal{O} = (Z=0, M=0, Y=1)$). Since this requires $Y_{0,0} = 1$, only the doomed type ($G = 1111$) is consistent with this evidence, yielding $\pi_{1111}(\mathcal{O}) = 100\%$. This indicates that cancer in unexposed individuals must stem from genetic susceptibility or unmeasured environmental factors.

Table \ref{tab:est-res-updated}(c) considers patients exposed to asbestos but not smoking who developed cancer ($\mathcal{O} = (Z=0, M=1, Y=1)$), requiring $Y_{0,1} = 1$. Three cases are compatible: asbestos-sensitive ($G = 0101$, 52.88\%), parallel pathways ($G = 0111$, 29.36\%), and doomed ($G = 1111$, 17.83\%). The large proportion of asbestos-sensitive cases suggests that most cancers in this scenario result from asbestos-specific mechanisms. 

Table \ref{tab:est-res-updated}(d) analyzes smoking-exposed but asbestos-free cancer patients ($\mathcal{O} = (Z=1, M=0, Y=1)$), requiring $Y_{1,0} = 1$. Here, smoking-sensitive ($G = 0011$, 66.95\%), parallel pathways ($G = 0111$, 20.57\%), and doomed ($G = 1111$, 12.49\%) phenotypes are consistent. The dominance of smoking-sensitive indicates that smoking-specific mechanisms primarily drive cancer development in this scenario.

Table \ref{tab:est-res-updated}(e) presents the dual-exposure scenario ($\mathcal{O} = (Z=1, M=1, Y=1)$), where $Y_{1,1} = 1$. All phenotypes except immune are consistent, yielding: synergistic ($G = 0001$, 71.01\%), smoking-sensitive ($G = 0011$, 14.16\%), asbestos-sensitive ($G = 0101$, 7.84\%), parallel pathways ($G = 0111$, 4.35\%), and doomed ($G = 1111$, 2.64\%). Notably, the clear dominance of synergistic attribution reveals that under dual exposure, 71.01\% of cancer cases are attributable to synergistic interactions between smoking and asbestos, rather than the effects of either factor alone.
\subsection{Further analysis with secondary outcome}

\begin{figure}[h]
    \centering
    \includegraphics[width=0.95\linewidth]{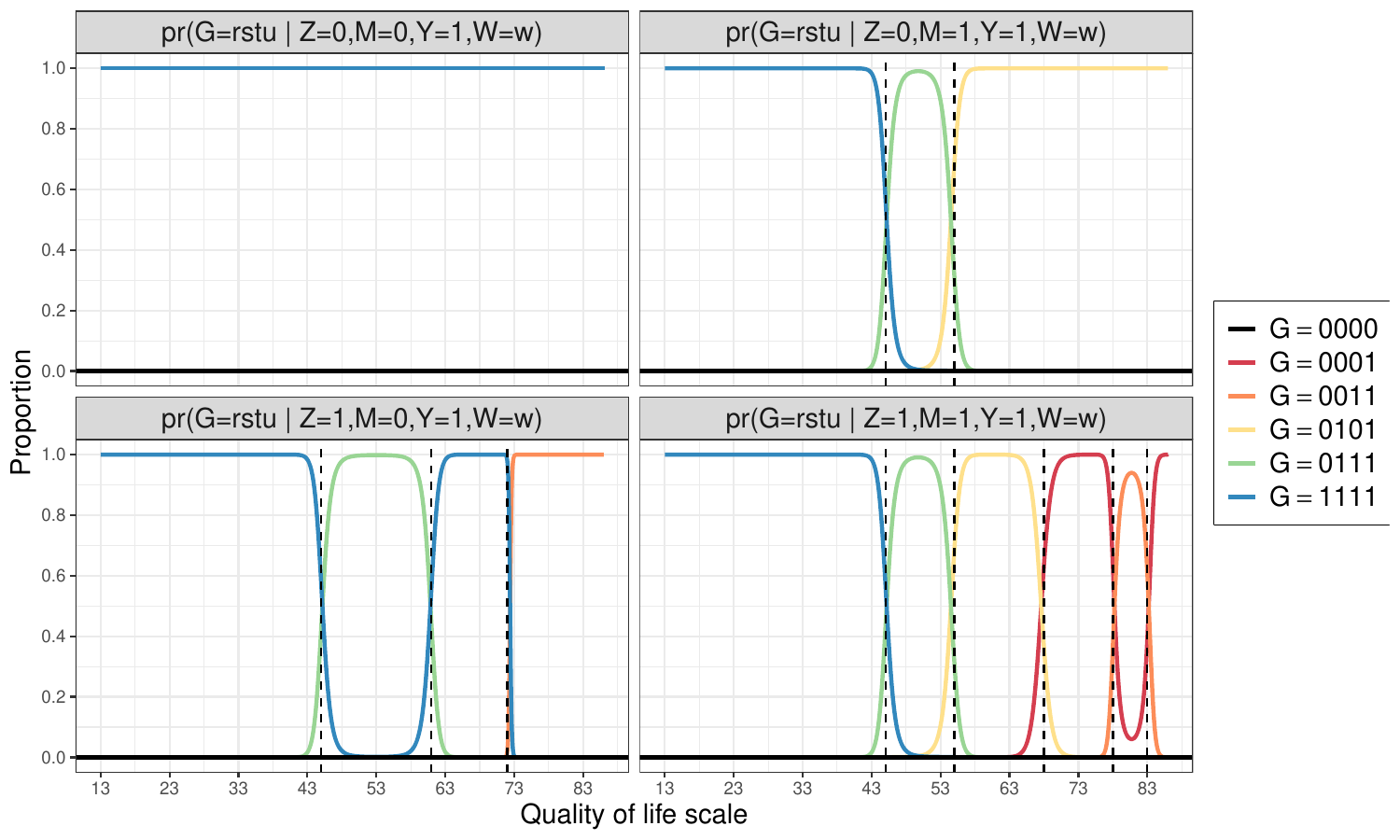}
    \caption{The posterior probabilities $\pr(G = rstu \mid Z = z, M = m, Y = 1, W)$ are plotted along the horizontal axis $W$ for four observed scenarios using QOLS as the secondary outcome. The vertical dashed lines mark the points where the two curves intersect. The thick black line represents the zero baseline on the vertical axis; for example, in the upper left panel, it covers all cases except the posterior probability $\pr(G = 1111 \mid Z = 0, M = 0, Y = 1, W)$.}
    \label{fig:cov-label-2}
\end{figure}

In this section, we further analyze the results incorporating the quality of life score $W$ as an additional secondary outcome variable in the evidence set for retrospective attribution analysis. For clarity, we plot the conditional probabilities $\pr(G={rstu} \mid Z=z, M=m, Y=1, W)$ for different $z, m, r, s, t, u \in \{0, 1\}$ in Figure \ref{fig:cov-label-2}, where the horizontal axis represents the quality of life score $W$.

In the upper left panel ($Z=0,M=0,Y=1$), $\pi_{1111}(\tilde{\mathcal{O}}) = 100\%$ remains constant across the entire quality of life score range, with no intersection points. This indicates that all individuals who developed lung cancer without exposure to smoking or asbestos belong to the doomed type ($G=1111$), with their cancer necessarily resulting from other unmeasured factors. Quality of life scores do not affect causal attribution for this patient group.

The upper right panel ($Z=0,M=1,Y=1$) shows two intersection points at $W=45$ and $W=55$. For $W<45$, $\pi_{1111}(\tilde{\mathcal{O}})$ dominates, indicating that asbestos-exposed patients with poor quality of life primarily belong to the doomed type, making them susceptible to multiple carcinogenic factors. In the interval $45<W<55$, $\pi_{0111}(\tilde{\mathcal{O}})$ becomes dominant, showing that patients with moderate quality of life primarily belong to the parallel pathways type, being sensitive to smoking, asbestos, or their interactions. When $W>55$, $\pi_{0101}(\tilde{\mathcal{O}})$ dominates, indicating that patients with better quality of life primarily exhibit asbestos-specific sensitivity, with cancer specifically attributed to asbestos exposure.

The lower left panel ($Z=1,M=0,Y=1$) presents three intersection points at $W=45$, $W=61$, and $W=72$. For $W<45$, $\pi_{1111}(\tilde{\mathcal{O}})$ dominates, indicating that smoking patients with poor quality of life primarily belong to the doomed type. In the $45<W<61$ interval, $\pi_{0111}(\tilde{\mathcal{O}})$ becomes dominant, showing that patients with moderate quality of life belong to the parallel pathways type. Interestingly, in the $61<W<72$ interval, $\pi_{1111}(\tilde{\mathcal{O}})$ regains dominance, and this bimodal pattern may reflect the complex nonlinear effects of quality of life on smoking carcinogenic mechanisms. When $W>72$, $\pi_{0011}(\tilde{\mathcal{O}})$ finally dominates, indicating that patients with high quality of life primarily exhibit smoking-specific sensitivity.

The lower right panel ($Z=1,M=1,Y=1$) displays the most complex pattern, with five intersection points at $W=45$, $W=55$, $W=68$, $W=78$, and $W=83$. For $W<45$, $\pi_{1111}(\tilde{\mathcal{O}})$ dominates, indicating that dual-exposed patients with very poor quality of life belong to the doomed type. As quality of life improves, the dominant attribution type transitions through parallel pathways ($\pi_{0111}(\tilde{\mathcal{O}})$ for $45<W<55$), asbestos-specific sensitivity ($\pi_{0101}(\tilde{\mathcal{O}})$ for $55<W<68$), and synergistic interactions ($\pi_{0001}(\tilde{\mathcal{O}})$ for $68<W<78$ and $W>83$). Notably, smoking-specific attribution ($\pi_{0011}(\tilde{\mathcal{O}})$) dominates in a narrow range $78<W<83$, suggesting that for patients with moderately high quality of life, cancer is primarily attributed to smoking effects alone.
  {\black \section{Additional Discussion of Interactive Causal Attribution}
\label{sec:practical-applications}
In this section, we provide additional discussion on interactive causal attribution that complements the introduction and discussion sections in the main text. The practical importance of interactive causal attribution extends far beyond theoretical interest, as the need for quantitative causal attribution in multi-exposure settings has been explicitly recognized in legal contexts involving occupational disease compensation. 

Prospective causal inference focuses on predicting what \emph{will happen} under alternative interventions in future populations, addressing questions such as: ``Should we implement workplace policies to eliminate both smoking and asbestos exposure to reduce future lung cancer incidence?'' This forward-looking perspective evaluates the effects of causes through standard treatment effect estimands, informing policy design and resource allocation for preventing future outcomes. In contrast, retrospective attribution analysis focuses on understanding what \emph{has happened} given observed evidence, addressing questions such as: ``Given that a worker has developed lung cancer after both smoking and asbestos exposure, what is the probability that the synergistic effect of these two factors caused it?'' This backward-looking perspective infers the causes of already occurred outcomes and is essential for legal liability determination, compensation allocation, and etiological understanding.

Notably, conclusions drawn from these two perspectives can differ substantially. 
In the field of labor economics, \citet{zhang2025identifying} analyzed the Lalonde data to study the causal effects of job training programs on income. From a prospective perspective, job training showed positive causal effects, increasing both the proportion of individuals achieving medium income and high income. However, retrospective attribution analysis revealed important heterogeneity: among individuals who achieved high income after training, the majority (83\%) would have achieved high income even without training, indicating that high income is not primarily attributed to the training.

Our interactive attribution method is also applicable to the following two practical scenarios.

The U.S.\ Black Lung Benefits Act, enacted in 1969 following advocacy by the United Mine Workers \citep{fox1980black}, provides monthly payments and medical benefits to coal miners totally disabled from pneumoconiosis arising from coal dust exposure. Under this Act, there exists a rebuttable presumption that pneumoconiosis resulted from coal mine employment for long-term miners. However, when miners have multiple occupational exposures (e.g., coal dust from different employers, or coal dust combined with smoking), retrospective attribution becomes critical for fair compensation allocation. The fairness of these administrative proceedings has been questioned due to the difficulty in accurately attributing responsibility when workers face multiple potential exposure sources. Our interactive attribution framework quantifies the probability that the disease results from synergistic effects of multiple exposures versus individual exposures, providing an objective basis for apportioning benefits among multiple responsible parties.

Similarly, asbestos litigation represents the longest-running mass tort in U.S.\ history, involving more than 8000 defendants and 700000 claimants \citep{maines2020asbestos}. As noted by British Member of Parliament Michael Wills \citep{hansard2006}, a central challenge is that ``Many of those who I see have worked in a number of workplaces and they could have been exposed to asbestos in each of them, but medical science is such that no one can identify which of them it is. As a result, there has been a long and complex history of legal discussion on how to apportion liability.'' When a worker was exposed to asbestos from multiple employers and also smoked, courts must retrospectively determine whether liability should be joint (if synergistic causation) or several (if independent causation), and how to apportion damages among defendants. Our method provides quantitative probabilities for these scenarios. This type of quantitative evidence can inform judicial decisions and provide a principled basis for liability apportionment that has been lacking in current legal practice.}
\end{document}